\documentclass[aps,prl,twocolumn,superscriptaddress,showpacs,aps,longbibliography]{revtex4-1}

\pdfoutput=1
\usepackage[titletoc,toc,title]{appendix}
\usepackage{amsmath,amssymb}
\usepackage{graphicx}
\usepackage{color}
\usepackage{rotating}
\usepackage{bm}
\usepackage{setspace}
\usepackage{hyperref}
\usepackage{float}
\usepackage{soul}
\usepackage[T1]{fontenc}
\usepackage{bbold}
\usepackage{color}
\usepackage{braket}
\usepackage{physics}
\usepackage{subcaption}
\usepackage{scrextend}

\hypersetup{
colorlinks=true,
urlcolor= blue,
citecolor=blue,
linkcolor= blue,
bookmarks=true,
bookmarksopen=false,
}

\renewcommand{\vec}[1]{\bm{#1}}

\newcommand{\vecl}[2]{\vec{#1}^{(#2)}}
\newcommand{\vecll}[1]{\vec{#1}^{(\ell)}}
\newcommand{\vecli}[1]{\vec{#1}^{(i)}}
\newcommand{\veclj}[1]{\vec{#1}^{(j)}}

\newcommand{\kpt}[1]{K_{\mathrm{L}{#1}}}

\begin{document}


\title{Twisted Trilayer Graphene: a Precisely Tunable Platform for Correlated Electrons}

\author{Ziyan Zhu}
\affiliation{Department of Physics, Harvard University, Cambridge, Massachusetts 02138, USA}
\author{Stephen Carr}
\affiliation{Department of Physics, Harvard University, Cambridge, Massachusetts 02138, USA}
\author{Daniel Massatt}
\affiliation{Department of Statistics, The University of Chicago, Chicago, Illinois 60637, USA}
\author{Mitchell Luskin}
\affiliation{School of Mathematics, University of Minnesota - Twin Cities, Minneapolis, Minnesota 55455, USA}
\author{Efthimios Kaxiras}
\affiliation{Department of Physics, Harvard University, Cambridge, Massachusetts 02138, USA}
\affiliation{John A. Paulson School of Engineering and Applied Sciences, Harvard University, Cambridge, Massachusetts 02138, USA}

\begin{abstract}
We introduce twisted trilayer graphene (tTLG) with two independent twist angles as an ideal system for the precise tuning of the electronic interlayer coupling to maximize the effect of correlated behaviors. 
As established by experiment and theory in the related twisted bilayer graphene system, 
van Hove singularities (VHS) in the density of states can be used as a proxy of the tendency for correlated behaviors. 
To explore the evolution of VHS in the twist-angle phase space of tTLG, 
we present a general low-energy electronic structure model for any pair of twist angles. 
We show that the basis of the model has infinite dimensions even at a finite energy 
cutoff and that no Brillouin zone exists even in the continuum limit. 
Using this model, we demonstrate that the tTLG system exhibits a wide range of magic angles at which VHS merge and the density of states has a sharp peak at the charge-neutrality point
through two distinct mechanisms: 
the incommensurate perturbation of twisted bilayer graphene's flatbands 
or the equal hybridization between two bilayer moir\'e superlattices. 
\end{abstract}

\maketitle

{\it Introduction. ---} Electronic properties in stacked graphene layers can be tuned by a small twist angle that modifies the interlayer interaction strength, an effect referred to as ``twistronics''~\cite{carr2017twistronics}. 
As the twist angle approaches a critical ``magic angle'' 
($\sim1.05^\circ$ in twisted bilayer graphene), the two van Hove singularities (VHS) 
in the density of states (DOS) of each monolayer merge, 
resulting in a sharp peak associated with flatbands, 
leading to the emergence of strongly correlated electronic phases~\cite{bistritzer2011moire}. 
The small twist angle gives rise to large-scale repeating patterns, known as moir\'e patterns. 
Unconventional correlated states have now been observed in many van der Waals (vdW) heterostructures with one twist angle, 
e.g., twisted bilayer graphene (tBLG) and twisted double bilayer graphene~\cite{cao2018correlated,cao2018unconventional,chen2018evidence,yankowitz2018dynamic,yankowitz2019tuning,shen2019observation,liu2019spin,cao2020electric,yankowitz2019tuning,william2019correlated,wang2019wse2,yu2019decoupling,zhang2020probing,liu2020tuning}. In these systems, electrons responsible for the correlation effects localize at the moir\'e scale~\cite{koshino2018maximally,po2019faithful,carr2019derivation}.

\begin{figure}[h]
    \centering
    \includegraphics[width=\linewidth]{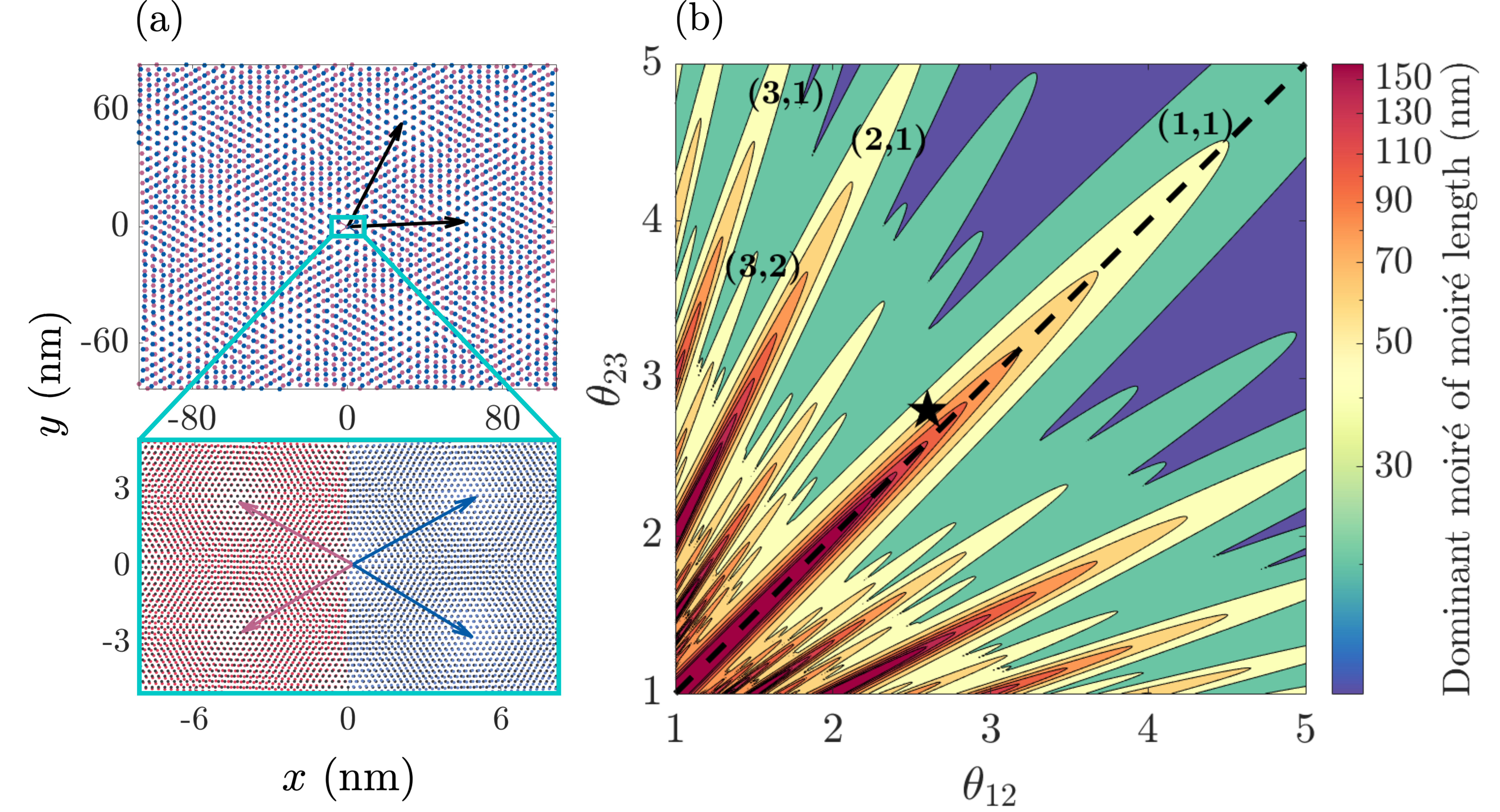}
    \caption{Illustration of 
moir\'e of moir\'e pattern in tTLG for 
$\theta_{12}=2.6^\circ$, $\theta_{23}=2.8^\circ$. 
Red and blue points represent the lattice points of the bilayer moir\'e supercells 
between L1-L2 and L2-L3 respectively.
Black arrows represent dominant moir\'e of moir\'e supercell lattice vectors. 
A blowup of the small boxed area is shown below, with 
points representing monolayer lattice points, 
for L1 and L2 on the left half and for L2 and L3 on the right half. 
The moir\'e lattice vectors (red and blue arrows)
 are slightly rotated and have different lattice constants.
(b) The dominant moir\'e of moir\'e length on a logarithmic color scale. The black star corresponds to the twist angle in (a), and $(m,n)$ labels the dominant moir\'e of moir\'e harmonic in the nearby lobe. Black dashed line represents $\theta_{12} = \theta_{23}$.}
    \label{fig:moire}
\end{figure}

The addition of a third layer introduces a new degree of freedom, the second twist angle, 
allowing for the further tuning of electron correlations. 
In twisted trilayer graphene (tTLG) with two consecutive twist angles, $\theta_{12}$ and $\theta_{23}$, the beating of two bilayer moir\'e patterns leads to higher-order patterns (moir\'e of moir\'e).
The length scale of these is orders of magnitude larger than the bilayer moir\'e [Fig.~\ref{fig:moire}(a)]~\cite{zhu2019moire,andelkovic2019double,leconte2020commensurate}. 
Unlike in tBLG where only the lowest-order moir\'e pattern dominates in the continuum limit, the dominant harmonic is twist-angle dependent in tTLG.
For a given moir\'e of moir\'e harmonic labeled by $(m, n)$, the primitive reciprocal lattice vectors are given as the column vectors of $G_{mn}^\mathrm{H} = mG_{12}-n G_{23}$, where the matrix $G_{ij}$ spans the bilayer moir\'e reciprocal space between layers $i$ (L$i$) and $j$ (L$j$). 
The real space moir\'e of moir\'e supercell $A_{mn}^\mathrm{H}$ is obtained by $A_{mn}^\mathrm{H}=\frac{1}{2\pi} (G_{mn}^\mathrm{H})^{-T}$, with the norm of its column vectors being the moir\'e of moir\'e length. 
Figure~\ref{fig:moire}b shows the dominant moir\'e of moir\'e length as a function of twist angles, in which each lobe corresponds to the region where a different harmonic $(m,n)$ dominates. The moir\'e of moir\'e patterns can be discerned visually only near the $(N,1)$ or $(1,N)$ lobes for $N \in \mathbb{Z}$. Generally, multiple harmonics have competing length scales
(see Supplemental Material Sec.~I~\cite{zhu2020sm}).
Therefore, tTLG cannot be approximated by two aligned tBLG and a general expression for the trilayer supercell does not exist, making it fundamentally different than multilayered vdW heterostructures with a single twist angle~\cite{li2019trilayer,carr2020ultraheavy,chen2020electrically,park2020gate}.
The lack of a supercell approximation and the large length scale pose many computational 
challenges to the theoretical modeling of tTLG. While there have been some theoretical studies 
of tTLG~\cite{amorim2018electronic,mora2019flat,zhu2019moire}, including 
an accurate treatment of any twist angles by~\citet{amorim2018electronic}, 
an electronic structure model incorporating both accuracy and efficiency is lacking; 
this severely restricts our ability to 
investigate its electronic properties and the potential for correlated phases, which have been observed recently in tTLG at the moir\'e of moir\'e scale~\cite{tsai2019correlated}. 

Here, we present tTLG as a platform to precisely tune twistronic correlations, 
using the VHS intensity as a proxy for strong correlations. 
We derive a general momentum-space model to study the electronic properties 
of the two-independent-twist-angle tTLG system 
using a low-energy $k \cdot p$ model that provides computational efficiency and removes the constraint on the twist angle in atomistic calculations with supercells. 
Using this model, we explore the tTLG phase space. We find that the two bilayer moir\'e superlattices hybridize when the two twist angles are equal, minimizing the separation between the two lowest VHS at a critical angle. 
At general twist angles, there exists a wide range of values 
at which the VHS merge and the DOS is sharply peaked at the charge-neutrality point (CNP). 
These magic angles can be understood as a tBLG magic angle modified 
by an incommensurate perturbative potential from the third layer. 
Our analysis is well suited to guide experimental searches for correlation effects 
and enables the interpretation of otherwise unclear experimental findings~\cite{tsai2019correlated}.

{\it Momentum-space Hamiltonian. ---} 
To obtain the electronic structure model for tTLG, 
we employ a momentum-space method by taking the Fourier 
transform of the real-space tight-binding model. 
At a momentum $\vec{k}$ (referred to as the center site), 
the model can be formally represented by a $3\times3$ block matrix: 
\begin{align}
&\mathcal{H} (\vec{k}) =\begin{bmatrix}
H^1(\vec{k}) & T^{12}  & 0 \\
T^{12 \dagger} & H^2(\vec{k}) & T^{23}  \\
0 & T^{23\dagger} & H^3(\vec{k})
\end{bmatrix}.
 \label{eqn:hamiltonian}
 \end{align}
The diagonal blocks are the monolayer graphene tight-binding Hamiltonians in the rotated basis~\cite{castro_neto2009electronic}, representing the intralayer hopping. 
The off-diagonal blocks represent the interlayer hopping. 
The interlayer terms that connect two momentum degrees of freedom 
$\vecli{k}$ and $\veclj{k}$ in L$i$ and L$j$ are given as,
\begin{align}
   & T_{\alpha\beta}^{ij} [\vecli{k}, \veclj{k}]
    = \frac{1}{|\Gamma|} \sum_{\vecli{G}, \veclj{G}}e^{i\vecli{G} \cdot \vecli{\tau}_\alpha}  \nonumber\\
   & \times \tilde{t}_{\alpha\beta}^{ij} [\vec{k}+\vecli{k}+\vecli{G}] e^{-i\veclj{G} \cdot \vecli{\tau}_\beta}
    \delta_{\vecli{k} - \vecli{G}, \veclj{k} - \veclj{G}},
    \label{eqn:inter}
\end{align}
where $|\Gamma|$ is the monolayer unit cell area, $\mathbf{\tau}_{\alpha}$ ($\mathbf{\tau}_{\beta}$) is the position of the sublattice $\alpha$ ($\beta$), 
$\vecll{G}$ is a reciprocal space lattice vector in L$\ell$, 
and $\tilde{t}^{ij}_{\alpha\beta}(\vec{p})$ is the momentum-space hopping parameter 
between sublattice $\alpha$ in L$i$ and sublattice $\beta$ in L$j$. 
The $\delta$ function imposes the constraint on the values of $\vecll{k}$, dictating the interlayer scattering selection rule. 
The above expressions are equivalent to a real-space tight-binding model in the Bloch basis
(See Supplemental Material, Sec.~IIA, for derivation~\cite{zhu2020sm}).

Unlike tBLG~\cite{nguyen2017lattice,carr2019exact,fang2019angle,guinea2019continuum,leconte2019relaxation}, the momentum-space basis in tTLG is infinitely dimensional and lacks a Brillouin zone even in the continuum limit.
In bilayers, coupled momentum states satisfy the selection rule $\vec{k}^{(1)} - \vec{k}^{(2)} =  \vec{G}^{(1)} - \vec{G}^{(2)}$~\cite{fang2019angle}. 
Note that for a given $\vecl{G}{1}=m\vecl{b}{1}_1 + n \vecl{b}{1}_2$ for $m,n\in\mathbb{Z}$, we also have $\vecl{G}{2}=m\vecl{b}{2}_1+n\vecl{b}{2}_2$ for the same $m,n$, where $\vecll{b}_i$ is the $i$-th component of the primitive reciprocal lattice vector of L$\ell$, since other hopping processes are much higher in energy.
As $|\vecll{G}|$ increases, the scattered momentum, $\vec{k}'$,
becomes farther away from the Dirac point.
Therefore, to implement a finite cutoff, we can simply constrain the magnitude of the scattered momentum 
$\vec{k}'=\vecl{G}{\ell}$ for $\ell=1,2$. 
Physically, $\vec{k}'$ is a monolayer reciprocal lattice vector 
that can scatter to a nearby momentum in the other layer. 
In contrast, in trilayers, the momentum states that form the basis of the Hamiltonian are connected in a more complicated way.
A given $\vecl{k}{1}$ can couple to a momentum state $\vecl{k}{2}$ that 
satisfies $\vecl{k}{2} = \vecl{k}{1} + \vecl{G}{2} - \vecl{G}{1}$, same as in bilayers. 
Each $\vecl{k}{2}$ can then couple to a momentum state $\vec{k}^{(3)}$ through the second selection rule [Eq.~\eqref{eqn:inter}], resulting in the following final momentum:
\begin{align}\label{eqn:selection}
\vecl{k}{3} = \vecl{k}{1} + [\vec{G}^{(2)} - \vec{G}^{(1)}] + [\vec{G}^{(3)} - \vec{G}'^{(2)}],
\end{align}
where the reciprocal lattice vectors satisfy $\vec{G}^{(2)} - \vec{G}^{(1)} = m\vecl{b}{12}_1 +n\vecl{b}{12}_2$ and $\vec{G}^{(3)} - \vec{G}'^{(2)} = m'\vecl{b}{23}_1 +n'\vecl{b}{23}_2$ for $m,n,n',m'\in \mathbb{Z}$, 
with $\vecl{b}{ij}_k=\vecl{b}{j}_k-\vecl{b}{i}_k$ being the bilayer moir\'e 
reciprocal space lattice vectors. 
Equation~\eqref{eqn:selection} suggests that L1 and L3 are coupled through L2, even though a direct interlayer hopping is not allowed. Unlike the simple 2D momentum crystal in bilayers, here the incommensuration between $\vec{b}_k^{(12)}$ and $\vec{b}_k^{(23)}$ creates for $\vec{k}^{(3)}$ a 4D structure that is projected onto 2D.

\begin{figure}[ht!]
\centering
\includegraphics[width=\linewidth]{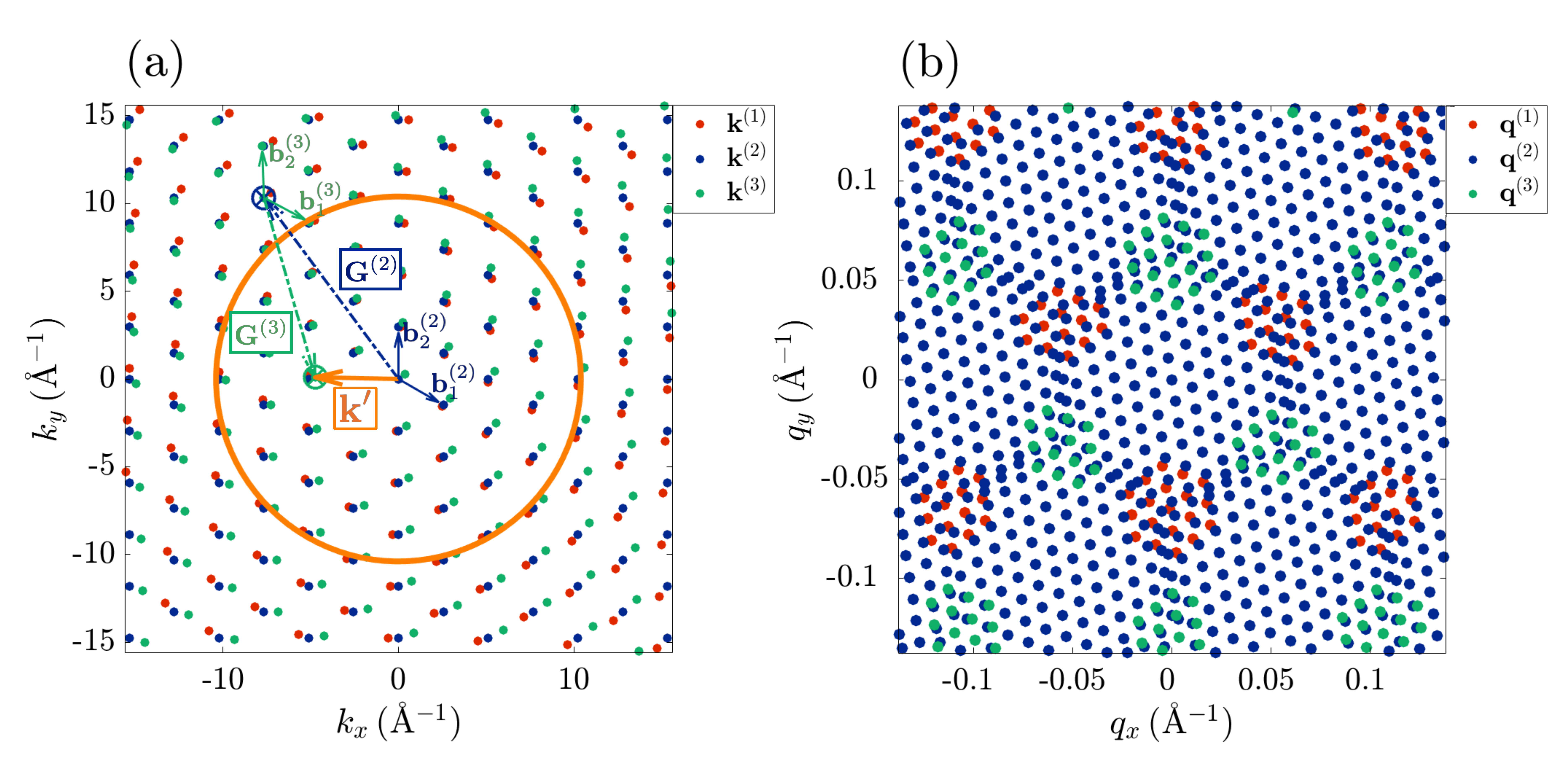}
\caption{Momentum degrees of freedom for tTLG at $\theta_{12} = 2.2^\circ, \theta_{23}=2^\circ$. Red, blue, and green are the reciprocal lattice vectors of L1, L2, and L3 respectively. The origin is the Dirac point of L2. (a) Extended zone scheme, with the orange circle indicating the cutoff in 
$|\vec{k}'|$. $\vec{k}' = \vec{G}^{(2)}+\vec{G}^{(3)}$ falls within the cutoff radius $10 \text{\AA}^{-1}$ despite both $|\vec{G}^{(2)}|$ and $|\vec{G}^{(3)}|$ being large. The momenta of L3 are centered at $\vec{G}^{(2)}.$
(b) Reduced zone scheme, folded back to the monolayer Dirac points, 
$\vecll{q} = \vecll{k} - K_\mathrm{L\ell}$. This basis corresponds to the same twist angle as (a) but with an additional constraint $|\vecll{G}| \leq k_c = 6 |\vecll{b}|$, leading to 26\,921 momenta. }
\label{fig:kdof}
\end{figure}

Equation~\eqref{eqn:selection} suggests that in L$\ell$ of tTLG, $\vec{k}'$ is given by 
$\vec{k}' = \vecli{G} + \veclj{G}$ for $\ell \neq i, j$.
To implement a cutoff, we should impose $|\vec{k}'| \leq k_{c}$ 
for some cutoff value $k_c$. 
However, the incommensurability of twisted trilayers suggests that
$|\vec{k}'|$ can be arbitrarily small and imposing 
$|\vec{k}'| \leq k_{c}$ still leads to an infinite basis. 
For example, in Fig.~\ref{fig:kdof}(a), even though $\vecl{G}{2}$ lies outside 
of the cutoff, the resulting $\vec{k}'$ 
is still a relevant low-energy degree of freedom, due to the two-step scattering process.
A similar construction can be made for all other $\vecl{G}{2}$ outside of the cutoff radius, which means within a finite cutoff, there are infinitely many coupled momentum states.
In practice, another set of cutoff conditions needs to be implemented, 
namely $|\vecl{G}{\ell}| \leq k_{c}$. With the constraint on $|\vecl{G}{\ell}|$, the $\vec{k}'$ in Fig.~\ref{fig:kdof}(a) is no longer allowed. 
In this way, we ignore the cases where $|\vecll{G}|$ is large 
but $|\vec{k}'|$ is small, leading to 
the neglect of some low-energy degrees of freedom and hence convergence is not guaranteed,
which merits future work (see Supplemental Material, Sec.~IID, for convergence study~\cite{zhu2020sm}). 
In this work, we choose $k_c = 4|\vecll{b}|$, with $\sim 5600$ momenta, such that the properties of interest (e.g., DOS maximum and the VHS location) 
do not change significantly as $k_c$ increases.

We take the low-energy limit by expanding around the Dirac point of each layer, $\kpt{\ell}$, letting $\vecll{k} = \vecll{q} + \kpt{\ell}$, which simplifies the model proposed by~\citet{amorim2018electronic}. 
The intralayer Hamiltonian becomes the rotated Dirac equation, $H^\ell = v_F \vec{q} \cdot (\sigma_x^{\theta_{\ell}}, -\sigma_y^{\theta_{\ell}})$, where
$\sigma_x^{\theta_\ell} = \sigma_x \cos \theta_\ell - \sigma_y \sin \theta_\ell$ and $\sigma^{\theta_\ell}_y = \sigma_x \sin \theta_\ell + \sigma_y \cos \theta_\ell$ are rotated Pauli matrices with $\theta_1 = \theta_{12}, \theta_2 = 0, \theta_3 = -\theta_{23}$,
$v_F = 0.8\times10^6$ cm/s is the Fermi velocity~\cite{fang2016electronic}, 
and $\vec{q}=\vec{k}+\vecll{k}-K_{L\ell}$.
For the interlayer hopping, we make the approximation that $\tilde{t}^{ij}_{\alpha\beta}[\vec{k}+\vecli{k}+\vecli{G}] \approx \tilde{t}^{ij}_{\alpha\beta} [\vecli{G}+\kpt{i}]$ since $|\vec{k}|, |\vecli{q}| \ll |\kpt{i}|,|\vecli{G}|$, for $\vec{k}$ near the Dirac point. 
Due to the rapid decay of $\tilde t_{\alpha\beta}^{ij}(\bm p)$ as $\bm p$ increases~\cite{bistritzer2011moire,fang2016electronic,catarina2019twisted}, we keep only the first shell in the summation in Eq.~\eqref{eqn:inter}:
\begin{align}
    T_{\alpha\beta}^{ij} [\vecli{q}, \veclj{q}]  = \sum_{n=1}^3 T_{n, \alpha\beta}^{ij} \delta_{\vecli{q}-\veclj{q}, -\vec{q}_n^{ij}}, 
\end{align}
where $\vec{q}_1^{ij} = \kpt{i} - \kpt{j}$, $\vec{q}_2^{ij} = \mathcal{R}^{-1}(2\pi/3) \vec{q}_1^{ij}$, and $\vec{q}_3^{ij} = \mathcal{R}(2\pi/3) \vec{q}_1^{ij}$ using a counterclockwise rotation matrix $\mathcal{R}(\theta)$. We include out-of-plane relaxation by letting $t^{ij}_{AA} = t^{ij}_{BB} = \omega_0 = 0.07 \, \mathrm{eV} $ and $t^{ij}_{AB} = t^{ij}_{BA} = \omega_1 = 0.11 \, \mathrm{eV}$ ~\cite{nam2017lattice,carr2019exact}. In the matrix form, 
\begin{align}
    T^{ij}_1 = \begin{bmatrix} 
    \omega_0 & \omega_1 \\
    \omega_1 & \omega_0
    \end{bmatrix},  
    T^{ij}_2 = \begin{bmatrix} 
    \omega_0 & \omega_1\bar{\phi} \\
    \omega_1 \phi & \omega_0
    \end{bmatrix}, 
  \; T^{ij}_3 = \bar{T}^{ij}_2,
\end{align}
where $\phi = \exp(i 2 \pi/3)$ and $\bar{z}$ indicates the complex conjugate of $z$.
In tBLG, with the low-energy expansion, momenta $\vecl{q}{1}$ and $\vecl{q}{2}$ form a hexagonal lattice with the neighboring hexagon corners representing states from alternating layers (a moir\'e momentum lattice)~\cite{bistritzer2011moire,fang2019angle}. 
In tTLG, on top of each lattice point of the L1-L2 moir\'e momentum lattice, the additional scattering process creates a copy of the L2-L3 moir\'e momentum lattice (Fig.~\ref{fig:kdof}b), 
suggesting the absence of a Brillouin zone.

{\it Density of states ---} 
We use Gaussian smearing to obtain the total DOS, summing over the two bilayer moir\'e Brillouin zones, each discretized using a $22\times22$ grid~\cite{massatt2017incommensurate}
(see Supplemental Material Sec.~IIC for the expression~\cite{zhu2020sm}). 
For normalization, we first calculate the DOS of only the intralayer 
Hamiltonian, which reduces to three independent copies of monolayer graphene~\cite{castro_neto2009electronic}. 
We then obtain the normalization constant by fixing the prefactor to the expected low-energy monolayer DOS and using the same constant for the DOS of the full Hamiltonian. We adapt the Gaussian FWHM, $\kappa$, based on the twist angle $\theta_{\ell,\ell+1}$: for $\theta_{\ell,\ell+1}\leq2^\circ$, $\kappa=0.35\,\mathrm{meV}$; for $\theta_{\ell,\ell+1}\in(2^\circ, 3.9^\circ], \kappa=1.2\,\mathrm{meV}$; for $\theta_{\ell,\ell+1}>3.9^\circ$, $\kappa=2.4\,\mathrm{meV}$.

\begin{figure}[ht!]
\centering
    \includegraphics[width=\linewidth]{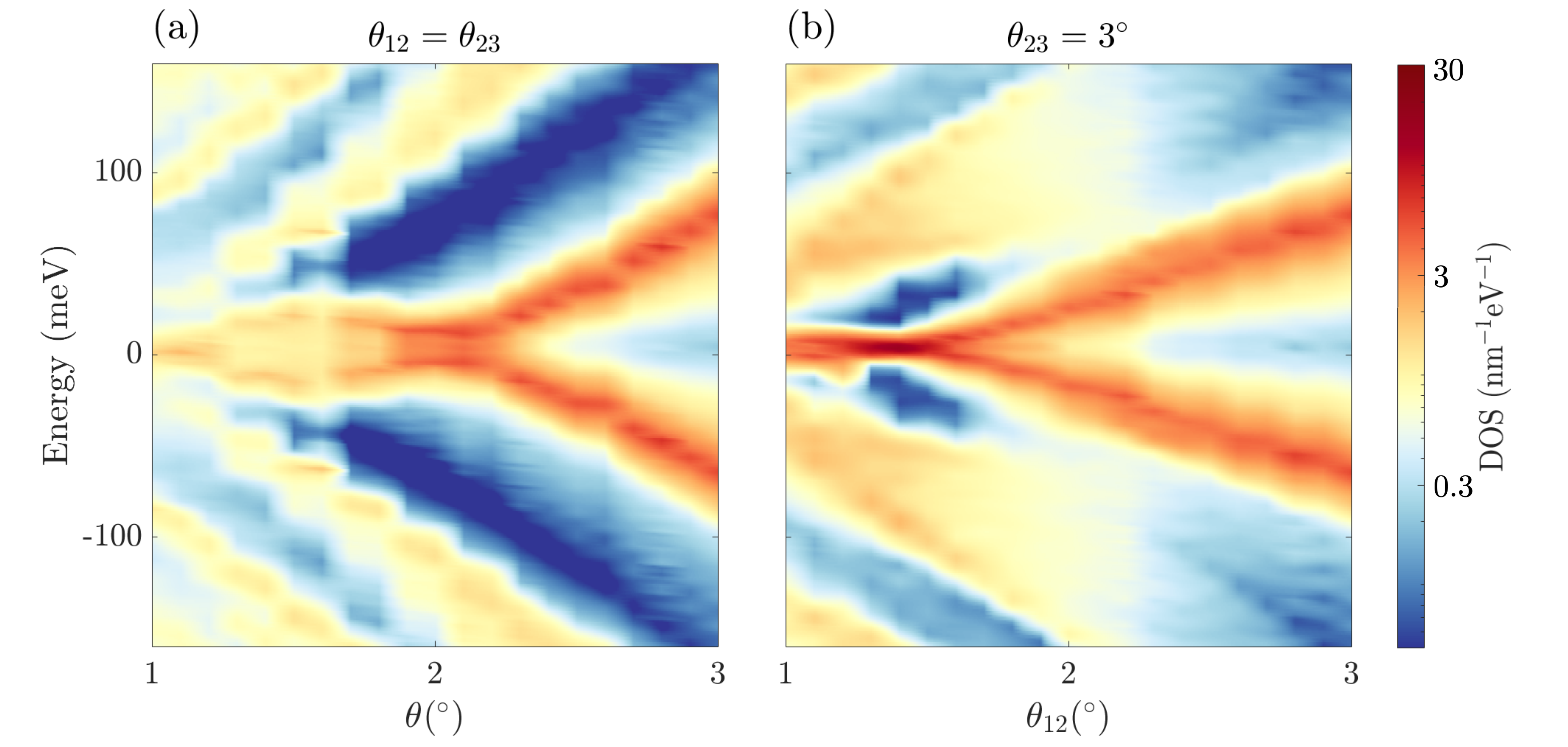}
    \caption{(a) DOS as a function of twist angle for $\theta_{12}=\theta_{23}$. (b) DOS as a function $\theta_{12}$ at $\theta_{23}=3^\circ$, both on a logarithmic color scale.
}
    \label{fig:dos}
\end{figure}

{\it Evolution of VHS. ---} 
We explore next the behavior of VHS as a function of twist angles in tTLG, 
by investigating the DOS enhancement and the narrowing of the separation 
between VHS (referred to as the VHS gap).
We define a magic angle approximately as 
a geometry where both features are achieved.
Figure~\ref{fig:dos}(a) shows the DOS of tTLG at $\theta_{12}=\theta_{23}$. The bright regions represent VHS. As the twist angle decreases, the VHS gap first decreases and then increases after reaching a minimum at $\sim 2.1^\circ$.
This behavior is similar to the evolution of VHS in tBLG in which changing the twist angle tunes the hybridization between two monolayer Dirac cones. 
In tTLG with $\theta_{12}=\theta_{23}$, varying the twist angle changes the hybridization strength between the two identical bilayer moir\'e superlattices. However, the two VHS can never merge at the CNP, with the minimum VHS gap being $\sim20$~meV at $2.1^\circ$. 
The DOS is also orders of magnitude lower than at the tBLG magic angle.
For general twist angles, Fig.~\ref{fig:dos}(b) shows the DOS as a function of $\theta_{12}$ with $\theta_{23}=3^\circ$. Unlike when $\theta_{12}=\theta_{23}$, the two VHS approach each other as the twist angle decreases and merge when $1.3^\circ \leq \theta_{12}\leq1.6^\circ$, resulting in a sharp DOS peak.

\begin{figure}[ht!]
\centering
    \includegraphics[width=\linewidth]{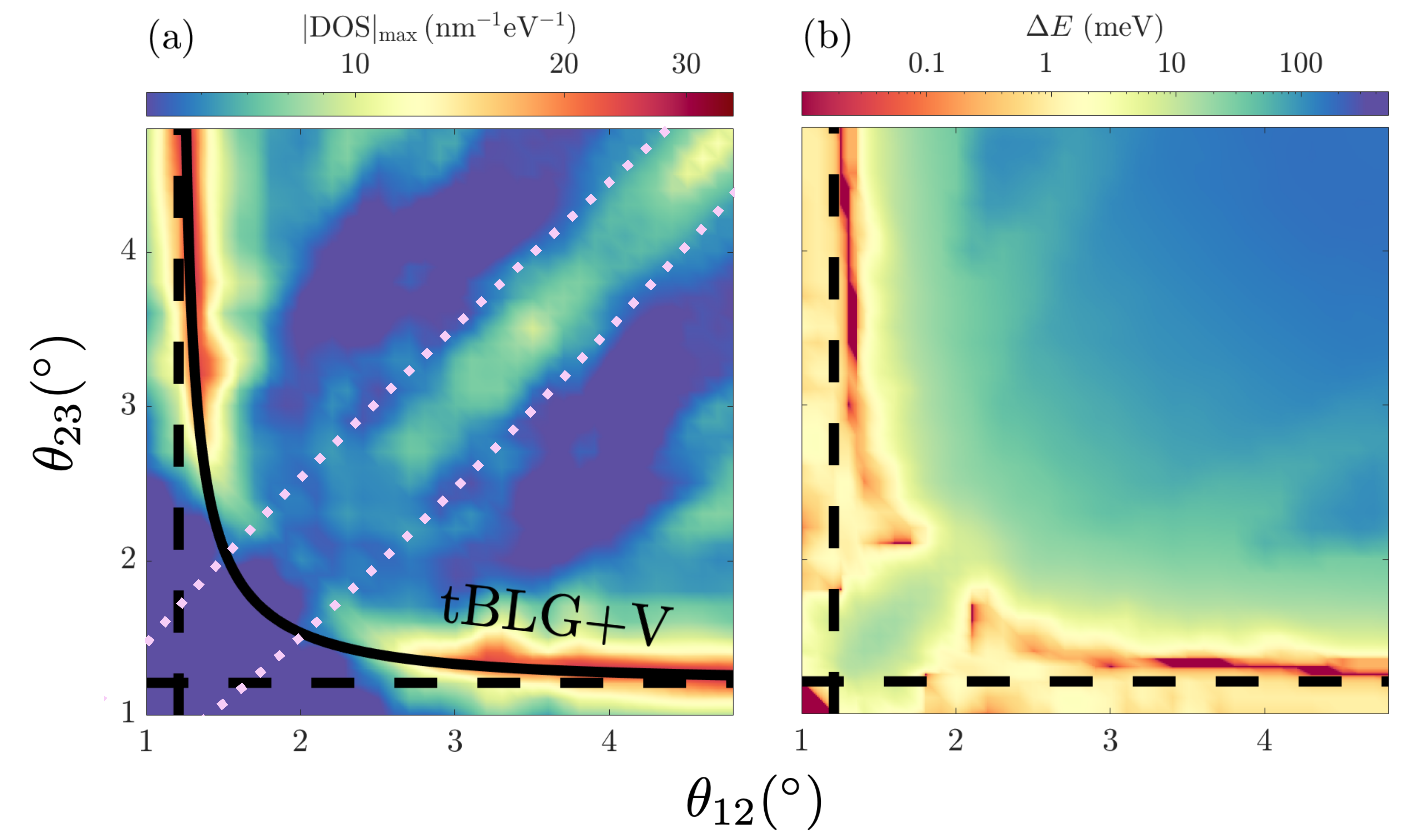}
    \caption{(a) DOS maximum and (b) VHS gap, $\Delta E$, as a function of twist angles on a logarithmic color scale. The black solid line follows the tTLG magic angles predicted by Eq.~\eqref{eqn:ma}. Vertical and horizontal black dashed lines correspond to $\theta_{12}$ and $\theta_{23}$ at the tBLG magic angle respectively. Within the dotted lines is roughly the region that can be understood as the hybridization between two bilayer moir\'e superlattices.
}
    \label{fig:sweep}
\end{figure}

To investigate the nature of DOS enhancements in tTLG, we performed calculations over an entire region of the $\theta_{12},\theta_{23}$ parameter space. Figure~\ref{fig:sweep} shows the DOS maximum and the VHS gap, $\Delta E$, as a function of both twist angles~\footnote{\label{note1}The data set consists of a $43\times43$ sampling. A Gaussian convolution kernel is applied for smoothening.}.
The magic-angle condition is met at a wide range of twist angles that follows a smooth curve but disappears near the diagonal. Although there is no significant DOS enhancement at $\theta_{12}=\theta_{23}$, the DOS maximum is higher compared to the nearby regions where $\theta_{12}$ and $\theta_{23}$ differ slightly [light yellow region within the dotted lines in Fig.~\ref{fig:sweep}(a)].

We now examine the magic angles away from the diagonal.
In the limit where $\theta_{12} \gg \theta_{23}$ or $\theta_{12} \ll \theta_{23}$, 
tTLG decomposes into a decoupled tBLG moir\'e supercell and a graphene monolayer; 
the monolayer does not contribute significantly to the low-energy features. 
Therefore, we observe that the tTLG magic-angle curve
asymptotically approaches the tBLG magic angle [dashed lines in Fig.~\ref{fig:sweep}(a)] for large $\theta_{12}$ or $\theta_{23}$.
We verified numerically that when one twist angle is very large, $\theta_{12} = 40^\circ$ for instance, the DOS maximum occurs exactly when $\theta_{23}$ is at the tBLG magic angle. The continuous curve and its asymptotic behavior suggest that these magic angles can be understood as the magic-angle tBLG modified by an effective potential, $V$, from the third layer. 
We can qualitatively analyze this argument using perturbation theory, by truncating the momentum space to the first shell, including one state from L2
and three states each from L1 and L3. 
We obtain the renormalized Fermi velocity $v^*_F$ by extracting the coefficient of the first-order effective Hamiltonian in $\vec{q}$ 
in the form of a Dirac Hamiltonian, given by
\begin{equation}\label{eqn:vstar}
    v^*_F = \frac{1-3(\alpha_{12}^2+\alpha_{23}^2)}{1+6(\alpha_{12}^2+\alpha_{23}^2)} v_F,
\end{equation}
where $\alpha_{ij} = \omega /(v_F k_{\theta_{ij}})$, $k_{\theta_{ij}} = 8\pi \sin(\theta_{ij}/2) / (3 a_G)$, 
assuming that $\omega_0 = \omega_1 = \omega$. The Hamiltonian and its derivation are provided in Section III of the Supplemental Material ~\cite{zhu2020sm}.
Magic angles occur when $v^*_F$ vanishes, leading to the following condition:
\begin{equation}
    \alpha_{12}^2+\alpha^2_{23} = \frac{1}{3}.\label{eqn:ma}
\end{equation}
The solid line in Fig.~\ref{fig:sweep}(a) corresponds to
$\theta_{12}$ and $\theta_{23}$ that satisfy Eq.~\eqref{eqn:ma}, which matches the DOS peaks and $\Delta E$ minima 
in Fig.~\ref{fig:sweep}(a), (b). 
Taking the large angle limit, for example, when $\theta_{23} \rightarrow \infty$, $\alpha_{23} \rightarrow 0$, Eq.~\eqref{eqn:ma} becomes 
$\alpha_{12}^2 = \frac{1}{3}$, which is the tBLG magic-angle condition~\cite{bistritzer2011moire}.
 
The evolution of VHS along the diagonal likely has a different origin than the magic angles for $\theta_{12}\neq \theta_{23}$. 
Perturbation theory predicts that $ v^*_F$ can reach 0 at $\theta_{12}=\theta_{23}=1.72^\circ$. In the numerical calculations, however, we do not observe $v^*_F=0$ at equal twist angles, and the twist angle with the minimal VHS gap ($2.1^\circ$) deviates from the perturbation theory prediction.
The discrepancy suggests that the perturbation argument does not apply to equal twist angles, since features near the diagonal are more aptly described by the hybridization between the two bilayer moir\'e superlattices with a shared middle layer, rather than between two independent unit cells as in tBLG.

{\it Moir\'e of moir\'e. ---}
In magic-angle tBLG, correlated states occur at the half-filling of the moir\'e supercell by filling two isolated flatbands~\cite{cao2018correlated,cao2018unconventional,koshino2018maximally,po2019faithful,carr2019derivation}. In tTLG, even though the origin of some magic angles is perturbed tBLG, filling each flatband corresponds to filling the moir\'e of moir\'e supercell rather than the bilayer moir\'e cell because the incommensurate effective potential modifies the relevant supercell area. 

We compare our results to a simplified model that approximates tTLG as two aligned moir\'e cells~\cite{mora2019flat}. While we observe similar qualitative behaviors, the simplified model fails to capture physics at the moir\'e of moir\'e scale and does not predict as drastic a DOS enhancement as our work. Moreover, the simplified model requires a new basis for different sets of twist angles, making it difficult to generalize -- limitations that our model overcomes. We include a comparison between the two models in Section~IV of the Supplemental Material IV~\cite{zhu2020sm}. 

In summary, we explore the rich electronic behavior of tTLG in its twist-angle phase space. We offer a general  low-energy momentum-space model to obtain electronic structure in tTLG. We show that the twisted trilayer momentum-space model does not have a Brillouin zone and has an infinitely sized basis. 
Although we do not predict correlation strengths directly, we can use the presence of 
VHS as a proxy for electronic correlation. 
We show that the tTLG system exhibits a wide range of magic angles with merging VHS at the CNP. 
Away from equal twist angles, the origin of the magic angles can be understood as tBLG in an incommensurate perturbative potential. At equal twist angles, the electronic properties are a result of the hybridization between two bilayer moir\'e superlattices that share the middle layer. Tuning the twist angle 
makes it possible to traverse between these two regimes.
Our MATLAB code for the model is openly available ~\cite{zhu2020github}.

\begin{acknowledgments}
We thank Ke Wang, Xi Zhang, Kan-Ting Tsai,
Francisco Guinea, Philip Kim, and Paul Cazeaux for helpful discussions. 
This work was supported by the STC Center for Integrated Quantum Materials, 
NSF Grant No. DMR-1231319, ARO MURI Grant No. W911NF-14-0247, and NSF DMREF Grant No. 1922165.
Calculations were performed on the Odyssey cluster supported by the FAS Division of Science, Research Computing Group at Harvard University.
\end{acknowledgments}

\bibliography{references}

\begin{thebibliography}{43}%
\makeatletter
\providecommand \@ifxundefined [1]{%
 \@ifx{#1\undefined}
}%
\providecommand \@ifnum [1]{%
 \ifnum #1\expandafter \@firstoftwo
 \else \expandafter \@secondoftwo
 \fi
}%
\providecommand \@ifx [1]{%
 \ifx #1\expandafter \@firstoftwo
 \else \expandafter \@secondoftwo
 \fi
}%
\providecommand \natexlab [1]{#1}%
\providecommand \enquote  [1]{``#1''}%
\providecommand \bibnamefont  [1]{#1}%
\providecommand \bibfnamefont [1]{#1}%
\providecommand \citenamefont [1]{#1}%
\providecommand \href@noop [0]{\@secondoftwo}%
\providecommand \href [0]{\begingroup \@sanitize@url \@href}%
\providecommand \@href[1]{\@@startlink{#1}\@@href}%
\providecommand \@@href[1]{\endgroup#1\@@endlink}%
\providecommand \@sanitize@url [0]{\catcode `\\12\catcode `\$12\catcode
  `\&12\catcode `\#12\catcode `\^12\catcode `\_12\catcode `\%12\relax}%
\providecommand \@@startlink[1]{}%
\providecommand \@@endlink[0]{}%
\providecommand \url  [0]{\begingroup\@sanitize@url \@url }%
\providecommand \@url [1]{\endgroup\@href {#1}{\urlprefix }}%
\providecommand \urlprefix  [0]{URL }%
\providecommand \Eprint [0]{\href }%
\providecommand \doibase [0]{http://dx.doi.org/}%
\providecommand \selectlanguage [0]{\@gobble}%
\providecommand \bibinfo  [0]{\@secondoftwo}%
\providecommand \bibfield  [0]{\@secondoftwo}%
\providecommand \translation [1]{[#1]}%
\providecommand \BibitemOpen [0]{}%
\providecommand \bibitemStop [0]{}%
\providecommand \bibitemNoStop [0]{.\EOS\space}%
\providecommand \EOS [0]{\spacefactor3000\relax}%
\providecommand \BibitemShut  [1]{\csname bibitem#1\endcsname}%
\let\auto@bib@innerbib\@empty
\bibitem [{\citenamefont {{Carr}}\ \emph {et~al.}(2017)\citenamefont {{Carr}},
  \citenamefont {{Massatt}}, \citenamefont {{Fang}}, \citenamefont {{Cazeaux}},
  \citenamefont {{Luskin}},\ and\ \citenamefont
  {{Kaxiras}}}]{carr2017twistronics}%
  \BibitemOpen
  \bibfield  {author} {\bibinfo {author} {\bibfnamefont {Stephen}\ \bibnamefont
  {{Carr}}}, \bibinfo {author} {\bibfnamefont {Daniel}\ \bibnamefont
  {{Massatt}}}, \bibinfo {author} {\bibfnamefont {Shiang}\ \bibnamefont
  {{Fang}}}, \bibinfo {author} {\bibfnamefont {Paul}\ \bibnamefont
  {{Cazeaux}}}, \bibinfo {author} {\bibfnamefont {Mitchell}\ \bibnamefont
  {{Luskin}}}, \ and\ \bibinfo {author} {\bibfnamefont {Efthimios}\
  \bibnamefont {{Kaxiras}}},\ }\bibfield  {title} {\enquote {\bibinfo {title}
  {{Twistronics: Manipulating the electronic properties of two-dimensional
  layered structures through their twist angle}},}\ }\href {\doibase
  10.1103/PhysRevB.95.075420} {\bibfield  {journal} {\bibinfo  {journal}
  {\prb}\ }\textbf {\bibinfo {volume} {95}},\ \bibinfo {eid} {075420} (\bibinfo
  {year} {2017})}\BibitemShut {NoStop}%
\bibitem [{\citenamefont {{Bistritzer}}\ and\ \citenamefont
  {{MacDonald}}(2011)}]{bistritzer2011moire}%
  \BibitemOpen
  \bibfield  {author} {\bibinfo {author} {\bibfnamefont {Rafi}\ \bibnamefont
  {{Bistritzer}}}\ and\ \bibinfo {author} {\bibfnamefont {Allan~H.}\
  \bibnamefont {{MacDonald}}},\ }\bibfield  {title} {\enquote {\bibinfo {title}
  {{Moir{\'e} bands in twisted double-layer graphene}},}\ }\href {\doibase
  10.1073/pnas.1108174108} {\bibfield  {journal} {\bibinfo  {journal}
  {Proceedings of the National Academy of Science}\ }\textbf {\bibinfo {volume}
  {108}},\ \bibinfo {pages} {12233--12237} (\bibinfo {year}
  {2011})}\BibitemShut {NoStop}%
\bibitem [{\citenamefont {{Cao}}\ \emph
  {et~al.}(2018{\natexlab{a}})\citenamefont {{Cao}}, \citenamefont {{Fatemi}},
  \citenamefont {{Demir}}, \citenamefont {{Fang}}, \citenamefont {{Tomarken}},
  \citenamefont {{Luo}}, \citenamefont {{Sanchez-Yamagishi}}, \citenamefont
  {{Watanabe}}, \citenamefont {{Taniguchi}}, \citenamefont {{Kaxiras}},
  \citenamefont {{Ashoori}},\ and\ \citenamefont
  {{Jarillo-Herrero}}}]{cao2018correlated}%
  \BibitemOpen
  \bibfield  {author} {\bibinfo {author} {\bibfnamefont {Yuan}\ \bibnamefont
  {{Cao}}}, \bibinfo {author} {\bibfnamefont {Valla}\ \bibnamefont {{Fatemi}}},
  \bibinfo {author} {\bibfnamefont {Ahmet}\ \bibnamefont {{Demir}}}, \bibinfo
  {author} {\bibfnamefont {Shiang}\ \bibnamefont {{Fang}}}, \bibinfo {author}
  {\bibfnamefont {Spencer~L.}\ \bibnamefont {{Tomarken}}}, \bibinfo {author}
  {\bibfnamefont {Jason~Y.}\ \bibnamefont {{Luo}}}, \bibinfo {author}
  {\bibfnamefont {Javier~D.}\ \bibnamefont {{Sanchez-Yamagishi}}}, \bibinfo
  {author} {\bibfnamefont {Kenji}\ \bibnamefont {{Watanabe}}}, \bibinfo
  {author} {\bibfnamefont {Takashi}\ \bibnamefont {{Taniguchi}}}, \bibinfo
  {author} {\bibfnamefont {Efthimios}\ \bibnamefont {{Kaxiras}}}, \bibinfo
  {author} {\bibfnamefont {Ray~C.}\ \bibnamefont {{Ashoori}}}, \ and\ \bibinfo
  {author} {\bibfnamefont {Pablo}\ \bibnamefont {{Jarillo-Herrero}}},\
  }\bibfield  {title} {\enquote {\bibinfo {title} {{Correlated insulator
  behaviour at half-filling in magic-angle graphene superlattices}},}\ }\href
  {\doibase 10.1038/nature26154} {\bibfield  {journal} {\bibinfo  {journal}
  {\nat}\ }\textbf {\bibinfo {volume} {556}},\ \bibinfo {pages} {80--84}
  (\bibinfo {year} {2018}{\natexlab{a}})}\BibitemShut {NoStop}%
\bibitem [{\citenamefont {{Cao}}\ \emph
  {et~al.}(2018{\natexlab{b}})\citenamefont {{Cao}}, \citenamefont {{Fatemi}},
  \citenamefont {{Fang}}, \citenamefont {{Watanabe}}, \citenamefont
  {{Taniguchi}}, \citenamefont {{Kaxiras}},\ and\ \citenamefont
  {{Jarillo-Herrero}}}]{cao2018unconventional}%
  \BibitemOpen
  \bibfield  {author} {\bibinfo {author} {\bibfnamefont {Yuan}\ \bibnamefont
  {{Cao}}}, \bibinfo {author} {\bibfnamefont {Valla}\ \bibnamefont {{Fatemi}}},
  \bibinfo {author} {\bibfnamefont {Shiang}\ \bibnamefont {{Fang}}}, \bibinfo
  {author} {\bibfnamefont {Kenji}\ \bibnamefont {{Watanabe}}}, \bibinfo
  {author} {\bibfnamefont {Takashi}\ \bibnamefont {{Taniguchi}}}, \bibinfo
  {author} {\bibfnamefont {Efthimios}\ \bibnamefont {{Kaxiras}}}, \ and\
  \bibinfo {author} {\bibfnamefont {Pablo}\ \bibnamefont {{Jarillo-Herrero}}},\
  }\bibfield  {title} {\enquote {\bibinfo {title} {{Unconventional
  superconductivity in magic-angle graphene superlattices}},}\ }\href {\doibase
  10.1038/nature26160} {\bibfield  {journal} {\bibinfo  {journal} {\nat}\
  }\textbf {\bibinfo {volume} {556}},\ \bibinfo {pages} {43--50} (\bibinfo
  {year} {2018}{\natexlab{b}})}\BibitemShut {NoStop}%
\bibitem [{\citenamefont {{Chen}}\ \emph {et~al.}(2019)\citenamefont {{Chen}},
  \citenamefont {{Jiang}}, \citenamefont {{Wu}}, \citenamefont {{Lyu}},
  \citenamefont {{Li}}, \citenamefont {{Chittari}}, \citenamefont {{Watanabe}},
  \citenamefont {{Taniguchi}}, \citenamefont {{Shi}}, \citenamefont {{Jung}},
  \citenamefont {{Zhang}},\ and\ \citenamefont {{Wang}}}]{chen2018evidence}%
  \BibitemOpen
  \bibfield  {author} {\bibinfo {author} {\bibfnamefont {Guorui}\ \bibnamefont
  {{Chen}}}, \bibinfo {author} {\bibfnamefont {Lili}\ \bibnamefont {{Jiang}}},
  \bibinfo {author} {\bibfnamefont {Shuang}\ \bibnamefont {{Wu}}}, \bibinfo
  {author} {\bibfnamefont {Bosai}\ \bibnamefont {{Lyu}}}, \bibinfo {author}
  {\bibfnamefont {Hongyuan}\ \bibnamefont {{Li}}}, \bibinfo {author}
  {\bibfnamefont {Bheema~Lingam}\ \bibnamefont {{Chittari}}}, \bibinfo {author}
  {\bibfnamefont {Kenji}\ \bibnamefont {{Watanabe}}}, \bibinfo {author}
  {\bibfnamefont {Takashi}\ \bibnamefont {{Taniguchi}}}, \bibinfo {author}
  {\bibfnamefont {Zhiwen}\ \bibnamefont {{Shi}}}, \bibinfo {author}
  {\bibfnamefont {Jeil}\ \bibnamefont {{Jung}}}, \bibinfo {author}
  {\bibfnamefont {Yuanbo}\ \bibnamefont {{Zhang}}}, \ and\ \bibinfo {author}
  {\bibfnamefont {Feng}\ \bibnamefont {{Wang}}},\ }\bibfield  {title} {\enquote
  {\bibinfo {title} {{Evidence of a gate-tunable Mott insulator in a trilayer
  graphene moir{\'e} superlattice}},}\ }\href {\doibase
  10.1038/s41567-018-0387-2} {\bibfield  {journal} {\bibinfo  {journal} {Nature
  Physics}\ }\textbf {\bibinfo {volume} {15}},\ \bibinfo {pages} {237--241}
  (\bibinfo {year} {2019})}\BibitemShut {NoStop}%
\bibitem [{\citenamefont {{Yankowitz}}\ \emph {et~al.}(2018)\citenamefont
  {{Yankowitz}}, \citenamefont {{Jung}}, \citenamefont {{Laksono}},
  \citenamefont {{Leconte}}, \citenamefont {{Chittari}}, \citenamefont
  {{Watanabe}}, \citenamefont {{Taniguchi}}, \citenamefont {{Adam}},
  \citenamefont {{Graf}},\ and\ \citenamefont {{Dean}}}]{yankowitz2018dynamic}%
  \BibitemOpen
  \bibfield  {author} {\bibinfo {author} {\bibfnamefont {Matthew}\ \bibnamefont
  {{Yankowitz}}}, \bibinfo {author} {\bibfnamefont {Jeil}\ \bibnamefont
  {{Jung}}}, \bibinfo {author} {\bibfnamefont {Evan}\ \bibnamefont
  {{Laksono}}}, \bibinfo {author} {\bibfnamefont {Nicolas}\ \bibnamefont
  {{Leconte}}}, \bibinfo {author} {\bibfnamefont {Bheema~L.}\ \bibnamefont
  {{Chittari}}}, \bibinfo {author} {\bibfnamefont {K.}~\bibnamefont
  {{Watanabe}}}, \bibinfo {author} {\bibfnamefont {T.}~\bibnamefont
  {{Taniguchi}}}, \bibinfo {author} {\bibfnamefont {Shaffique}\ \bibnamefont
  {{Adam}}}, \bibinfo {author} {\bibfnamefont {David}\ \bibnamefont {{Graf}}},
  \ and\ \bibinfo {author} {\bibfnamefont {Cory~R.}\ \bibnamefont {{Dean}}},\
  }\bibfield  {title} {\enquote {\bibinfo {title} {{Dynamic band-structure
  tuning of graphene moir{\'e} superlattices with pressure}},}\ }\href
  {\doibase 10.1038/s41586-018-0107-1} {\bibfield  {journal} {\bibinfo
  {journal} {\nat}\ }\textbf {\bibinfo {volume} {557}},\ \bibinfo {pages}
  {404--408} (\bibinfo {year} {2018})}\BibitemShut {NoStop}%
\bibitem [{\citenamefont {{Yankowitz}}\ \emph {et~al.}(2019)\citenamefont
  {{Yankowitz}}, \citenamefont {{Chen}}, \citenamefont {{Polshyn}},
  \citenamefont {{Zhang}}, \citenamefont {{Watanabe}}, \citenamefont
  {{Taniguchi}}, \citenamefont {{Graf}}, \citenamefont {{Young}},\ and\
  \citenamefont {{Dean}}}]{yankowitz2019tuning}%
  \BibitemOpen
  \bibfield  {author} {\bibinfo {author} {\bibfnamefont {Matthew}\ \bibnamefont
  {{Yankowitz}}}, \bibinfo {author} {\bibfnamefont {Shaowen}\ \bibnamefont
  {{Chen}}}, \bibinfo {author} {\bibfnamefont {Hryhoriy}\ \bibnamefont
  {{Polshyn}}}, \bibinfo {author} {\bibfnamefont {Yuxuan}\ \bibnamefont
  {{Zhang}}}, \bibinfo {author} {\bibfnamefont {K.}~\bibnamefont {{Watanabe}}},
  \bibinfo {author} {\bibfnamefont {T.}~\bibnamefont {{Taniguchi}}}, \bibinfo
  {author} {\bibfnamefont {David}\ \bibnamefont {{Graf}}}, \bibinfo {author}
  {\bibfnamefont {Andrea~F.}\ \bibnamefont {{Young}}}, \ and\ \bibinfo {author}
  {\bibfnamefont {Cory~R.}\ \bibnamefont {{Dean}}},\ }\bibfield  {title}
  {\enquote {\bibinfo {title} {{Tuning superconductivity in twisted bilayer
  graphene}},}\ }\href {\doibase 10.1126/science.aav1910} {\bibfield  {journal}
  {\bibinfo  {journal} {Science}\ }\textbf {\bibinfo {volume} {363}},\ \bibinfo
  {pages} {1059--1064} (\bibinfo {year} {2019})}\BibitemShut {NoStop}%
\bibitem [{\citenamefont {{Shen}}\ \emph {et~al.}(2020)\citenamefont {{Shen}},
  \citenamefont {{Li}}, \citenamefont {{Wang}}, \citenamefont {{Zhao}},
  \citenamefont {{Tang}}, \citenamefont {{Liu}}, \citenamefont {{Tian}},
  \citenamefont {{Chu}}, \citenamefont {{Watanabe}}, \citenamefont
  {{Taniguchi}}, \citenamefont {{Yang}}, \citenamefont {{Meng}}, \citenamefont
  {{Shi}},\ and\ \citenamefont {{Zhang}}}]{shen2019observation}%
  \BibitemOpen
  \bibfield  {author} {\bibinfo {author} {\bibfnamefont {Cheng}\ \bibnamefont
  {{Shen}}}, \bibinfo {author} {\bibfnamefont {Na}~\bibnamefont {{Li}}},
  \bibinfo {author} {\bibfnamefont {Shuopei}\ \bibnamefont {{Wang}}}, \bibinfo
  {author} {\bibfnamefont {Yanchong}\ \bibnamefont {{Zhao}}}, \bibinfo {author}
  {\bibfnamefont {Jian}\ \bibnamefont {{Tang}}}, \bibinfo {author}
  {\bibfnamefont {Jieying}\ \bibnamefont {{Liu}}}, \bibinfo {author}
  {\bibfnamefont {Jinpeng}\ \bibnamefont {{Tian}}}, \bibinfo {author}
  {\bibfnamefont {Yanbang}\ \bibnamefont {{Chu}}}, \bibinfo {author}
  {\bibfnamefont {Kenji}\ \bibnamefont {{Watanabe}}}, \bibinfo {author}
  {\bibfnamefont {Takashi}\ \bibnamefont {{Taniguchi}}}, \bibinfo {author}
  {\bibfnamefont {Rong}\ \bibnamefont {{Yang}}}, \bibinfo {author}
  {\bibfnamefont {Zi~Yang}\ \bibnamefont {{Meng}}}, \bibinfo {author}
  {\bibfnamefont {Dongxia}\ \bibnamefont {{Shi}}}, \ and\ \bibinfo {author}
  {\bibfnamefont {Guangyu}\ \bibnamefont {{Zhang}}},\ }\bibfield  {title}
  {\enquote {\bibinfo {title} {Correlated states in twisted double bilayer
  graphene},}\ }\href {https://doi.org/10.1038/s41567-020-0825-9} {\bibfield
  {journal} {\bibinfo  {journal} {Nature Physics}\ }\textbf {\bibinfo {volume}
  {16}},\ \bibinfo {pages} {520--525} (\bibinfo {year} {2020})}\BibitemShut
  {NoStop}%
\bibitem [{\citenamefont {Liu}\ \emph {et~al.}(2020)\citenamefont {Liu},
  \citenamefont {Hao}, \citenamefont {Khalaf}, \citenamefont {Lee},
  \citenamefont {Ronen}, \citenamefont {Yoo}, \citenamefont {Najafabadi},
  \citenamefont {Watanabe}, \citenamefont {Taniguchi}, \citenamefont
  {Vishwanath} \emph {et~al.}}]{liu2019spin}%
  \BibitemOpen
  \bibfield  {author} {\bibinfo {author} {\bibfnamefont {Xiaomeng}\
  \bibnamefont {Liu}}, \bibinfo {author} {\bibfnamefont {Zeyu}\ \bibnamefont
  {Hao}}, \bibinfo {author} {\bibfnamefont {Eslam}\ \bibnamefont {Khalaf}},
  \bibinfo {author} {\bibfnamefont {Jong~Yeon}\ \bibnamefont {Lee}}, \bibinfo
  {author} {\bibfnamefont {Yuval}\ \bibnamefont {Ronen}}, \bibinfo {author}
  {\bibfnamefont {Hyobin}\ \bibnamefont {Yoo}}, \bibinfo {author}
  {\bibfnamefont {Danial~Haei}\ \bibnamefont {Najafabadi}}, \bibinfo {author}
  {\bibfnamefont {Kenji}\ \bibnamefont {Watanabe}}, \bibinfo {author}
  {\bibfnamefont {Takashi}\ \bibnamefont {Taniguchi}}, \bibinfo {author}
  {\bibfnamefont {Ashvin}\ \bibnamefont {Vishwanath}},  \emph {et~al.},\ }\href
  {https://doi.org/10.1038/s41586-020-2458-7} {\bibfield  {journal} {\bibinfo
  {journal} {Nature}\ }\textbf {\bibinfo {volume} {583}},\ \bibinfo {pages}
  {221--225} (\bibinfo {year} {2020})}\BibitemShut {NoStop}%
\bibitem [{\citenamefont {{Cao}}\ \emph {et~al.}(2020)\citenamefont {{Cao}},
  \citenamefont {{Rodan-Legrain}}, \citenamefont {{Rubies-Bigorda}},
  \citenamefont {{Park}}, \citenamefont {{Watanabe}}, \citenamefont
  {{Taniguchi}},\ and\ \citenamefont {{Jarillo-Herrero}}}]{cao2020electric}%
  \BibitemOpen
  \bibfield  {author} {\bibinfo {author} {\bibfnamefont {Yuan}\ \bibnamefont
  {{Cao}}}, \bibinfo {author} {\bibfnamefont {Daniel}\ \bibnamefont
  {{Rodan-Legrain}}}, \bibinfo {author} {\bibfnamefont {Oriol}\ \bibnamefont
  {{Rubies-Bigorda}}}, \bibinfo {author} {\bibfnamefont {Jeong~Min}\
  \bibnamefont {{Park}}}, \bibinfo {author} {\bibfnamefont {Kenji}\
  \bibnamefont {{Watanabe}}}, \bibinfo {author} {\bibfnamefont {Takashi}\
  \bibnamefont {{Taniguchi}}}, \ and\ \bibinfo {author} {\bibfnamefont {Pablo}\
  \bibnamefont {{Jarillo-Herrero}}},\ }\bibfield  {title} {\enquote {\bibinfo
  {title} {{Electric Field Tunable Correlated States and Magnetic Phase
  Transitions in Twisted Bilayer-Bilayer Graphene}},}\ }\href
  {https://doi.org/10.1038/s41586-020-2260-6} {\bibfield  {journal} {\bibinfo
  {journal} {Nature}\ ,\ \bibinfo {pages} {1--6}} (\bibinfo {year}
  {2020})}\BibitemShut {NoStop}%
\bibitem [{\citenamefont {Burg}\ \emph {et~al.}(2019)\citenamefont {Burg},
  \citenamefont {Zhu}, \citenamefont {Taniguchi}, \citenamefont {Watanabe},
  \citenamefont {MacDonald},\ and\ \citenamefont
  {Tutuc}}]{william2019correlated}%
  \BibitemOpen
  \bibfield  {author} {\bibinfo {author} {\bibfnamefont {G.~William}\
  \bibnamefont {Burg}}, \bibinfo {author} {\bibfnamefont {Jihang}\ \bibnamefont
  {Zhu}}, \bibinfo {author} {\bibfnamefont {Takashi}\ \bibnamefont
  {Taniguchi}}, \bibinfo {author} {\bibfnamefont {Kenji}\ \bibnamefont
  {Watanabe}}, \bibinfo {author} {\bibfnamefont {Allan~H.}\ \bibnamefont
  {MacDonald}}, \ and\ \bibinfo {author} {\bibfnamefont {Emanuel}\ \bibnamefont
  {Tutuc}},\ }\bibfield  {title} {\enquote {\bibinfo {title} {Correlated
  insulating states in twisted double bilayer graphene},}\ }\href {\doibase
  10.1103/PhysRevLett.123.197702} {\bibfield  {journal} {\bibinfo  {journal}
  {Phys. Rev. Lett.}\ }\textbf {\bibinfo {volume} {123}},\ \bibinfo {pages}
  {197702} (\bibinfo {year} {2019})}\BibitemShut {NoStop}%
\bibitem [{\citenamefont {{Wang}}\ \emph {et~al.}(2019)\citenamefont {{Wang}},
  \citenamefont {{Shih}}, \citenamefont {{Ghiotto}}, \citenamefont {{Xian}},
  \citenamefont {{Rhodes}}, \citenamefont {{Tan}}, \citenamefont {{Claassen}},
  \citenamefont {{Kennes}}, \citenamefont {{Bai}}, \citenamefont {{Kim}},
  \citenamefont {{Watanabe}}, \citenamefont {{Taniguchi}}, \citenamefont
  {{Zhu}}, \citenamefont {{Hone}}, \citenamefont {{Rubio}}, \citenamefont
  {{Pasupathy}},\ and\ \citenamefont {{Dean}}}]{wang2019wse2}%
  \BibitemOpen
  \bibfield  {author} {\bibinfo {author} {\bibfnamefont {Lei}\ \bibnamefont
  {{Wang}}}, \bibinfo {author} {\bibfnamefont {En-Min}\ \bibnamefont {{Shih}}},
  \bibinfo {author} {\bibfnamefont {Augusto}\ \bibnamefont {{Ghiotto}}},
  \bibinfo {author} {\bibfnamefont {Lede}\ \bibnamefont {{Xian}}}, \bibinfo
  {author} {\bibfnamefont {Daniel~A.}\ \bibnamefont {{Rhodes}}}, \bibinfo
  {author} {\bibfnamefont {Cheng}\ \bibnamefont {{Tan}}}, \bibinfo {author}
  {\bibfnamefont {Martin}\ \bibnamefont {{Claassen}}}, \bibinfo {author}
  {\bibfnamefont {Dante~M.}\ \bibnamefont {{Kennes}}}, \bibinfo {author}
  {\bibfnamefont {Yusong}\ \bibnamefont {{Bai}}}, \bibinfo {author}
  {\bibfnamefont {Bumho}\ \bibnamefont {{Kim}}}, \bibinfo {author}
  {\bibfnamefont {Kenji}\ \bibnamefont {{Watanabe}}}, \bibinfo {author}
  {\bibfnamefont {Takashi}\ \bibnamefont {{Taniguchi}}}, \bibinfo {author}
  {\bibfnamefont {Xiaoyang}\ \bibnamefont {{Zhu}}}, \bibinfo {author}
  {\bibfnamefont {James}\ \bibnamefont {{Hone}}}, \bibinfo {author}
  {\bibfnamefont {Angel}\ \bibnamefont {{Rubio}}}, \bibinfo {author}
  {\bibfnamefont {Abhay}\ \bibnamefont {{Pasupathy}}}, \ and\ \bibinfo {author}
  {\bibfnamefont {Cory~R.}\ \bibnamefont {{Dean}}},\ }\bibfield  {title}
  {\enquote {\bibinfo {title} {{Magic continuum in twisted bilayer WSe2}},}\
  }\href@noop {} {\bibfield  {journal} {\bibinfo  {journal} {arXiv e-prints}\
  ,\ \bibinfo {eid} {arXiv:1910.12147}} (\bibinfo {year} {2019})},\ \Eprint
  {http://arxiv.org/abs/1910.12147} {arXiv:1910.12147 [cond-mat.mes-hall]}
  \BibitemShut {NoStop}%
\bibitem [{\citenamefont {{Saito}}\ \emph {et~al.}(2019)\citenamefont
  {{Saito}}, \citenamefont {{Ge}}, \citenamefont {{Watanabe}}, \citenamefont
  {{Taniguchi}},\ and\ \citenamefont {{Young}}}]{yu2019decoupling}%
  \BibitemOpen
  \bibfield  {author} {\bibinfo {author} {\bibfnamefont {Yu}~\bibnamefont
  {{Saito}}}, \bibinfo {author} {\bibfnamefont {Jingyuan}\ \bibnamefont
  {{Ge}}}, \bibinfo {author} {\bibfnamefont {Kenji}\ \bibnamefont
  {{Watanabe}}}, \bibinfo {author} {\bibfnamefont {Takashi}\ \bibnamefont
  {{Taniguchi}}}, \ and\ \bibinfo {author} {\bibfnamefont {Andrea~F.}\
  \bibnamefont {{Young}}},\ }\bibfield  {title} {\enquote {\bibinfo {title}
  {{Decoupling superconductivity and correlated insulators in twisted bilayer
  graphene}},}\ }\href@noop {} {\bibfield  {journal} {\bibinfo  {journal}
  {arXiv e-prints}\ ,\ \bibinfo {eid} {arXiv:1911.13302}} (\bibinfo {year}
  {2019})},\ \Eprint {http://arxiv.org/abs/1911.13302} {arXiv:1911.13302
  [cond-mat.mes-hall]} \BibitemShut {NoStop}%
\bibitem [{\citenamefont {{Zhang}}\ \emph {et~al.}(2020)\citenamefont
  {{Zhang}}, \citenamefont {{Myers}}, \citenamefont {{Watanabe}}, \citenamefont
  {{Taniguchi}},\ and\ \citenamefont {{LeRoy}}}]{zhang2020probing}%
  \BibitemOpen
  \bibfield  {author} {\bibinfo {author} {\bibfnamefont {Zhiming}\ \bibnamefont
  {{Zhang}}}, \bibinfo {author} {\bibfnamefont {Rachel}\ \bibnamefont
  {{Myers}}}, \bibinfo {author} {\bibfnamefont {Kenji}\ \bibnamefont
  {{Watanabe}}}, \bibinfo {author} {\bibfnamefont {Takashi}\ \bibnamefont
  {{Taniguchi}}}, \ and\ \bibinfo {author} {\bibfnamefont {Brian~J.}\
  \bibnamefont {{LeRoy}}},\ }\bibfield  {title} {\enquote {\bibinfo {title}
  {{Probing the wavefunctions of correlated states in magic angle graphene}},}\
  }\href@noop {} {\bibfield  {journal} {\bibinfo  {journal} {arXiv e-prints}\
  ,\ \bibinfo {eid} {arXiv:2003.09482}} (\bibinfo {year} {2020})},\ \Eprint
  {http://arxiv.org/abs/2003.09482} {arXiv:2003.09482 [cond-mat.mes-hall]}
  \BibitemShut {NoStop}%
\bibitem [{\citenamefont {{Liu}}\ \emph {et~al.}(2020)\citenamefont {{Liu}},
  \citenamefont {{Wang}}, \citenamefont {{Watanabe}}, \citenamefont
  {{Taniguchi}}, \citenamefont {{Vafek}},\ and\ \citenamefont
  {{Li}}}]{liu2020tuning}%
  \BibitemOpen
  \bibfield  {author} {\bibinfo {author} {\bibfnamefont {Xiaoxue}\ \bibnamefont
  {{Liu}}}, \bibinfo {author} {\bibfnamefont {Zhi}\ \bibnamefont {{Wang}}},
  \bibinfo {author} {\bibfnamefont {K.}~\bibnamefont {{Watanabe}}}, \bibinfo
  {author} {\bibfnamefont {T.}~\bibnamefont {{Taniguchi}}}, \bibinfo {author}
  {\bibfnamefont {Oskar}\ \bibnamefont {{Vafek}}}, \ and\ \bibinfo {author}
  {\bibfnamefont {J.~I.~A.}\ \bibnamefont {{Li}}},\ }\bibfield  {title}
  {\enquote {\bibinfo {title} {{Tuning electron correlation in magic-angle
  twisted bilayer graphene using Coulomb screening}},}\ }\href@noop {}
  {\bibfield  {journal} {\bibinfo  {journal} {arXiv e-prints}\ ,\ \bibinfo
  {eid} {arXiv:2003.11072}} (\bibinfo {year} {2020})},\ \Eprint
  {http://arxiv.org/abs/2003.11072} {arXiv:2003.11072 [cond-mat.mes-hall]}
  \BibitemShut {NoStop}%
\bibitem [{\citenamefont {{Koshino}}\ \emph {et~al.}(2018)\citenamefont
  {{Koshino}}, \citenamefont {{Yuan}}, \citenamefont {{Koretsune}},
  \citenamefont {{Ochi}}, \citenamefont {{Kuroki}},\ and\ \citenamefont
  {{Fu}}}]{koshino2018maximally}%
  \BibitemOpen
  \bibfield  {author} {\bibinfo {author} {\bibfnamefont {Mikito}\ \bibnamefont
  {{Koshino}}}, \bibinfo {author} {\bibfnamefont {Noah F.~Q.}\ \bibnamefont
  {{Yuan}}}, \bibinfo {author} {\bibfnamefont {Takashi}\ \bibnamefont
  {{Koretsune}}}, \bibinfo {author} {\bibfnamefont {Masayuki}\ \bibnamefont
  {{Ochi}}}, \bibinfo {author} {\bibfnamefont {Kazuhiko}\ \bibnamefont
  {{Kuroki}}}, \ and\ \bibinfo {author} {\bibfnamefont {Liang}\ \bibnamefont
  {{Fu}}},\ }\bibfield  {title} {\enquote {\bibinfo {title} {{Maximally
  Localized Wannier Orbitals and the Extended Hubbard Model for Twisted Bilayer
  Graphene}},}\ }\href {\doibase 10.1103/PhysRevX.8.031087} {\bibfield
  {journal} {\bibinfo  {journal} {Physical Review X}\ }\textbf {\bibinfo
  {volume} {8}},\ \bibinfo {eid} {031087} (\bibinfo {year} {2018})}\BibitemShut
  {NoStop}%
\bibitem [{\citenamefont {{Po}}\ \emph {et~al.}(2019)\citenamefont {{Po}},
  \citenamefont {{Zou}}, \citenamefont {{Senthil}},\ and\ \citenamefont
  {{Vishwanath}}}]{po2019faithful}%
  \BibitemOpen
  \bibfield  {author} {\bibinfo {author} {\bibfnamefont {Hoi~Chun}\
  \bibnamefont {{Po}}}, \bibinfo {author} {\bibfnamefont {Liujun}\ \bibnamefont
  {{Zou}}}, \bibinfo {author} {\bibfnamefont {T.}~\bibnamefont {{Senthil}}}, \
  and\ \bibinfo {author} {\bibfnamefont {Ashvin}\ \bibnamefont
  {{Vishwanath}}},\ }\bibfield  {title} {\enquote {\bibinfo {title} {{Faithful
  tight-binding models and fragile topology of magic-angle bilayer
  graphene}},}\ }\href {\doibase 10.1103/PhysRevB.99.195455} {\bibfield
  {journal} {\bibinfo  {journal} {\prb}\ }\textbf {\bibinfo {volume} {99}},\
  \bibinfo {eid} {195455} (\bibinfo {year} {2019})}\BibitemShut {NoStop}%
\bibitem [{\citenamefont {{Carr}}\ \emph
  {et~al.}(2019{\natexlab{a}})\citenamefont {{Carr}}, \citenamefont {{Fang}},
  \citenamefont {{Po}}, \citenamefont {{Vishwanath}},\ and\ \citenamefont
  {{Kaxiras}}}]{carr2019derivation}%
  \BibitemOpen
  \bibfield  {author} {\bibinfo {author} {\bibfnamefont {Stephen}\ \bibnamefont
  {{Carr}}}, \bibinfo {author} {\bibfnamefont {Shiang}\ \bibnamefont {{Fang}}},
  \bibinfo {author} {\bibfnamefont {Hoi~Chun}\ \bibnamefont {{Po}}}, \bibinfo
  {author} {\bibfnamefont {Ashvin}\ \bibnamefont {{Vishwanath}}}, \ and\
  \bibinfo {author} {\bibfnamefont {Efthimios}\ \bibnamefont {{Kaxiras}}},\
  }\bibfield  {title} {\enquote {\bibinfo {title} {{Derivation of Wannier
  orbitals and minimal-basis tight-binding Hamiltonians for twisted bilayer
  graphene: First-principles approach}},}\ }\href {\doibase
  10.1103/PhysRevResearch.1.033072} {\bibfield  {journal} {\bibinfo  {journal}
  {Physical Review Research}\ }\textbf {\bibinfo {volume} {1}},\ \bibinfo {eid}
  {033072} (\bibinfo {year} {2019}{\natexlab{a}})}\BibitemShut {NoStop}%
\bibitem [{\citenamefont {Zhu}\ \emph {et~al.}(2020)\citenamefont {Zhu},
  \citenamefont {Cazeaux}, \citenamefont {Luskin},\ and\ \citenamefont
  {Kaxiras}}]{zhu2019moire}%
  \BibitemOpen
  \bibfield  {author} {\bibinfo {author} {\bibfnamefont {Ziyan}\ \bibnamefont
  {Zhu}}, \bibinfo {author} {\bibfnamefont {Paul}\ \bibnamefont {Cazeaux}},
  \bibinfo {author} {\bibfnamefont {Mitchell}\ \bibnamefont {Luskin}}, \ and\
  \bibinfo {author} {\bibfnamefont {Efthimios}\ \bibnamefont {Kaxiras}},\
  }\bibfield  {title} {\enquote {\bibinfo {title} {Modeling mechanical
  relaxation in incommensurate trilayer van der waals heterostructures},}\
  }\href {\doibase 10.1103/PhysRevB.101.224107} {\bibfield  {journal} {\bibinfo
   {journal} {Phys. Rev. B}\ }\textbf {\bibinfo {volume} {101}},\ \bibinfo
  {pages} {224107} (\bibinfo {year} {2020})}\BibitemShut {NoStop}%
\bibitem [{\citenamefont {{An{\dj}elkovi{\'c}}}\ \emph
  {et~al.}(2020)\citenamefont {{An{\dj}elkovi{\'c}}}, \citenamefont
  {{Milovanovi{\'c}}}, \citenamefont {{Covaci}},\ and\ \citenamefont
  {{Peeters}}}]{andelkovic2019double}%
  \BibitemOpen
  \bibfield  {author} {\bibinfo {author} {\bibfnamefont {M.}~\bibnamefont
  {{An{\dj}elkovi{\'c}}}}, \bibinfo {author} {\bibfnamefont {S.~P.}\
  \bibnamefont {{Milovanovi{\'c}}}}, \bibinfo {author} {\bibfnamefont
  {L.}~\bibnamefont {{Covaci}}}, \ and\ \bibinfo {author} {\bibfnamefont
  {F.~M.}\ \bibnamefont {{Peeters}}},\ }\bibfield  {title} {\enquote {\bibinfo
  {title} {{Double moir{\'e} with a twist: super-moir{\'e} in encapsulated
  graphene}},}\ }\href {https://doi.org/10.1021/acs.nanolett.9b04058}
  {\bibfield  {journal} {\bibinfo  {journal} {Nano Letters}\ }\textbf {\bibinfo
  {volume} {20}},\ \bibinfo {pages} {979--988} (\bibinfo {year}
  {2020})}\BibitemShut {NoStop}%
\bibitem [{\citenamefont {Leconte}\ and\ \citenamefont
  {Jung}(2020)}]{leconte2020commensurate}%
  \BibitemOpen
  \bibfield  {author} {\bibinfo {author} {\bibfnamefont {N}~\bibnamefont
  {Leconte}}\ and\ \bibinfo {author} {\bibfnamefont {J}~\bibnamefont {Jung}},\
  }\bibfield  {title} {\enquote {\bibinfo {title} {Commensurate and
  incommensurate double moire interference in graphene encapsulated by
  hexagonal boron nitride},}\ }\href
  {https://doi.org/10.1088%2F2053-1583%2Fab891a} {\bibfield  {journal}
  {\bibinfo  {journal} {2D Materials}\ }\textbf {\bibinfo {volume} {7}},\
  \bibinfo {pages} {031005} (\bibinfo {year} {2020})}\BibitemShut {NoStop}%
\bibitem [{zhu()}]{zhu2020sm}%
  \BibitemOpen
  \href@noop {} {}\bibinfo {note} {See Supplemental Material for the
  calculation of moir\'e of moir\'e length, a detailed derivation of the
  momentum space model and convergence test, the derivation of the analytical
  expression of the tTLG magic angles, and comparisons with other models, which
  includes Refs.~\cite{yankowitz2012emergence,zuo2018scanning}}\BibitemShut
  {NoStop}%
\bibitem [{\citenamefont {{Li}}\ \emph {et~al.}(2019)\citenamefont {{Li}},
  \citenamefont {{Wu}},\ and\ \citenamefont {{MacDonald}}}]{li2019trilayer}%
  \BibitemOpen
  \bibfield  {author} {\bibinfo {author} {\bibfnamefont {Xiao}\ \bibnamefont
  {{Li}}}, \bibinfo {author} {\bibfnamefont {Fengcheng}\ \bibnamefont {{Wu}}},
  \ and\ \bibinfo {author} {\bibfnamefont {Allan~H.}\ \bibnamefont
  {{MacDonald}}},\ }\bibfield  {title} {\enquote {\bibinfo {title} {{Electronic
  Structure of Single-Twist Trilayer Graphene}},}\ }\href@noop {} {\bibfield
  {journal} {\bibinfo  {journal} {arXiv e-prints}\ ,\ \bibinfo {eid}
  {arXiv:1907.12338}} (\bibinfo {year} {2019})},\ \Eprint
  {http://arxiv.org/abs/1907.12338} {arXiv:1907.12338 [cond-mat.mtrl-sci]}
  \BibitemShut {NoStop}%
\bibitem [{\citenamefont {Carr}\ \emph {et~al.}(2020)\citenamefont {Carr},
  \citenamefont {Li}, \citenamefont {Zhu}, \citenamefont {Kaxiras},
  \citenamefont {Sachdev},\ and\ \citenamefont
  {Kruchkov}}]{carr2020ultraheavy}%
  \BibitemOpen
  \bibfield  {author} {\bibinfo {author} {\bibfnamefont {Stephen}\ \bibnamefont
  {Carr}}, \bibinfo {author} {\bibfnamefont {Chenyuan}\ \bibnamefont {Li}},
  \bibinfo {author} {\bibfnamefont {Ziyan}\ \bibnamefont {Zhu}}, \bibinfo
  {author} {\bibfnamefont {Efthimios}\ \bibnamefont {Kaxiras}}, \bibinfo
  {author} {\bibfnamefont {Subir}\ \bibnamefont {Sachdev}}, \ and\ \bibinfo
  {author} {\bibfnamefont {Alexander}\ \bibnamefont {Kruchkov}},\ }\bibfield
  {title} {\enquote {\bibinfo {title} {Ultraheavy and ultrarelativistic dirac
  quasiparticles in sandwiched graphenes},}\ }\href {\doibase
  https://doi.org/10.1021/acs.nanolett.9b04979} {\bibfield  {journal} {\bibinfo
   {journal} {Nano Letters}\ }\textbf {\bibinfo {volume} {20}},\ \bibinfo
  {pages} {3030–3038} (\bibinfo {year} {2020})}\BibitemShut {NoStop}%
\bibitem [{\citenamefont {{Chen}}\ \emph {et~al.}(2020)\citenamefont {{Chen}},
  \citenamefont {{He}}, \citenamefont {{Zhang}}, \citenamefont {{Hsieh}},
  \citenamefont {{Fei}}, \citenamefont {{Watanabe}}, \citenamefont
  {{Taniguchi}}, \citenamefont {{Cobden}}, \citenamefont {{Xu}}, \citenamefont
  {{Dean}},\ and\ \citenamefont {{Yankowitz}}}]{chen2020electrically}%
  \BibitemOpen
  \bibfield  {author} {\bibinfo {author} {\bibfnamefont {Shaowen}\ \bibnamefont
  {{Chen}}}, \bibinfo {author} {\bibfnamefont {Minhao}\ \bibnamefont {{He}}},
  \bibinfo {author} {\bibfnamefont {Ya-Hui}\ \bibnamefont {{Zhang}}}, \bibinfo
  {author} {\bibfnamefont {Valerie}\ \bibnamefont {{Hsieh}}}, \bibinfo {author}
  {\bibfnamefont {Zaiyao}\ \bibnamefont {{Fei}}}, \bibinfo {author}
  {\bibfnamefont {K.}~\bibnamefont {{Watanabe}}}, \bibinfo {author}
  {\bibfnamefont {T.}~\bibnamefont {{Taniguchi}}}, \bibinfo {author}
  {\bibfnamefont {David~H.}\ \bibnamefont {{Cobden}}}, \bibinfo {author}
  {\bibfnamefont {Xiaodong}\ \bibnamefont {{Xu}}}, \bibinfo {author}
  {\bibfnamefont {Cory~R.}\ \bibnamefont {{Dean}}}, \ and\ \bibinfo {author}
  {\bibfnamefont {Matthew}\ \bibnamefont {{Yankowitz}}},\ }\bibfield  {title}
  {\enquote {\bibinfo {title} {{Electrically tunable correlated and topological
  states in twisted monolayer-bilayer graphene}},}\ }\href@noop {} {\bibfield
  {journal} {\bibinfo  {journal} {arXiv e-prints}\ ,\ \bibinfo {eid}
  {arXiv:2004.11340}} (\bibinfo {year} {2020})},\ \Eprint
  {http://arxiv.org/abs/2004.11340} {arXiv:2004.11340 [cond-mat.mes-hall]}
  \BibitemShut {NoStop}%
\bibitem [{\citenamefont {Park}\ \emph {et~al.}(2020)\citenamefont {Park},
  \citenamefont {Chittari},\ and\ \citenamefont {Jung}}]{park2020gate}%
  \BibitemOpen
  \bibfield  {author} {\bibinfo {author} {\bibfnamefont {Youngju}\ \bibnamefont
  {Park}}, \bibinfo {author} {\bibfnamefont {Bheema~Lingam}\ \bibnamefont
  {Chittari}}, \ and\ \bibinfo {author} {\bibfnamefont {Jeil}\ \bibnamefont
  {Jung}},\ }\bibfield  {title} {\enquote {\bibinfo {title} {Gate-tunable
  topological flat bands in twisted monolayer-bilayer graphene},}\ }\href
  {\doibase 10.1103/PhysRevB.102.035411} {\bibfield  {journal} {\bibinfo
  {journal} {Phys. Rev. B}\ }\textbf {\bibinfo {volume} {102}},\ \bibinfo
  {pages} {035411} (\bibinfo {year} {2020})}\BibitemShut {NoStop}%
\bibitem [{\citenamefont {{Amorim}}\ and\ \citenamefont
  {{Castro}}(2018)}]{amorim2018electronic}%
  \BibitemOpen
  \bibfield  {author} {\bibinfo {author} {\bibfnamefont {B.}~\bibnamefont
  {{Amorim}}}\ and\ \bibinfo {author} {\bibfnamefont {Eduardo~V.}\ \bibnamefont
  {{Castro}}},\ }\bibfield  {title} {\enquote {\bibinfo {title} {{Electronic
  spectral properties of incommensurate twisted trilayer graphene}},}\
  }\href@noop {} {\bibfield  {journal} {\bibinfo  {journal} {arXiv e-prints}\
  ,\ \bibinfo {eid} {arXiv:1807.11909}} (\bibinfo {year} {2018})},\ \Eprint
  {http://arxiv.org/abs/1807.11909} {arXiv:1807.11909 [cond-mat.mes-hall]}
  \BibitemShut {NoStop}%
\bibitem [{\citenamefont {{Mora}}\ \emph {et~al.}(2019)\citenamefont {{Mora}},
  \citenamefont {{Regnault}},\ and\ \citenamefont {{Bernevig}}}]{mora2019flat}%
  \BibitemOpen
  \bibfield  {author} {\bibinfo {author} {\bibfnamefont {Christophe}\
  \bibnamefont {{Mora}}}, \bibinfo {author} {\bibfnamefont {Nicolas}\
  \bibnamefont {{Regnault}}}, \ and\ \bibinfo {author} {\bibfnamefont
  {B.~Andrei}\ \bibnamefont {{Bernevig}}},\ }\bibfield  {title} {\enquote
  {\bibinfo {title} {{Flatbands and Perfect Metal in Trilayer Moir{\'e}
  Graphene}},}\ }\href {\doibase 10.1103/PhysRevLett.123.026402} {\bibfield
  {journal} {\bibinfo  {journal} {\prl}\ }\textbf {\bibinfo {volume} {123}},\
  \bibinfo {eid} {026402} (\bibinfo {year} {2019})}\BibitemShut {NoStop}%
\bibitem [{\citenamefont {{Tsai}}\ \emph {et~al.}(2019)\citenamefont {{Tsai}},
  \citenamefont {{Zhang}}, \citenamefont {{Zhu}}, \citenamefont {{Luo}},
  \citenamefont {{Carr}}, \citenamefont {{Luskin}}, \citenamefont {{Kaxiras}},\
  and\ \citenamefont {{Wang}}}]{tsai2019correlated}%
  \BibitemOpen
  \bibfield  {author} {\bibinfo {author} {\bibfnamefont {Kan-Ting}\
  \bibnamefont {{Tsai}}}, \bibinfo {author} {\bibfnamefont {Xi}~\bibnamefont
  {{Zhang}}}, \bibinfo {author} {\bibfnamefont {Ziyan}\ \bibnamefont {{Zhu}}},
  \bibinfo {author} {\bibfnamefont {Yujie}\ \bibnamefont {{Luo}}}, \bibinfo
  {author} {\bibfnamefont {Stephen}\ \bibnamefont {{Carr}}}, \bibinfo {author}
  {\bibfnamefont {Mitchell}\ \bibnamefont {{Luskin}}}, \bibinfo {author}
  {\bibfnamefont {Efthimios}\ \bibnamefont {{Kaxiras}}}, \ and\ \bibinfo
  {author} {\bibfnamefont {Ke}~\bibnamefont {{Wang}}},\ }\bibfield  {title}
  {\enquote {\bibinfo {title} {{Correlated Superconducting and Insulating
  States in Twisted Trilayer Graphene Moir\'e of Moir\'e Superlattices}},}\
  }\href@noop {} {\bibfield  {journal} {\bibinfo  {journal} {arXiv e-prints}\
  ,\ \bibinfo {eid} {arXiv:1912.03375}} (\bibinfo {year} {2019})},\ \Eprint
  {http://arxiv.org/abs/1912.03375} {arXiv:1912.03375 [cond-mat.mes-hall]}
  \BibitemShut {NoStop}%
\bibitem [{\citenamefont {{Castro Neto}}\ \emph {et~al.}(2009)\citenamefont
  {{Castro Neto}}, \citenamefont {{Guinea}}, \citenamefont {{Peres}},
  \citenamefont {{Novoselov}},\ and\ \citenamefont
  {{Geim}}}]{castro_neto2009electronic}%
  \BibitemOpen
  \bibfield  {author} {\bibinfo {author} {\bibfnamefont {A.~H.}\ \bibnamefont
  {{Castro Neto}}}, \bibinfo {author} {\bibfnamefont {F.}~\bibnamefont
  {{Guinea}}}, \bibinfo {author} {\bibfnamefont {N.~M.~R.}\ \bibnamefont
  {{Peres}}}, \bibinfo {author} {\bibfnamefont {K.~S.}\ \bibnamefont
  {{Novoselov}}}, \ and\ \bibinfo {author} {\bibfnamefont {A.~K.}\ \bibnamefont
  {{Geim}}},\ }\bibfield  {title} {\enquote {\bibinfo {title} {{The electronic
  properties of graphene}},}\ }\href {\doibase 10.1103/RevModPhys.81.109}
  {\bibfield  {journal} {\bibinfo  {journal} {Reviews of Modern Physics}\
  }\textbf {\bibinfo {volume} {81}},\ \bibinfo {pages} {109--162} (\bibinfo
  {year} {2009})}\BibitemShut {NoStop}%
\bibitem [{\citenamefont {{Nam}}\ and\ \citenamefont
  {{Koshino}}(2017{\natexlab{a}})}]{nguyen2017lattice}%
  \BibitemOpen
  \bibfield  {author} {\bibinfo {author} {\bibfnamefont {Nguyen N.~T.}\
  \bibnamefont {{Nam}}}\ and\ \bibinfo {author} {\bibfnamefont {Mikito}\
  \bibnamefont {{Koshino}}},\ }\bibfield  {title} {\enquote {\bibinfo {title}
  {{Lattice relaxation and energy band modulation in twisted bilayer
  graphene}},}\ }\href {\doibase 10.1103/PhysRevB.96.075311} {\bibfield
  {journal} {\bibinfo  {journal} {\prb}\ }\textbf {\bibinfo {volume} {96}},\
  \bibinfo {eid} {075311} (\bibinfo {year} {2017}{\natexlab{a}})}\BibitemShut
  {NoStop}%
\bibitem [{\citenamefont {{Carr}}\ \emph
  {et~al.}(2019{\natexlab{b}})\citenamefont {{Carr}}, \citenamefont {{Fang}},
  \citenamefont {{Zhu}},\ and\ \citenamefont {{Kaxiras}}}]{carr2019exact}%
  \BibitemOpen
  \bibfield  {author} {\bibinfo {author} {\bibfnamefont {Stephen}\ \bibnamefont
  {{Carr}}}, \bibinfo {author} {\bibfnamefont {Shiang}\ \bibnamefont {{Fang}}},
  \bibinfo {author} {\bibfnamefont {Ziyan}\ \bibnamefont {{Zhu}}}, \ and\
  \bibinfo {author} {\bibfnamefont {Efthimios}\ \bibnamefont {{Kaxiras}}},\
  }\bibfield  {title} {\enquote {\bibinfo {title} {{Exact continuum model for
  low-energy electronic states of twisted bilayer graphene}},}\ }\href
  {\doibase 10.1103/PhysRevResearch.1.013001} {\bibfield  {journal} {\bibinfo
  {journal} {Physical Review Research}\ }\textbf {\bibinfo {volume} {1}},\
  \bibinfo {eid} {013001} (\bibinfo {year} {2019}{\natexlab{b}})}\BibitemShut
  {NoStop}%
\bibitem [{\citenamefont {{Fang}}\ \emph {et~al.}(2019)\citenamefont {{Fang}},
  \citenamefont {{Carr}}, \citenamefont {{Zhu}}, \citenamefont {{Massatt}},\
  and\ \citenamefont {{Kaxiras}}}]{fang2019angle}%
  \BibitemOpen
  \bibfield  {author} {\bibinfo {author} {\bibfnamefont {Shiang}\ \bibnamefont
  {{Fang}}}, \bibinfo {author} {\bibfnamefont {Stephen}\ \bibnamefont
  {{Carr}}}, \bibinfo {author} {\bibfnamefont {Ziyan}\ \bibnamefont {{Zhu}}},
  \bibinfo {author} {\bibfnamefont {Daniel}\ \bibnamefont {{Massatt}}}, \ and\
  \bibinfo {author} {\bibfnamefont {Efthimios}\ \bibnamefont {{Kaxiras}}},\
  }\bibfield  {title} {\enquote {\bibinfo {title} {{Angle-Dependent {\it Ab
  initio} Low-Energy Hamiltonians for a Relaxed Twisted Bilayer Graphene
  Heterostructure}},}\ }\href@noop {} {\bibfield  {journal} {\bibinfo
  {journal} {arXiv e-prints}\ ,\ \bibinfo {eid} {arXiv:1908.00058}} (\bibinfo
  {year} {2019})},\ \Eprint {http://arxiv.org/abs/1908.00058} {arXiv:1908.00058
  [cond-mat.mes-hall]} \BibitemShut {NoStop}%
\bibitem [{\citenamefont {{Guinea}}\ and\ \citenamefont
  {{Walet}}(2019)}]{guinea2019continuum}%
  \BibitemOpen
  \bibfield  {author} {\bibinfo {author} {\bibfnamefont {Francisco}\
  \bibnamefont {{Guinea}}}\ and\ \bibinfo {author} {\bibfnamefont {Niels~R.}\
  \bibnamefont {{Walet}}},\ }\bibfield  {title} {\enquote {\bibinfo {title}
  {{Continuum models for twisted bilayer graphene: Effect of lattice
  deformation and hopping parameters}},}\ }\href {\doibase
  10.1103/PhysRevB.99.205134} {\bibfield  {journal} {\bibinfo  {journal}
  {\prb}\ }\textbf {\bibinfo {volume} {99}},\ \bibinfo {eid} {205134} (\bibinfo
  {year} {2019})}\BibitemShut {NoStop}%
\bibitem [{\citenamefont {{Leconte}}\ \emph {et~al.}(2019)\citenamefont
  {{Leconte}}, \citenamefont {{Javvaji}}, \citenamefont {{An}},\ and\
  \citenamefont {{Jung}}}]{leconte2019relaxation}%
  \BibitemOpen
  \bibfield  {author} {\bibinfo {author} {\bibfnamefont {Nicolas}\ \bibnamefont
  {{Leconte}}}, \bibinfo {author} {\bibfnamefont {Srivani}\ \bibnamefont
  {{Javvaji}}}, \bibinfo {author} {\bibfnamefont {Jiaqi}\ \bibnamefont {{An}}},
  \ and\ \bibinfo {author} {\bibfnamefont {Jeil}\ \bibnamefont {{Jung}}},\
  }\bibfield  {title} {\enquote {\bibinfo {title} {{Relaxation Effects in
  Twisted Bilayer Graphene: a Multi-Scale Approach}},}\ }\href@noop {}
  {\bibfield  {journal} {\bibinfo  {journal} {arXiv e-prints}\ ,\ \bibinfo
  {eid} {arXiv:1910.12805}} (\bibinfo {year} {2019})},\ \Eprint
  {http://arxiv.org/abs/1910.12805} {arXiv:1910.12805 [cond-mat.mes-hall]}
  \BibitemShut {NoStop}%
\bibitem [{\citenamefont {{Fang}}\ and\ \citenamefont
  {{Kaxiras}}(2016)}]{fang2016electronic}%
  \BibitemOpen
  \bibfield  {author} {\bibinfo {author} {\bibfnamefont {Shiang}\ \bibnamefont
  {{Fang}}}\ and\ \bibinfo {author} {\bibfnamefont {Efthimios}\ \bibnamefont
  {{Kaxiras}}},\ }\bibfield  {title} {\enquote {\bibinfo {title} {{Electronic
  structure theory of weakly interacting bilayers}},}\ }\href {\doibase
  10.1103/PhysRevB.93.235153} {\bibfield  {journal} {\bibinfo  {journal}
  {\prb}\ }\textbf {\bibinfo {volume} {93}},\ \bibinfo {eid} {235153} (\bibinfo
  {year} {2016})}\BibitemShut {NoStop}%
\bibitem [{\citenamefont {Catarina}\ \emph {et~al.}(2019)\citenamefont
  {Catarina}, \citenamefont {Amorim}, \citenamefont {Castro}, \citenamefont
  {Lopes},\ and\ \citenamefont {Peres}}]{catarina2019twisted}%
  \BibitemOpen
  \bibfield  {author} {\bibinfo {author} {\bibfnamefont {Gon{\c{c}}alo}\
  \bibnamefont {Catarina}}, \bibinfo {author} {\bibfnamefont {Bruno}\
  \bibnamefont {Amorim}}, \bibinfo {author} {\bibfnamefont {Eduardo~V}\
  \bibnamefont {Castro}}, \bibinfo {author} {\bibfnamefont {Jo{\~a}o~MVP}\
  \bibnamefont {Lopes}}, \ and\ \bibinfo {author} {\bibfnamefont {Nuno}\
  \bibnamefont {Peres}},\ }\bibfield  {title} {\enquote {\bibinfo {title}
  {Twisted bilayer graphene: Low-energy physics, electronic and optical
  properties},}\ }\href@noop {} {\bibfield  {journal} {\bibinfo  {journal}
  {Handbook of Graphene, Volume 3: Graphene-like 2D Materials}\ ,\ \bibinfo
  {pages} {177}} (\bibinfo {year} {2019})}\BibitemShut {NoStop}%
\bibitem [{\citenamefont {{Nam}}\ and\ \citenamefont
  {{Koshino}}(2017{\natexlab{b}})}]{nam2017lattice}%
  \BibitemOpen
  \bibfield  {author} {\bibinfo {author} {\bibfnamefont {Nguyen N.~T.}\
  \bibnamefont {{Nam}}}\ and\ \bibinfo {author} {\bibfnamefont {Mikito}\
  \bibnamefont {{Koshino}}},\ }\bibfield  {title} {\enquote {\bibinfo {title}
  {{Lattice relaxation and energy band modulation in twisted bilayer
  graphene}},}\ }\href {\doibase 10.1103/PhysRevB.96.075311} {\bibfield
  {journal} {\bibinfo  {journal} {\prb}\ }\textbf {\bibinfo {volume} {96}},\
  \bibinfo {eid} {075311} (\bibinfo {year} {2017}{\natexlab{b}})}\BibitemShut
  {NoStop}%
\bibitem [{\citenamefont {Massatt}\ \emph {et~al.}(2018)\citenamefont
  {Massatt}, \citenamefont {Carr}, \citenamefont {Luskin},\ and\ \citenamefont
  {Ortner}}]{massatt2017incommensurate}%
  \BibitemOpen
  \bibfield  {author} {\bibinfo {author} {\bibfnamefont {Daniel}\ \bibnamefont
  {Massatt}}, \bibinfo {author} {\bibfnamefont {Stephen}\ \bibnamefont {Carr}},
  \bibinfo {author} {\bibfnamefont {Mitchell}\ \bibnamefont {Luskin}}, \ and\
  \bibinfo {author} {\bibfnamefont {Christoph}\ \bibnamefont {Ortner}},\
  }\bibfield  {title} {\enquote {\bibinfo {title} {Incommensurate
  heterostructures in momentum space},}\ }\href {\doibase 10.1137/17M1141035}
  {\bibfield  {journal} {\bibinfo  {journal} {Multiscale Modeling \&
  Simulation}\ }\textbf {\bibinfo {volume} {16}},\ \bibinfo {pages} {429--451}
  (\bibinfo {year} {2018})}\BibitemShut {NoStop}%
\bibitem [{Note1()}]{Note1}%
  \BibitemOpen
  \bibinfo {note} {\label {note1}The data set consists of a $43\times 43$
  sampling. A Gaussian convolution kernel is applied for
  smoothening.}\BibitemShut {Stop}%
\bibitem [{\citenamefont {{Zhu}}\ \emph {et~al.}(2020)\citenamefont {{Zhu}},
  \citenamefont {Stephen}, \citenamefont {Daniel}, \citenamefont {Mitchell},\
  and\ \citenamefont {Efthimios}}]{zhu2020github}%
  \BibitemOpen
  \bibfield  {author} {\bibinfo {author} {\bibfnamefont {Ziyan}\ \bibnamefont
  {{Zhu}}}, \bibinfo {author} {\bibfnamefont {{Carr}}\ \bibnamefont {Stephen}},
  \bibinfo {author} {\bibfnamefont {{Massatt}}\ \bibnamefont {Daniel}},
  \bibinfo {author} {\bibfnamefont {{Luskin}}\ \bibnamefont {Mitchell}}, \ and\
  \bibinfo {author} {\bibfnamefont {{Kaxiras}}\ \bibnamefont {Efthimios}},\
  }\href {https://github.com/ziyanzzhu/ttlg} {\enquote {\bibinfo {title}
  {{Model for Twisted Trilayer Graphene: a precisely tunable platform for
  correlated electrons: https://github.com/ziyanzzhu/ttlg}},}\ } (\bibinfo
  {year} {2020})\BibitemShut {NoStop}%
\bibitem [{\citenamefont {{Yankowitz}}\ \emph {et~al.}(2012)\citenamefont
  {{Yankowitz}}, \citenamefont {{Xue}}, \citenamefont {{Cormode}},
  \citenamefont {{Sanchez-Yamagishi}}, \citenamefont {{Watanabe}},
  \citenamefont {{Taniguchi}}, \citenamefont {{Jarillo-Herrero}}, \citenamefont
  {{Jacquod}},\ and\ \citenamefont {{Leroy}}}]{yankowitz2012emergence}%
  \BibitemOpen
  \bibfield  {author} {\bibinfo {author} {\bibfnamefont {Matthew}\ \bibnamefont
  {{Yankowitz}}}, \bibinfo {author} {\bibfnamefont {Jiamin}\ \bibnamefont
  {{Xue}}}, \bibinfo {author} {\bibfnamefont {Daniel}\ \bibnamefont
  {{Cormode}}}, \bibinfo {author} {\bibfnamefont {Javier~D.}\ \bibnamefont
  {{Sanchez-Yamagishi}}}, \bibinfo {author} {\bibfnamefont {K.}~\bibnamefont
  {{Watanabe}}}, \bibinfo {author} {\bibfnamefont {T.}~\bibnamefont
  {{Taniguchi}}}, \bibinfo {author} {\bibfnamefont {Pablo}\ \bibnamefont
  {{Jarillo-Herrero}}}, \bibinfo {author} {\bibfnamefont {Philippe}\
  \bibnamefont {{Jacquod}}}, \ and\ \bibinfo {author} {\bibfnamefont
  {Brian~J.}\ \bibnamefont {{Leroy}}},\ }\bibfield  {title} {\enquote {\bibinfo
  {title} {{Emergence of superlattice Dirac points in graphene on hexagonal
  boron nitride}},}\ }\href {\doibase 10.1038/nphys2272} {\bibfield  {journal}
  {\bibinfo  {journal} {Nature Physics}\ }\textbf {\bibinfo {volume} {8}},\
  \bibinfo {pages} {382--386} (\bibinfo {year} {2012})}\BibitemShut {NoStop}%
\bibitem [{\citenamefont {Zuo}\ \emph {et~al.}(2018)\citenamefont {Zuo},
  \citenamefont {Qiao}, \citenamefont {Ma}, \citenamefont {Yin}, \citenamefont
  {Sun}, \citenamefont {Zhang}, \citenamefont {Guan},\ and\ \citenamefont
  {He}}]{zuo2018scanning}%
  \BibitemOpen
  \bibfield  {author} {\bibinfo {author} {\bibfnamefont {Wei-Jie}\ \bibnamefont
  {Zuo}}, \bibinfo {author} {\bibfnamefont {Jia-Bin}\ \bibnamefont {Qiao}},
  \bibinfo {author} {\bibfnamefont {Dong-Lin}\ \bibnamefont {Ma}}, \bibinfo
  {author} {\bibfnamefont {Long-Jing}\ \bibnamefont {Yin}}, \bibinfo {author}
  {\bibfnamefont {Gan}\ \bibnamefont {Sun}}, \bibinfo {author} {\bibfnamefont
  {Jun-Yang}\ \bibnamefont {Zhang}}, \bibinfo {author} {\bibfnamefont
  {Li-Yang}\ \bibnamefont {Guan}}, \ and\ \bibinfo {author} {\bibfnamefont
  {Lin}\ \bibnamefont {He}},\ }\bibfield  {title} {\enquote {\bibinfo {title}
  {Scanning tunneling microscopy and spectroscopy of twisted trilayer
  graphene},}\ }\href {\doibase PhysRevB.97.035440} {\bibfield  {journal}
  {\bibinfo  {journal} {Physical Review B}\ }\textbf {\bibinfo {volume} {97}},\
  \bibinfo {pages} {035440} (\bibinfo {year} {2018})}\BibitemShut {NoStop}%
\end{thebibliography}%


\begin{thebibliography}{12}%
\makeatletter
\providecommand \@ifxundefined [1]{%
 \@ifx{#1\undefined}
}%
\providecommand \@ifnum [1]{%
 \ifnum #1\expandafter \@firstoftwo
 \else \expandafter \@secondoftwo
 \fi
}%
\providecommand \@ifx [1]{%
 \ifx #1\expandafter \@firstoftwo
 \else \expandafter \@secondoftwo
 \fi
}%
\providecommand \natexlab [1]{#1}%
\providecommand \enquote  [1]{``#1''}%
\providecommand \bibnamefont  [1]{#1}%
\providecommand \bibfnamefont [1]{#1}%
\providecommand \citenamefont [1]{#1}%
\providecommand \href@noop [0]{\@secondoftwo}%
\providecommand \href [0]{\begingroup \@sanitize@url \@href}%
\providecommand \@href[1]{\@@startlink{#1}\@@href}%
\providecommand \@@href[1]{\endgroup#1\@@endlink}%
\providecommand \@sanitize@url [0]{\catcode `\\12\catcode `\$12\catcode
  `\&12\catcode `\#12\catcode `\^12\catcode `\_12\catcode `\%12\relax}%
\providecommand \@@startlink[1]{}%
\providecommand \@@endlink[0]{}%
\providecommand \url  [0]{\begingroup\@sanitize@url \@url }%
\providecommand \@url [1]{\endgroup\@href {#1}{\urlprefix }}%
\providecommand \urlprefix  [0]{URL }%
\providecommand \Eprint [0]{\href }%
\providecommand \doibase [0]{http://dx.doi.org/}%
\providecommand \selectlanguage [0]{\@gobble}%
\providecommand \bibinfo  [0]{\@secondoftwo}%
\providecommand \bibfield  [0]{\@secondoftwo}%
\providecommand \translation [1]{[#1]}%
\providecommand \BibitemOpen [0]{}%
\providecommand \bibitemStop [0]{}%
\providecommand \bibitemNoStop [0]{.\EOS\space}%
\providecommand \EOS [0]{\spacefactor3000\relax}%
\providecommand \BibitemShut  [1]{\csname bibitem#1\endcsname}%
\let\auto@bib@innerbib\@empty
\bibitem [{\citenamefont {{Mora}}\ \emph {et~al.}(2019)\citenamefont {{Mora}},
  \citenamefont {{Regnault}},\ and\ \citenamefont {{Bernevig}}}]{mora2019flat}%
  \BibitemOpen
  \bibfield  {author} {\bibinfo {author} {\bibfnamefont {Christophe}\
  \bibnamefont {{Mora}}}, \bibinfo {author} {\bibfnamefont {Nicolas}\
  \bibnamefont {{Regnault}}}, \ and\ \bibinfo {author} {\bibfnamefont
  {B.~Andrei}\ \bibnamefont {{Bernevig}}},\ }\bibfield  {title} {\enquote
  {\bibinfo {title} {{Flatbands and Perfect Metal in Trilayer Moir{\'e}
  Graphene}},}\ }\href {\doibase 10.1103/PhysRevLett.123.026402} {\bibfield
  {journal} {\bibinfo  {journal} {\prl}\ }\textbf {\bibinfo {volume} {123}},\
  \bibinfo {eid} {026402} (\bibinfo {year} {2019})}\BibitemShut {NoStop}%
\bibitem [{\citenamefont {{Amorim}}\ and\ \citenamefont
  {{Castro}}(2018)}]{amorim2018electronic}%
  \BibitemOpen
  \bibfield  {author} {\bibinfo {author} {\bibfnamefont {B.}~\bibnamefont
  {{Amorim}}}\ and\ \bibinfo {author} {\bibfnamefont {Eduardo~V.}\ \bibnamefont
  {{Castro}}},\ }\bibfield  {title} {\enquote {\bibinfo {title} {{Electronic
  spectral properties of incommensurate twisted trilayer graphene}},}\
  }\href@noop {} {\bibfield  {journal} {\bibinfo  {journal} {arXiv e-prints}\
  ,\ \bibinfo {eid} {arXiv:1807.11909}} (\bibinfo {year} {2018})},\ \Eprint
  {http://arxiv.org/abs/1807.11909} {arXiv:1807.11909 [cond-mat.mes-hall]}
  \BibitemShut {NoStop}%
\bibitem [{\citenamefont {{Yankowitz}}\ \emph {et~al.}(2012)\citenamefont
  {{Yankowitz}}, \citenamefont {{Xue}}, \citenamefont {{Cormode}},
  \citenamefont {{Sanchez-Yamagishi}}, \citenamefont {{Watanabe}},
  \citenamefont {{Taniguchi}}, \citenamefont {{Jarillo-Herrero}}, \citenamefont
  {{Jacquod}},\ and\ \citenamefont {{Leroy}}}]{yankowitz2012emergence}%
  \BibitemOpen
  \bibfield  {author} {\bibinfo {author} {\bibfnamefont {Matthew}\ \bibnamefont
  {{Yankowitz}}}, \bibinfo {author} {\bibfnamefont {Jiamin}\ \bibnamefont
  {{Xue}}}, \bibinfo {author} {\bibfnamefont {Daniel}\ \bibnamefont
  {{Cormode}}}, \bibinfo {author} {\bibfnamefont {Javier~D.}\ \bibnamefont
  {{Sanchez-Yamagishi}}}, \bibinfo {author} {\bibfnamefont {K.}~\bibnamefont
  {{Watanabe}}}, \bibinfo {author} {\bibfnamefont {T.}~\bibnamefont
  {{Taniguchi}}}, \bibinfo {author} {\bibfnamefont {Pablo}\ \bibnamefont
  {{Jarillo-Herrero}}}, \bibinfo {author} {\bibfnamefont {Philippe}\
  \bibnamefont {{Jacquod}}}, \ and\ \bibinfo {author} {\bibfnamefont
  {Brian~J.}\ \bibnamefont {{Leroy}}},\ }\bibfield  {title} {\enquote {\bibinfo
  {title} {{Emergence of superlattice Dirac points in graphene on hexagonal
  boron nitride}},}\ }\href {\doibase 10.1038/nphys2272} {\bibfield  {journal}
  {\bibinfo  {journal} {Nature Physics}\ }\textbf {\bibinfo {volume} {8}},\
  \bibinfo {pages} {382--386} (\bibinfo {year} {2012})}\BibitemShut {NoStop}%
\bibitem [{\citenamefont {Carr}\ \emph {et~al.}(2020)\citenamefont {Carr},
  \citenamefont {Li}, \citenamefont {Zhu}, \citenamefont {Kaxiras},
  \citenamefont {Sachdev},\ and\ \citenamefont
  {Kruchkov}}]{carr2020ultraheavy}%
  \BibitemOpen
  \bibfield  {author} {\bibinfo {author} {\bibfnamefont {Stephen}\ \bibnamefont
  {Carr}}, \bibinfo {author} {\bibfnamefont {Chenyuan}\ \bibnamefont {Li}},
  \bibinfo {author} {\bibfnamefont {Ziyan}\ \bibnamefont {Zhu}}, \bibinfo
  {author} {\bibfnamefont {Efthimios}\ \bibnamefont {Kaxiras}}, \bibinfo
  {author} {\bibfnamefont {Subir}\ \bibnamefont {Sachdev}}, \ and\ \bibinfo
  {author} {\bibfnamefont {Alexander}\ \bibnamefont {Kruchkov}},\ }\bibfield
  {title} {\enquote {\bibinfo {title} {Ultraheavy and ultrarelativistic dirac
  quasiparticles in sandwiched graphenes},}\ }\href {\doibase
  https://doi.org/10.1021/acs.nanolett.9b04979} {\bibfield  {journal} {\bibinfo
   {journal} {Nano Letters}\ }\textbf {\bibinfo {volume} {20}},\ \bibinfo
  {pages} {3030–3038} (\bibinfo {year} {2020})}\BibitemShut {NoStop}%
\bibitem [{\citenamefont {{Castro Neto}}\ \emph {et~al.}(2009)\citenamefont
  {{Castro Neto}}, \citenamefont {{Guinea}}, \citenamefont {{Peres}},
  \citenamefont {{Novoselov}},\ and\ \citenamefont
  {{Geim}}}]{castro_neto2009electronic}%
  \BibitemOpen
  \bibfield  {author} {\bibinfo {author} {\bibfnamefont {A.~H.}\ \bibnamefont
  {{Castro Neto}}}, \bibinfo {author} {\bibfnamefont {F.}~\bibnamefont
  {{Guinea}}}, \bibinfo {author} {\bibfnamefont {N.~M.~R.}\ \bibnamefont
  {{Peres}}}, \bibinfo {author} {\bibfnamefont {K.~S.}\ \bibnamefont
  {{Novoselov}}}, \ and\ \bibinfo {author} {\bibfnamefont {A.~K.}\ \bibnamefont
  {{Geim}}},\ }\bibfield  {title} {\enquote {\bibinfo {title} {{The electronic
  properties of graphene}},}\ }\href {\doibase 10.1103/RevModPhys.81.109}
  {\bibfield  {journal} {\bibinfo  {journal} {Reviews of Modern Physics}\
  }\textbf {\bibinfo {volume} {81}},\ \bibinfo {pages} {109--162} (\bibinfo
  {year} {2009})}\BibitemShut {NoStop}%
\bibitem [{\citenamefont {{Carr}}\ \emph {et~al.}(2019)\citenamefont {{Carr}},
  \citenamefont {{Fang}}, \citenamefont {{Zhu}},\ and\ \citenamefont
  {{Kaxiras}}}]{carr2019exact}%
  \BibitemOpen
  \bibfield  {author} {\bibinfo {author} {\bibfnamefont {Stephen}\ \bibnamefont
  {{Carr}}}, \bibinfo {author} {\bibfnamefont {Shiang}\ \bibnamefont {{Fang}}},
  \bibinfo {author} {\bibfnamefont {Ziyan}\ \bibnamefont {{Zhu}}}, \ and\
  \bibinfo {author} {\bibfnamefont {Efthimios}\ \bibnamefont {{Kaxiras}}},\
  }\bibfield  {title} {\enquote {\bibinfo {title} {{Exact continuum model for
  low-energy electronic states of twisted bilayer graphene}},}\ }\href
  {\doibase 10.1103/PhysRevResearch.1.013001} {\bibfield  {journal} {\bibinfo
  {journal} {Physical Review Research}\ }\textbf {\bibinfo {volume} {1}},\
  \bibinfo {eid} {013001} (\bibinfo {year} {2019})}\BibitemShut {NoStop}%
\bibitem [{\citenamefont {{Fang}}\ \emph {et~al.}(2019)\citenamefont {{Fang}},
  \citenamefont {{Carr}}, \citenamefont {{Zhu}}, \citenamefont {{Massatt}},\
  and\ \citenamefont {{Kaxiras}}}]{fang2019angle}%
  \BibitemOpen
  \bibfield  {author} {\bibinfo {author} {\bibfnamefont {Shiang}\ \bibnamefont
  {{Fang}}}, \bibinfo {author} {\bibfnamefont {Stephen}\ \bibnamefont
  {{Carr}}}, \bibinfo {author} {\bibfnamefont {Ziyan}\ \bibnamefont {{Zhu}}},
  \bibinfo {author} {\bibfnamefont {Daniel}\ \bibnamefont {{Massatt}}}, \ and\
  \bibinfo {author} {\bibfnamefont {Efthimios}\ \bibnamefont {{Kaxiras}}},\
  }\bibfield  {title} {\enquote {\bibinfo {title} {{Angle-Dependent {\it Ab
  initio} Low-Energy Hamiltonians for a Relaxed Twisted Bilayer Graphene
  Heterostructure}},}\ }\href@noop {} {\bibfield  {journal} {\bibinfo
  {journal} {arXiv e-prints}\ ,\ \bibinfo {eid} {arXiv:1908.00058}} (\bibinfo
  {year} {2019})},\ \Eprint {http://arxiv.org/abs/1908.00058} {arXiv:1908.00058
  [cond-mat.mes-hall]} \BibitemShut {NoStop}%
\bibitem [{\citenamefont {{Bistritzer}}\ and\ \citenamefont
  {{MacDonald}}(2011)}]{bistritzer2011moire}%
  \BibitemOpen
  \bibfield  {author} {\bibinfo {author} {\bibfnamefont {Rafi}\ \bibnamefont
  {{Bistritzer}}}\ and\ \bibinfo {author} {\bibfnamefont {Allan~H.}\
  \bibnamefont {{MacDonald}}},\ }\bibfield  {title} {\enquote {\bibinfo {title}
  {{Moir{\'e} bands in twisted double-layer graphene}},}\ }\href {\doibase
  10.1073/pnas.1108174108} {\bibfield  {journal} {\bibinfo  {journal}
  {Proceedings of the National Academy of Science}\ }\textbf {\bibinfo {volume}
  {108}},\ \bibinfo {pages} {12233--12237} (\bibinfo {year}
  {2011})}\BibitemShut {NoStop}%
\bibitem [{\citenamefont {{Nam}}\ and\ \citenamefont
  {{Koshino}}(2017)}]{nam2017lattice}%
  \BibitemOpen
  \bibfield  {author} {\bibinfo {author} {\bibfnamefont {Nguyen N.~T.}\
  \bibnamefont {{Nam}}}\ and\ \bibinfo {author} {\bibfnamefont {Mikito}\
  \bibnamefont {{Koshino}}},\ }\bibfield  {title} {\enquote {\bibinfo {title}
  {{Lattice relaxation and energy band modulation in twisted bilayer
  graphene}},}\ }\href {\doibase 10.1103/PhysRevB.96.075311} {\bibfield
  {journal} {\bibinfo  {journal} {\prb}\ }\textbf {\bibinfo {volume} {96}},\
  \bibinfo {eid} {075311} (\bibinfo {year} {2017})}\BibitemShut {NoStop}%
\bibitem [{\citenamefont {{Carr}}\ \emph {et~al.}(2017)\citenamefont {{Carr}},
  \citenamefont {{Massatt}}, \citenamefont {{Fang}}, \citenamefont {{Cazeaux}},
  \citenamefont {{Luskin}},\ and\ \citenamefont
  {{Kaxiras}}}]{carr2017twistronics}%
  \BibitemOpen
  \bibfield  {author} {\bibinfo {author} {\bibfnamefont {Stephen}\ \bibnamefont
  {{Carr}}}, \bibinfo {author} {\bibfnamefont {Daniel}\ \bibnamefont
  {{Massatt}}}, \bibinfo {author} {\bibfnamefont {Shiang}\ \bibnamefont
  {{Fang}}}, \bibinfo {author} {\bibfnamefont {Paul}\ \bibnamefont
  {{Cazeaux}}}, \bibinfo {author} {\bibfnamefont {Mitchell}\ \bibnamefont
  {{Luskin}}}, \ and\ \bibinfo {author} {\bibfnamefont {Efthimios}\
  \bibnamefont {{Kaxiras}}},\ }\bibfield  {title} {\enquote {\bibinfo {title}
  {{Twistronics: Manipulating the electronic properties of two-dimensional
  layered structures through their twist angle}},}\ }\href {\doibase
  10.1103/PhysRevB.95.075420} {\bibfield  {journal} {\bibinfo  {journal}
  {\prb}\ }\textbf {\bibinfo {volume} {95}},\ \bibinfo {eid} {075420} (\bibinfo
  {year} {2017})}\BibitemShut {NoStop}%
\bibitem [{\citenamefont {Massatt}\ \emph {et~al.}(2018)\citenamefont
  {Massatt}, \citenamefont {Carr}, \citenamefont {Luskin},\ and\ \citenamefont
  {Ortner}}]{massatt2017incommensurate}%
  \BibitemOpen
  \bibfield  {author} {\bibinfo {author} {\bibfnamefont {Daniel}\ \bibnamefont
  {Massatt}}, \bibinfo {author} {\bibfnamefont {Stephen}\ \bibnamefont {Carr}},
  \bibinfo {author} {\bibfnamefont {Mitchell}\ \bibnamefont {Luskin}}, \ and\
  \bibinfo {author} {\bibfnamefont {Christoph}\ \bibnamefont {Ortner}},\
  }\bibfield  {title} {\enquote {\bibinfo {title} {Incommensurate
  heterostructures in momentum space},}\ }\href {\doibase 10.1137/17M1141035}
  {\bibfield  {journal} {\bibinfo  {journal} {Multiscale Modeling \&
  Simulation}\ }\textbf {\bibinfo {volume} {16}},\ \bibinfo {pages} {429--451}
  (\bibinfo {year} {2018})}\BibitemShut {NoStop}%
\bibitem [{\citenamefont {Zuo}\ \emph {et~al.}(2018)\citenamefont {Zuo},
  \citenamefont {Qiao}, \citenamefont {Ma}, \citenamefont {Yin}, \citenamefont
  {Sun}, \citenamefont {Zhang}, \citenamefont {Guan},\ and\ \citenamefont
  {He}}]{zuo2018scanning}%
  \BibitemOpen
  \bibfield  {author} {\bibinfo {author} {\bibfnamefont {Wei-Jie}\ \bibnamefont
  {Zuo}}, \bibinfo {author} {\bibfnamefont {Jia-Bin}\ \bibnamefont {Qiao}},
  \bibinfo {author} {\bibfnamefont {Dong-Lin}\ \bibnamefont {Ma}}, \bibinfo
  {author} {\bibfnamefont {Long-Jing}\ \bibnamefont {Yin}}, \bibinfo {author}
  {\bibfnamefont {Gan}\ \bibnamefont {Sun}}, \bibinfo {author} {\bibfnamefont
  {Jun-Yang}\ \bibnamefont {Zhang}}, \bibinfo {author} {\bibfnamefont
  {Li-Yang}\ \bibnamefont {Guan}}, \ and\ \bibinfo {author} {\bibfnamefont
  {Lin}\ \bibnamefont {He}},\ }\bibfield  {title} {\enquote {\bibinfo {title}
  {Scanning tunneling microscopy and spectroscopy of twisted trilayer
  graphene},}\ }\href {\doibase PhysRevB.97.035440} {\bibfield  {journal}
  {\bibinfo  {journal} {Physical Review B}\ }\textbf {\bibinfo {volume} {97}},\
  \bibinfo {pages} {035440} (\bibinfo {year} {2018})}\BibitemShut {NoStop}%
\end{thebibliography}%

\end{document}


\title{Supplemental Material for ``Twisted Trilayer Graphene: a Precisely Tunable Platform for Correlated Electrons'' }

\author{Ziyan Zhu}
\affiliation{Department of Physics, Harvard University, Cambridge, Massachusetts 02138, USA}
\author{Stephen Carr}
\affiliation{Department of Physics, Harvard University, Cambridge, Massachusetts 02138, USA}
\author{Daniel Massatt}
\affiliation{Department of Statistics, The University of Chicago, Chicago, Illinois 60637, USA}
\author{Mitchell Luskin}
\affiliation{School of Mathematics, University of Minnesota - Twin Cities, Minneapolis, Minnesota 55455, USA}
\author{Efthimios Kaxiras}
\affiliation{Department of Physics, Harvard University, Cambridge, Massachusetts 02138, USA}
\affiliation{John A. Paulson School of Engineering and Applied Sciences, Harvard University, Cambridge, Massachusetts 02138, USA}
\maketitle
The Supplemental Material includes four sections. In Section~\ref{sec:geom}, we discuss the geometry of the twisted trilayer graphene (tTLG) and calculate the higher-order moir\'e of moir\'e lengths. In Section~\ref{sec:model}, we present a detailed derivation of the momentum-space model and test its convergence. In Section~\ref{sec:mattlg}, we derive analytically the magic angles in tTLG. Finally, in Section~\ref{sec:bernevig}, we compare and contrast our results with a simplified model proposed by \citet{mora2019flat} as well as results obtained with a full model without the low-energy expansion proposed by \citet{amorim2018electronic}.

\section{Calculation of moir\'e of moir\'e lengths} \label{sec:geom}
\begin{figure}[ht!]
\centering
\begin{minipage}{0.4\linewidth}
    \includegraphics[width=1.0\linewidth]{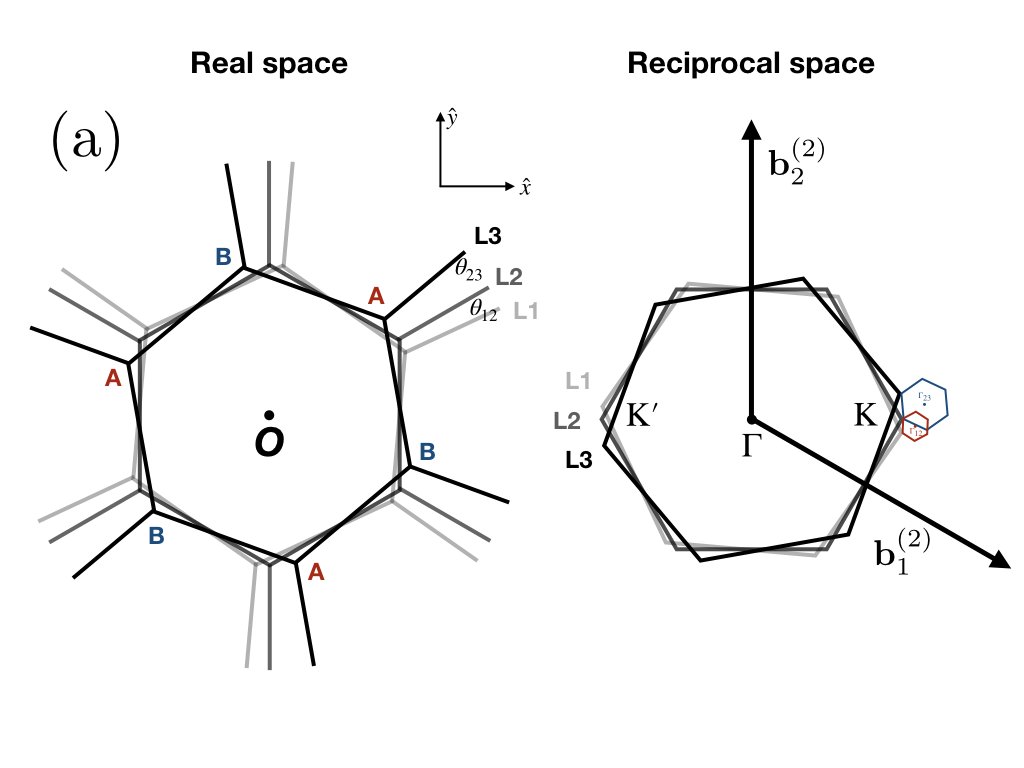} 
\end{minipage}
\begin{minipage}{0.4\linewidth}
    \includegraphics[width=1.0\linewidth]{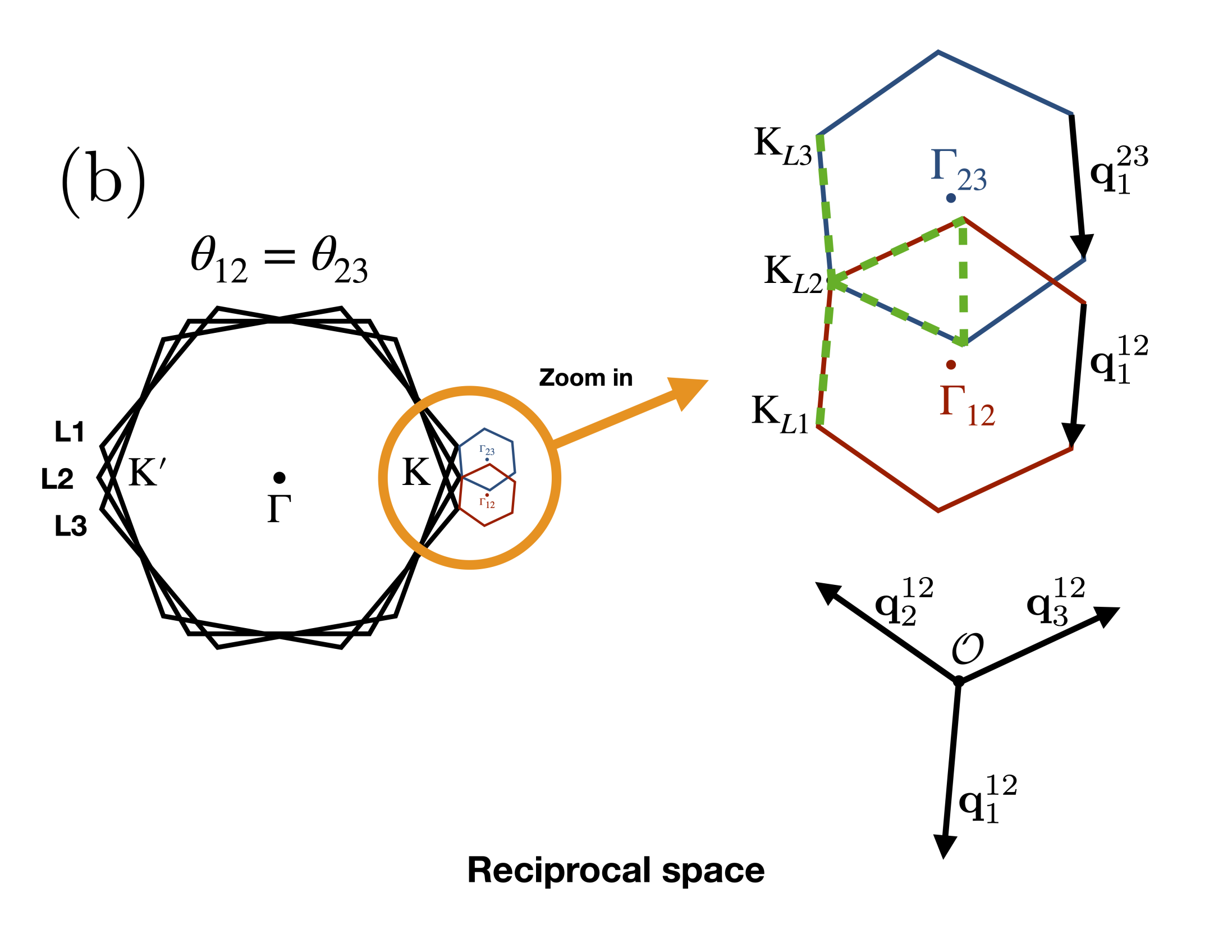} 
\end{minipage}
\caption{Lattice geometries of the tTLG system. (a) The twisted trilayer graphene system in real space (left) and momentum space with their original monolayer reciprocal lattice vectors (right). (b) Bilayer moir\'e Brillouin zone for $\theta_{12} = \theta_{23}$ with high symmetry points }
\label{fig:geom_demo}
\end{figure}

The atomic and reciprocal space geometry of tTLG with two independent twist angles are shown in Fig.~\ref{fig:geom_demo}(a).
The monolayer lattice vectors are defined as the column vectors of the following matrix: 
\begin{align}
A_0 = a_G \begin{bmatrix} 1 & 1/2 \\
0 & \sqrt{3}/2
\end{bmatrix} = \begin{bmatrix} \vec{a}_1 & \vec{a}_2
\end{bmatrix}, 
\end{align}
where $a_G = 2.4768\,$\AA \ is the graphene lattice constant (as obtained from DFT). The $l$-th layer will be referred to as L$\ell$. We assume that L2 is unrotated, with L1 rotated clockwise by $\theta_{12}$ and L3 rotated counterclockwise by $\theta_{23}$. Defining the counterclockwise rotation matrix
\begin{align}
\mathcal{R}(\theta)=
    \begin{bmatrix}
        \cos \theta  & -\sin \theta \\
        \sin \theta  & \cos \theta
    \end{bmatrix},
\end{align}
the lattice vectors of the three layers can be written as $A_1 = \mathcal{R}(-\theta_{12}) A_0$, $A_2 = A_0$, and $A_3 = \mathcal{R} (\theta_{23}) A_0$ respectively, with the column vectors denoted as $\vecll{a}_i$ for $\ell = 1, 2, 3$ and $i = 1, 2$. The monolayer reciprocal lattice vectors are given by the columns of $G_{\ell}=2\pi A_\ell^{-T}$. For example, the reciprocal lattice vectors of L2 are $\vecl{b}{2}_1 = 2\pi/a_G(1, -\sqrt{3}/3)$ and $\vecl{b}{2}_2 = 2\pi/a_G (0, 2/\sqrt{3})$. The K point of L2 is given as $\kpt{2} =  (2 \vecl{b}{2} + \vecl{b}{1}) /3= (4\pi/(3a_G), 0).$ The reciprocal lattice vectors of layers L1 and L3 can be obtained by acting $\mathcal{R}(-\theta_{12})$ and $\mathcal{R}(\theta_{23})$ on $\vecl{b}{i}$ for $i=1,2$. We also denote the monolayer unit cell of layer $\ell$ to be $\ucll$ and the reciprocal space to be $\bzll$.

The twisted trilayer system exhibits higher order moir\'e of moir\'e patterns due to the interference between the two bilayer moir\'e patterns.
To the lowest order, the bilayer moir\'e length between layers $i$ and $j$ is given by $a_M^{ij} = a_G / \sin \theta_{ij}$. We denote the bilayer moir\'e superlattice between layers $ij$ to be $\Gamma^{ij}$, spanned by the column vectors of matrix $A_{ij}=(\vec{a}^{(ij)}_1 \quad \vec{a}^{(ij)}_2)$.
The bilayer moir\'e Brillouin zone between layers $i$ and $j$ are given by the column vectors of $G_{ij} = G_j - G_i = 2\pi(A_j^{-T}  - A_i^{-T}).$ The lattice vector of the moir\'e superlattice is the column vectors of $A_{ij} = 2\pi G_{ij}^{-T}$. 
After some algebra, we obtain the lattice vectors for the two bilayer supercells, $\vec{a}_i^{12}, \vec{a}_i^{23}$ for $i=1,2$: 
\begin{align}
    \vec{a}^{12}_1 &= \frac{a_G}{2(1-\cos\theta_{12})}  \left[  (1-\cos \theta_{12}) \hat{x} - \sin\theta_{12} \hat{y}  \right],  \nonumber \\
    \vec{a}^{12}_2 &= \frac{a_G}{2(1-\cos\theta_{12})} \Bigg[ \left( \frac{1-\cos\theta_{12}}{2} +\frac{\sqrt{3}}{2}\sin\theta_{12}\right)\hat{x} 
    + \left( -\frac{1}{2} \sin\theta_{12} + \frac{\sqrt{3}}{2}(1-\cos\theta_{12}) \right)  \hat{y} \Bigg], \nonumber \\
    \vec{a}^{23}_1 &= \frac{a_G}{2(1-\cos\theta_{23})}  \left[  (1-\cos \theta_{23}) \hat{x} + \sin\theta_{23} \hat{y}  \right],  \nonumber \\
    \vec{a}^{23}_2& = \frac{a_G}{2(1-\cos\theta_{23})} \Bigg[ \left( \frac{1-\cos\theta_{23}}{2} -\frac{\sqrt{3}}{2}\sin\theta_{23}\right)\hat{x}
    + \left( \frac{1}{2} \sin\theta_{23} + \frac{\sqrt{3}}{2}(1-\cos\theta_{23})  \right)\hat{y} \Bigg].
\end{align} 
Note that there is a small twist angle between the two bilayer moir\'e superlattices. 
Moreover, for general twist angles $\theta_{12} \neq \theta_{23}$, the two bilayer moir\'e cells have a lattice mismatch. The twist angle $\phi$ and the lattice mismatch $\delta$ between the two bilayer moir\'e patterns give rise to higher-order moir\'e of moir\'e lengths. The primitive reciprocal lattice vectors of a given harmonic $(m, n)$ are given as the column vectors of $G_{mn}^\mathrm{H} = mG_{12}-n G_{23}$. Inverting $G_{mn}^\mathrm{H}$, we obtain the moir\'e of moir\'e supercell in real space $A_{mn}^\mathrm{H}=\frac{1}{2\pi} (G_{mn}^\mathrm{H})^{-T}$. The norm of the column vectors are the moir\'e of moir\'e lengths, denoted as $\lambda_{mn}^\mathrm{H}$.
For $m=n=1$, the moir\'e of moir\'e length $\lambda_{11}^\mathrm{H}$ is explicitly given as
\begin{equation}
    \lambda_{11}^\mathrm{H} = \frac{(1+\delta)a_M^{12}}{\sqrt{2(1+\delta)(1-\cos\phi)+\delta^2}}, \label{eqn:lambda11}
\end{equation}
where $\phi = \cos^{-1} \left( \frac{\vec{a}^{(12)}_1\cdot \vec{a}^{23}_1}{|\vec{a}^{(12)}_1||\vec{a}^{(23)}_1|} \right)$ is the twist angle between the bilayer moir\'e supercells and $\delta = \frac{\sin\theta_{23}}{\sin\theta_{12}}-1$ is the lattice mismatch between the two bilayer moir\'e supercells such that $a^{23}_M = (1+\delta)a^{12}_M$.
Equation~\eqref{eqn:lambda11} agrees with the first-order approximation for the moir\'e length for a twisted bilayer with a lattice mismatch~\cite{yankowitz2012emergence}, with the lattice constant being the bilayer moir\'e length between L1 and L2.

\begin{figure}[ht!]
    \centering
    \includegraphics[width=\linewidth]{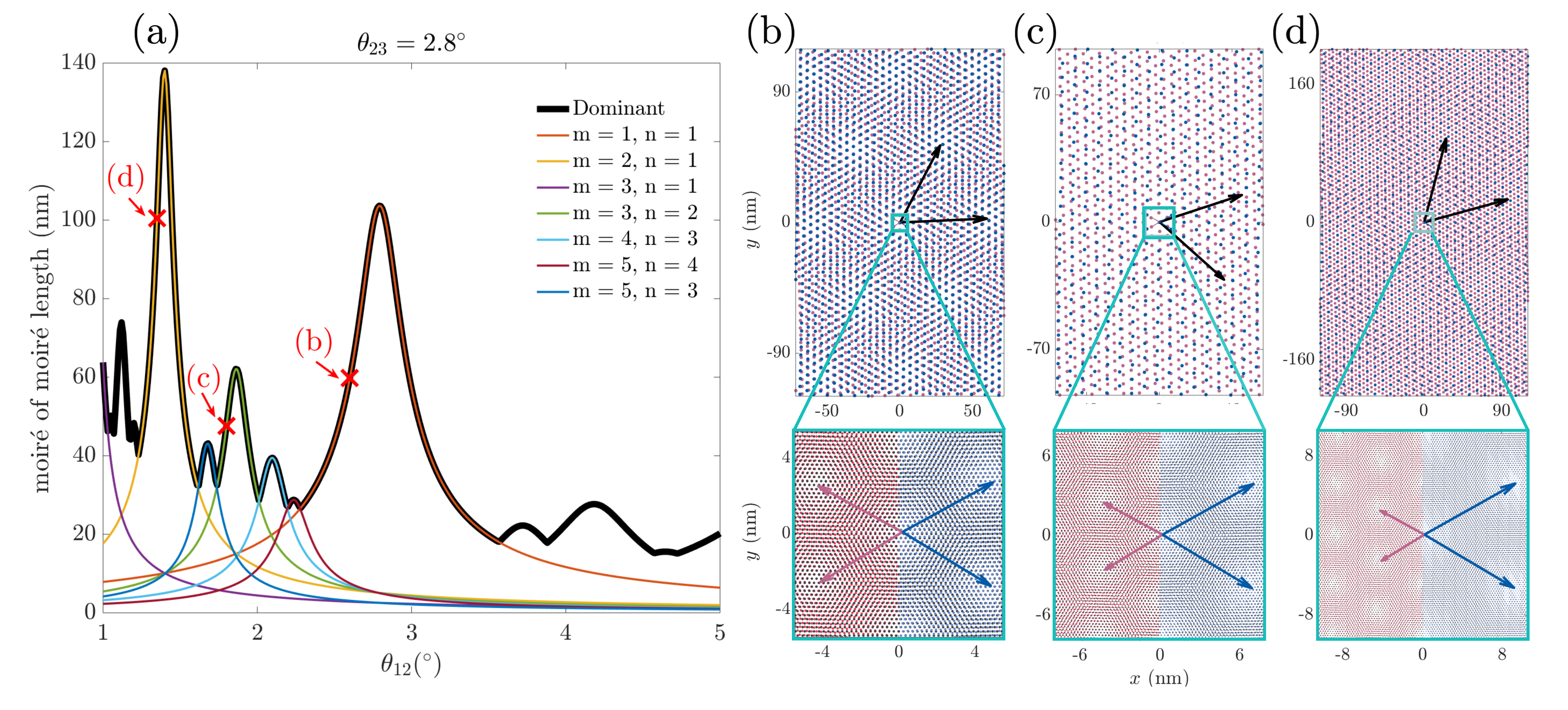}
    \caption{(a) Moir\'e of moir\'e of lengths $\lambda_{mn}^\mathrm{H}$ as a function of $\theta_{12}$ for $\theta_{23}=2.8^\circ$. Each color corresponds to a different set of $(m,n)$. The thick black line indicates the dominant length. (b)-(d) Examples moir\'e of moir\'e geometries, corresponding to the red crosses in (a). Top: red and blue scattered points are the lattice points of the bilayer moir\'e supercells between L1-L2 and L2-L3 respectively. Due to the different twist angles, the moir\'e lattice vectors are slightly rotated and have different lattice constants. Black vectors indicate estimated dominant moir\'e of moir\'e supercell lattice vectors. A blowup of the small boxed area is shown below, with 
    points representing the atomic positions of each monolayer graphene, 
    for L1-L2 on the left half and for L2-L3 on the right half. Red and blue vectors are the bilayer moir\'e lattice vectors of L1-L2 and L2-L3 respectively.}
    \label{fig:moire}
\end{figure}

A dominant moir\'e of moir\'e length does not necessarily exist nor evolve smoothly under the continuous change of the twist angle. To see this, we will consider different harmonics of the higher-order moir\'e patterns. 
To find the dominant harmonic for an arbitrary pair of twist angles, we calculate $A_{mn}^\mathrm{H}$ for $|m|,|n|\leq15$ numerically and find the $(m,n)$ such that the norm of $G_{mn}^\mathrm{H}$ is the smallest, or, equivalently, that the moir\'e of moir\'e length $\lambda_{mn}$ is largest. 
We are neglecting the cases where higher order harmonics dominate, such as the cases where $\theta_{12}$ and $\theta_{23}$ are different by more than a factor of 15. In those cases, the two bilayers moir\'e supercells have very different sizes and become essentially decoupled, which is not the focus of our study.
Figure~\ref{fig:moire}(a) shows the moir\'e of moir\'e harmonics for varying $\theta_{12}$ at a fixed $\theta_{23} = 2.8^\circ$, indicating the non-smooth dependence of the dominant moir\'e of moir\'e length on the twist angle. 
In tTLG, there exists a supercell approximation when there is a clear dominant harmonic, that is when $\theta_{12} \approx N \theta_{23}$ or $\theta_{23} \approx N \theta_{12}$ for $N \in \mathbb{Z}$.
For example, at $\theta_{12}=2.6^\circ, \theta_{23}=2.8^\circ$, the dominant harmonic is $(1,1)$ [see Fig.~\ref{fig:moire}(b)] and at $\theta_{12}=1.35^\circ, \theta_{23}=2.8^\circ$, the dominant harmonic is $(2,1)$ [see Fig.~\ref{fig:moire}(d)]. However, there are cases where there is no clear dominant moir\'e of moir\'e. For example, in Fig.~\ref{fig:moire}(c), it is difficult to visually discern a large repeating pattern and the estimated moir\'e of moir\'e lattice vectors fail to capture the relevant length scale. This is because near $\theta_{12}=1.8^\circ$, many harmonics, such as (3,2), (5,4), and (5,3), all have comparable lengths [see Fig.~\ref{fig:moire}(a) the corresponding point].

\section{Momentum-space model}\label{sec:model}

In this section, we offer a detailed derivation of our momentum-space model and density of states formalism, and study the convergence as a function of the momentum space cutoff radius. 

\subsection{Detailed derivation of momentum-space model} 

To model the electronic structure of the tTLG system, we start from a tight-binding approximation for each individual layer; we take into account the interlayer hopping in a transverse tight-binding approximation between nearest neighbors. We start by writing the Hamiltonian for the trilayer as a sum of the following terms 
\begin{equation}
    H = \sum_{l=1}^3 H^\ell + \sum_{l=1,2}\left(V^{\ell,\ell+1} + V^{\ell+1, \ell}\right), 
\end{equation} 
where $H^\ell$ is the Hamiltonian for the $\ell$-th layer and $V^{ij}$ describes the interlayer hopping. 
For simplicity, we only consider the interlayer couplings between adjacent layers. DFT calculations predict that the interlayer coupling between L1 and L3 is roughly 10 times smaller than the coupling between adjacent layers [e.g., between L1 and L2]~\cite{carr2020ultraheavy}.
In a second quantized notation, $H^\ell$ can be written as 
\begin{align}
H^\ell = -t \sum_{\R{\ell}} c_{\ell, A}^\dagger (\R{\ell})[c_{\ell, B} (\R{\ell}) + c_{\ell, B} (\R{\ell}-\vecl{a}{\ell}_{1})+c_{\ell,B} (\R{\ell}-\vecl{a}{\ell}_{2})] + h.c.,
 \end{align}
where $c_{\ell,\alpha}^\dagger$ and $c_{\ell,\alpha}$ are the creation and annihilation fermionic operators of the orbital $\alpha$ in layer $l$, $\vecl{a}{\ell}_{1,2}$ are the lattice vectors of layer $l$, and $t$ is the hopping parameter between nearest neighbors. 
As for the interlayer coupling, we define the following overlap matrix element in the tight-binding basis
\begin{equation}
    t_{\alpha\beta}^{ij} (\vecli{R}, \veclj{R}) = \bra{i, \vec{R}^{(i)}, \alpha} H \ket{j, \vec{R}^{(j)}, \beta}, 
\end{equation}
where $\alpha$ and $\beta$ denotes the sublattice degree of freedom. The interlayer Hamiltonian in the second-quantized notation is
\begin{equation}
    V^{ij} = \sum_{\vec{R}^{(i)}, \alpha, \vec{R}^{(j)}, \beta} c_{i, \alpha}^\dagger (\vec{R}^{(i)}) t_{\alpha \beta}^{ij}(\vec{R}^{(i)}, \vec{R}^{(j)}) c_{j,\beta} (\vec{R}^{(j)}). 
\end{equation}
We obtain the Hamiltonian in the momentum basis at a center site momentum $\vec{k}$. Defining $\vecll{Q} = \vecll{k} + \vec{k}$ for $\vecll{k}\in \bzll$. We perform the Fourier transform as follows
\begin{align}
    c^\dagger_{\ell,\alpha} (\vecl{R}{\ell}) = \frac{1}{\sqrt{|\bzll|}} \intkbz e^{i \vecll{Q} \cdot (\vecl{R}{\ell} + \vecl{\tau}{\ell}_{\alpha})} c^\dagger_{\ell, \vecl{k}{\ell}, \alpha}, \nonumber \\
    c_{\ell,\alpha} (\vecl{R}{\ell}) = \frac{1}{\sqrt{|\bzll|}} \intkbz e^{-i \vecll{Q} \cdot (\vecl{R}{\ell} + \vecl{\tau}{\ell}_{\alpha})} c_{\ell, \vecl{k}{\ell}, \alpha},
    \label{eqn:decompr}
\end{align}
where the integral is over the Brillouin zone of the $\ell$-th layer, $\vecl{\tau}{\ell}_A = \vec{0}$, $\vecl{\tau}{\ell}_B = 1/3 (\vecll{a}_1 + \vecll{a}_2)$. The inverse of the transform in Eq.~(\ref{eqn:decompr}) is 
\begin{align}
    c^\dagger_{\ell, \vecl{k}{\ell}, \alpha} = \frac{1}{\sqrt{|\bzll|}} \sum_{\vecll{R}}   e^{-i \vecll{Q} \cdot (\vecll{R} + \vecll{\tau}_\alpha) }c^\dagger_{\ell,\alpha} (\vecll{R}), \nonumber \\
    c_{\ell, \vecl{k}{\ell}, \alpha} = \frac{1}{\sqrt{|\bzll|}} \sum_{\vecll{R}}   e^{i \vecll{Q} \cdot (\vecll{R} + \vecll{\tau}_\alpha) }c_{\ell,\alpha} (\vecll{R}),
    \label{eqn:decompk}
\end{align} 
where $|\Gamma^{(\ell)*}|$ is the area of the Brillouin zone in the $\ell$-th layer.
The intralayer Hamiltonian in the Bloch basis can now be written as follows 
\begin{align}
    H^\ell &= -\frac{t}{|\bzll|} \sum_{\vecl{R}{\ell}}  \intkbz  \int_{\bzll} \mathrm{d}\vec{k}'^{(\ell)} e^{i (\vecl{k}{\ell} - \vec{k}'^{(\ell)}) \cdot \vecl{R}{\ell}} \sum_{\vecl{s}{\ell}_i} e^{i \vec{k} \cdot \vecl{s}{\ell}_i} c^\dagger_{\ell,\vec{k},A} c_{\ell,\vec{k}'^{(\ell)},B} \nonumber \\
    &= - t \intkbz \sum_{\vecl{s}{\ell}} e^{i \vecll{Q} \cdot \vecl{s}{\ell}_i} c^\dagger_{\ell,\vecll{k},A} c_{\ell,\vecll{k},B}, 
\end{align}
where we use the Poisson summation formula, $\sum_{\vecl{R}{\ell}} e^{i \vecll{k}\cdot \vecll{R}} = |\bzll| \sum_{\vecll{G}} \delta_{\vecll{k},\vecll{G}}$. We also define $\vecll{s}_i$ to describe the nearest neighbor separation between $A$ and $B$ sublattices, which are given as $\vecll{s}_1 = 1/3(\vecll{a}_1 + \vecll{a}_2), \vecll{s}_2 = 1/3 (-2 \vecll{a}_1 + \vecll{a}_2), \vecll{s}_3 = 1/3 (\vecll{a}_1 - 2\vecll{a}_2)$. The intralayer Hamiltonian in the basis of $c_{\ell,\vecll{k}, \alpha}$ can then be written as
\begin{equation}
    H^\ell (\vecll{Q}) = -t \begin{bmatrix}
    0 & f_\ell (\vecll{Q}) \\
    f_\ell^* (\vecll{Q}) & 0 
    \end{bmatrix}, 
\end{equation}
where $f_\ell (\vecll{Q}) = \sum_{\vecll{s}_i} e^{i \vecll{Q} \cdot \vecll{s}_i}$. The Hamiltonian is equivalent to the monolayer graphene tight-binding model at a given momentum $\vecll{Q}$ ~\cite{castro_neto2009electronic}. For the intralayer Hamiltonian, there is no constraint on $\vecll{Q}$.

Similarly, we write the interlayer Hamiltonian in the $c^\dagger_{\ell, \vecll{k}, \alpha}$ basis
\begin{align}
    V^{ij}  = \int_{\bzl{i}} \mathrm{d} \vecli{k} \int_{\bzl{j}} \mathrm{d} \veclj{k} \sum_{\alpha \beta} c^\dagger_{i,\vecl{k}{i},\alpha} T^{\alpha \beta}_{ij} (\vecl{k}{i}, \vecl{k}{j}) c_{j,\vecl{k}{j},\beta},
\end{align}
where we use Eq.~(\ref{eqn:decompr}) and 
\begin{align}
    T_{\alpha\beta}^{ij} (\vecli{k}, \veclj{k}) =  \frac{1}{\sqrt{|\bzl{i}||\bzl{j}|}}\sum_{\vecl{R}{i},\vecl{R}{j}} e^{i \vecli{Q}\cdot (\vecli{R} +\vecli{\tau}_\alpha)}  t_{\alpha\beta}^{ij} (\vecli{R},\veclj{R})  e^{-i \veclj{Q}\cdot (\veclj{R} +\veclj{\tau}_\beta)}.
    \label{eqn:Treal}
\end{align}
We now apply the two center approximation
\begin{equation}
    t_{\alpha\beta}^{ij} (\vecli{R},\veclj{R}) = t_{\alpha\beta}^{ij} (\vecli{R}+\vecli{\tau}_\alpha - \veclj{R} - \veclj{\tau}_\beta),
\end{equation}
and write the interlayer coupling in terms of a two-dimensional Fourier Transform 
\begin{align}
    t_{\alpha\beta}^{ij} (\vecli{R},\veclj{R}) = t^{ij}_{\alpha\beta} (\vecli{R}+\vecli{\tau}_\alpha - \veclj{R} - \veclj{\tau}_\beta) \nonumber \\
 = \intp e^{i\vec{p} \cdot (\vecli{R}+\vecli{\tau}_\alpha - \veclj{R} - \veclj{\tau}_\beta) } \tilde t^{ij}_{\alpha\beta} (\bm p). 
    \label{eqn:tft}
\end{align}
Plugging Eq.~(\ref{eqn:tft}) into Eq.~(\ref{eqn:Treal}), the interlayer coupling matrix element in momentum space is
\begin{align}
   T_{\alpha\beta}^{ij} (\vecli{k}, \veclj{k})  &=\frac{1}{\sqrt{|\bzl{i}||\bzl{j}|}} \intp \sum_{\vecli{R},\veclj{R}} e^{i (\vecli{Q} + \vec{p}) \cdot (\vecli{R}+\vecli{\tau}_\alpha) } t_{\alpha\beta}^{ij} (\bm p) e^{-i (\veclj{Q} + \vec{p}) \cdot (\veclj{R}+\veclj{\tau}_\beta) } \nonumber \\
    &= \sqrt{|\bzl{i}||\bzl{j}|} \sum_{\vecli{G}, \veclj{G}}\intp e^{i\vecli{G} \cdot \vecli{\tau}_\alpha}\tilde t^{ij}_{\alpha\beta}(\bm p)   e^{-i\veclj{G} \cdot \vecli{\tau}_\beta}\delta_{\vec{k}+\vecli{k}-\bm p, \vecli{G}} \delta_{\vec{k}+\veclj{k}-\bm p, \veclj{G}}\nonumber\\
   & =  \frac{1}{|\Gamma|}
    \sum_{\vecli{G}, \veclj{G}}
    e^{i \vecli{G} \cdot \vecli{\tau}_\alpha}
    \tilde{t}^{ij}_{\alpha \beta} (\vecli{k}+\vec{k}-\vecli{G})
    e^{-i \veclj{G} \cdot \veclj{\tau}_\beta}
    \delta_{\vecli{k} - \vecli{G}, \veclj{k} - \veclj{G}}.
    \label{eqn:inter_hopping}
\end{align} In the last step, we use the Possion summation rule and $|\bzll| = 4\pi^2 |\Gamma|^{-1}$, where $|\Gamma|$ is the monolayer unit cell area. We have obtained the scattering selection rule $\vecli{k} - \vecli{G} = \veclj{k} - \veclj{G}$ for $i = j \pm 1$, which imposes the constraint on the values of allowed $\vecll{k}$.

Combining the intralayer and interlayer terms, the Hamiltonian in the $c^\dagger_{\ell,\vecll{k},\alpha}$ basis can be represented as a $3\times3$ block given in Eq.~(1) of the main text.

\subsection{Low-energy limit}\label{sec:low_energy}
We can greatly simplify the model by taking the low-energy limit. 
Each $H^\ell$ can be expanded around its Dirac point, $\vecll{k} = K_{\mathrm{L}\ell} + \vec{q}^{(\ell)}$, as a rotated Dirac Hamlitonian $H_D^\ell (\vec{q})$ for $\vec{q} = \vec{k} + \vecll{k} - K_\mathrm{L\ell}$:
\begin{align}
    H^1 (\vec{k}) \approx H_D^1 (\vec{q}) = v_F\begin{bmatrix}
    0 & e^{i \theta_{12}} q_+ \\ 
    e^{-i \theta_{12}} q_- & 0
    \end{bmatrix}, \nonumber \\ 
    H^2 (\vec{k}) \approx H_D^2 (\vec{q}) = v_F \begin{bmatrix}
    0 & q_+ \\ 
   q_-  & 0
    \end{bmatrix}, \nonumber \\
    H^3 (\vec{k}) \approx H_D^3 (\vec{q}) = v_F  \begin{bmatrix}
    0 & e^{-i \theta_{23}} q_+ \\ 
    e^{i \theta_{23}} q_-  & 0
    \end{bmatrix},
\end{align}
where $q_\pm = q_x \pm i q_y$.
For the interlayer coupling, we substitute $\vecll{k} = \vecll{q} + \kpt{\ell}$ into Eq.~\eqref{eqn:inter_hopping}, 
\begin{align}\label{eqn:inter_q}
   & T_{\alpha\beta}^{ij} (\vecli{q}, \veclj{q}) = \frac{1}{|\Gamma|} \sum_{\vecli{G}, \veclj{G}} e^{i \vecli{G} \cdot \vecli{\tau}_\alpha} \tilde t_{\alpha\beta}^{ij} (\vec{k}+\kpt{i} +\vecli{q} +\vecli{G}) e^{-i \veclj{G} \cdot \veclj{\tau}_\alpha} \delta_{\vecli{q}+\kpt{i}-\vecli{G}, \veclj{q}+\kpt{j}-\veclj{G}}.
\end{align}
For momenta near the Dirac point, since $|\vecli{q}|, |\vec{k}| \ll |\kpt{i}|, |\vecli{G}|$ we can approximate $\tilde t_{\alpha\beta}^{ij}(\vec{k} + \kpt{i} + \vecli{q} + \vecli{G}) \approx \tilde t_{\alpha\beta}^{ij} (\kpt{i} + \vecli{G}).$ This approximation can lead to the suppression of particle-hole asymmetry in the tight-binding model~\cite{carr2019exact,fang2019angle}. Due to the rapid decay of the hopping parameter $\tilde t(\bm p)$ as $\bm p$ increases~\cite{bistritzer2011moire}, we keep only the first shell in the summation in Eq.~\eqref{eqn:inter_q}:
\begin{align}
    T_{\alpha\beta}^{ij} (\vecli{q}, \veclj{q})  = \sum_{n=1}^3 T_{n, \alpha\beta}^{ij} \delta_{\vecli{q}-\veclj{q}, -\vec{q}_n^{ij}}, 
\end{align}
where $\vec{q}_1^{ij} = \kpt{i} - \kpt{j}$, $\vec{q}_2^{ij} = \mathcal{R}^{-1}(2\pi/3) \vec{q}_1^{ij}$, and $\vec{q}_3^{ij} = \mathcal{R}(2\pi/3) \vec{q}_1^{ij}$ (see Fig.~\ref{fig:geom_demo}b). We include out-of-plane relaxation by letting $t^{ij}_{AA} = t^{ij}_{BB} = \omega_0 = 0.07 \, \mathrm{eV} $ and $t^{ij}_{AB} = t^{ij}_{BA} = \omega_1 = 0.11 \, \mathrm{eV},$ which matches with the interlayer coupling in \citet{nam2017lattice} and \citet{carr2019exact}. In matrix form, 
\begin{align}
    T^{ij}_1 = \begin{bmatrix} 
    \omega_0 & \omega_1 \\
    \omega_1 & \omega_0
    \end{bmatrix},  
    T^{ij}_2 = \begin{bmatrix} 
    \omega_0 & \omega_1\bar{\phi} \\
    \omega_1 \phi & \omega_0
    \end{bmatrix}, 
    T^{ij}_3 = \begin{bmatrix} 
    \omega_0 & \omega_1\phi \\
    \omega_1\bar{\phi} & \omega_0
    \end{bmatrix},
\end{align}
where $\phi = \exp(i\frac{2\pi}{3})$, $\bar{\phi} = \exp(-i\frac{2\pi}{3})$.

\begin{figure*}[t!]
    \centering
    \includegraphics[width=\linewidth]{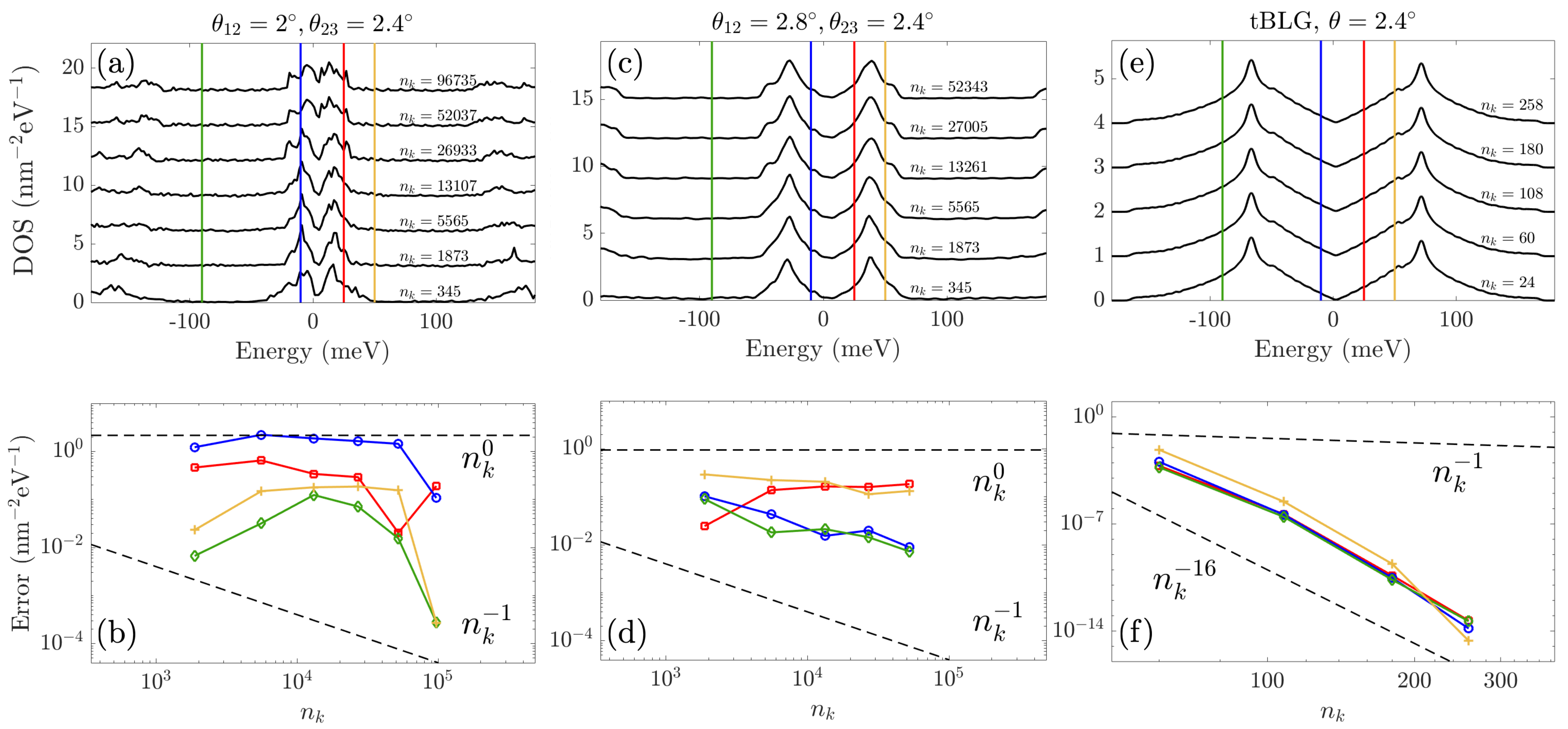}
    \caption{Top: DOS obtained using different sizes of momentum-space basis, $n_k$. Bottom: errors in the DOS corresponding to the vertical lines on the top with the same color. The error is defined as the difference between the DOS value at the given energy and the DOS at the largest cutoff shown on the top. The dashed lines are guides to the eyes showing power law scaling of the error as a function of $n_k$. }
    \label{fig:convergence}
\end{figure*}

\subsection{Density of States}
The DOS at a given energy $\epsilon$, $\mathcal{D} (\epsilon)$, for an incommensurate tight-binding model is defined as~\cite{carr2017twistronics}
\begin{align}
    \mathcal{D} (\epsilon) = \sum_r \frac{1}{N} \sum_{n = 1}^N \delta (\epsilon - \epsilon_n) |\psi_n (r)|^2, 
\end{align}
where the $r$ sum is over all real space lattice positions, $n$ is the band index, and $\psi_n (r)$ is the corresponding eigenfunction.
To obtain the DOS numerically, we use a Gaussian function $\phi_{\epsilon, \kappa} (x) = \frac{2\sqrt{\ln 2}}{\sqrt{\pi} \kappa} \exp[-4\ln 2\frac{(x-\epsilon)^2}{\kappa^2}]$ to approximate the $\delta$ function, and $\kappa$ is the full-width-half-maximum of the Gaussian, which determines the energy resolution of the DOS~\cite{massatt2017incommensurate}. We can transform the DOS equation to momentum space:
\begin{align}
    \mathcal{D} (\epsilon) = \frac{\mathcal{N}}{2} \sum_{\alpha=A,B} \sum_{\ell = 1, 2} \int_{\bzl{\ell,\ell+1}} \sum_n \phi_{\epsilon, \kappa}(\epsilon_{n,\vec{k}}) |\psi_{n,\vec{k}} |^2\  \mathrm{d} \vec{k},
\end{align}
where $\mathcal{N}$ is a normalization constant, $\epsilon_{n,\vec{k}}$ is an energy within the energy window $[\epsilon - \Delta \epsilon/2, \epsilon + \Delta \epsilon/2),$ $\Delta \epsilon$ is the energy interval, $\psi_{n, \vec{k}}$ and $\epsilon_{n,\vec{k}}$ is an eigen-pair of the Hamiltonian $\mathcal{H} (\vec{k})$ in Eq.~(1) of the main text associated with the center site $\vec{k}$ and band $n$. The integral is evaluated over the bilayer moir\'e Brillouin zone between layers $\ell$ and $\ell+1$, $\bzl{\ell, \ell+1}$, and we discretize $\bzl{\ell, \ell+1}$ using a $22\times22$ grid to evaluate the integral. We adapt $\kappa$ based on the area of the integration domain $\bzl{\ell, \ell+1}$ as $\theta_{\ell,\ell+1}$ changes. 

In order to make a direct comparison between the DOS at different twist angles, we need to properly normalize the DOS. For a given cutoff radius, we first calculate the DOS of the intralayer Hamiltonian only, which reduces to three independent copies of monolayer graphene. Near the charge-neutrality point, the DOS per $\mathrm{eV}$ per $\mathrm{nm}^2$ is given by~\cite{castro_neto2009electronic}
\begin{equation}
    \mathcal{D} (\epsilon) = \frac{6}{\pi}\frac{|\epsilon|}{v_F^2} , 
    \label{eqn:dos_slope}
\end{equation}
where the prefactor includes a factor 3 from the number of layers as well as a factor of 4 from spin and valley degeneracies. 
We then obtain a normalization constant by fixing the prefactor to the expected slope given in Eq.~(\ref{eqn:dos_slope}) and use the same constant for the DOS of the full Hamiltonian.

\subsection{Convergence} \label{sec:convergence} 
The incommensurability of the tTLG system leads to an infinite number of coupled momenta within any finite cutoff radius. Due to the additional constraints we impose on the magnitude of $\vecl{G}{\ell}$, we neglect degrees of freedom that can contribute to the low energy states. As a result, there is no guaranteed convergence. Figure~\ref{fig:convergence}(a)-(d) shows the DOS and the corresponding errors for different numbers of momentum degrees of freedom for tTLG with two different sets of twist angles. In both cases, as the cutoff increases, the error does not decay significantly. Note that in the case of $\theta_{12}=2^\circ, \theta_{23}=2.4^\circ$, the drop in error is most likely a numerical artifact and further increasing the cutoff will not likely to reduce the error. However, the physically relevant features, such as the magnitude of the DOS maximum and the positions of the VHS, are relatively stable as the cutoff increases. In contrast, Fig.~\ref{fig:convergence}(e), (f) shows the fast convergence of the DOS in tBLG as a function of cutoff radius. This is because in tBLG, increasing the cutoff radius does not change the number of relevant low-energy degrees of freedom. In this work, we choose a cutoff at the 4$^\textrm{th}$ honeycomb shell (i.e., $k_c=4|\vecll{b}|$ corresponding to $\sim 5\,600$ momenta). This choice was made by considering both computational efficiency and the accuracy of physical properties of interest. 

\section{Effective Hamiltonian and renormalized Fermi velocity}\label{sec:mattlg}
\begin{figure*}[ht!]
    \centering
    \includegraphics[width=0.8\linewidth]{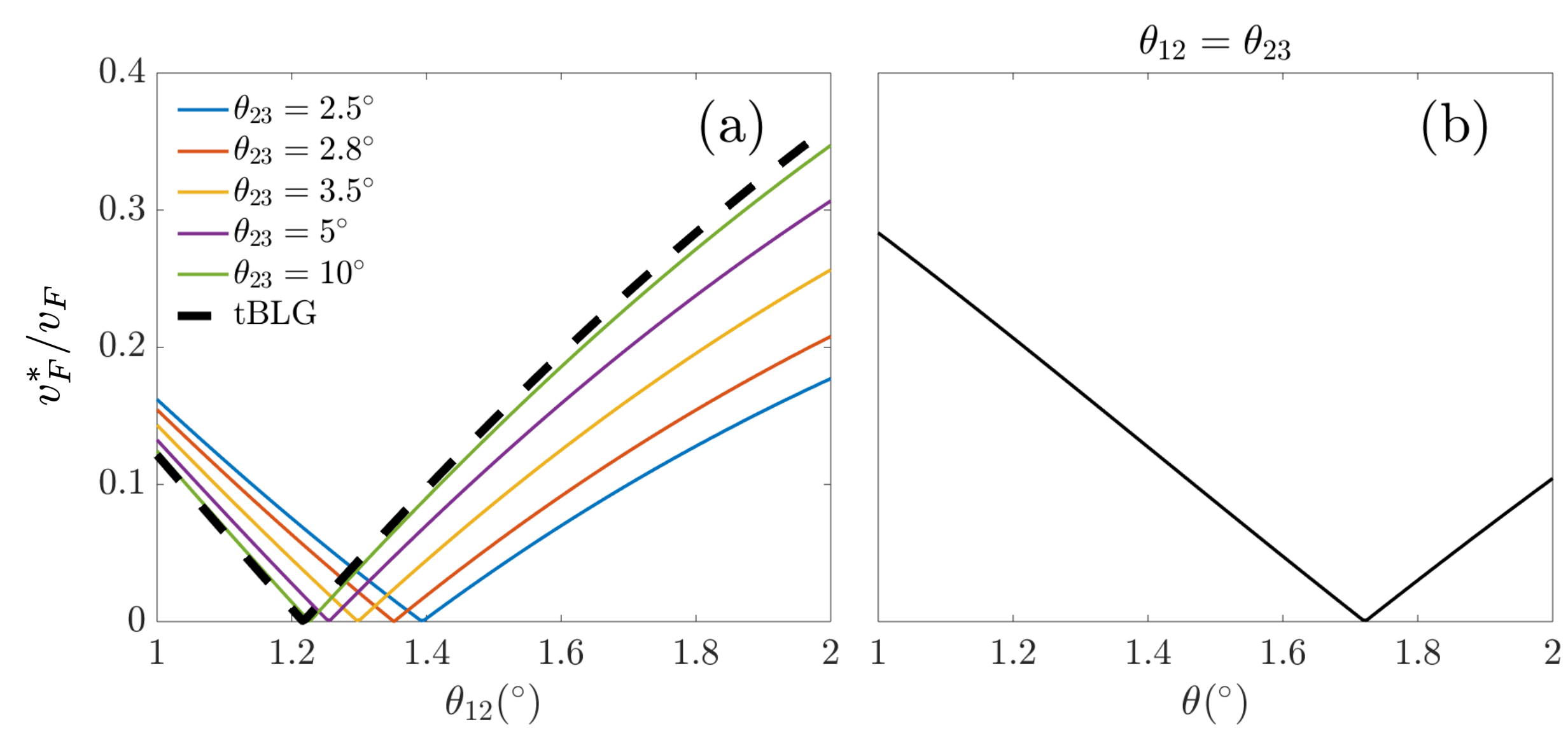}
    \caption{(a) The ratio of renormalized Fermi velocity $v_F^*$ to the monolayer Fermi velocity $v_F$ as a function of $\theta_{12}$ for given values of $\theta_{23}$. Black dashed line shows the tBLG $v_F^*/v_F$ ratio. (b) $v_F^*/v_F$ ratio as a function of twist angle for $\theta_{12}=\theta_{23}$.}
    \label{fig:vstar_vs_theta}
\end{figure*}

We examine the limit in which the momentum-space is truncated at the first honeycomb shell. The truncation gives rise to the following $14\times14$ Hamiltonian: 
\begin{equation}
    \mathcal{H} (\vec{q}) =
        \begin{bmatrix}
        H^1_D (\vec{q}+\vec{q}^{12}_1) &  &  & (T_1^{12})^\dagger &  &  &  \\
         & H^1_D (\vec{q}+\vec{q}^{12}_2) &  & (T_2^{12})^\dagger &  &  & \\
         &  & H^1_D (\vec{q}+\vec{q}^{12}_3)  & (T_3^{12})^\dagger &  &  & \\
        T_1^{12}& T_2^{12} & T_3^{12} & H^2_D (\vec{q}) & T_1^{23} & T_2^{23} & T_3^{23} \\
         &  &  & (T_1^{23})^\dagger & H^3_D (\vec{q} + \vec{q}^{23}_1) \\
         &  &  & (T_2^{23})^\dagger &  & H^3_D (\vec{q} +  \vec{q}^{23}_2)  \\
         &  &  & (T_3^{23})^\dagger &  &  & H^3_D (\vec{q} + \vec{q}^{23}_3)
        \end{bmatrix}. \label{eqn:heff}
\end{equation}
This Hamiltonian acts on seven two-component spinors
$\Psi = (\psi_{11}, \psi_{12}, \psi_{13}, \psi_{20}, \psi_{31}, \psi_{32}, \psi_{33})$, where $\ell$ and $j$ in $\psi_{\ell j}$ denote the layer and the momentum basis index respectively. Using this Hamiltonian, we can derive an expression for the renormalized Fermi velocity $v_F^*$. We first define the dimensionless quantities $\alpha_{12} = \omega/v_F k_{\theta_{12}}$ and $\alpha_{23} = \omega/v_F k_{\theta_{23}}$, where $k_{\theta_{ij}} = \frac{8\pi \sin{ (\theta_{ij} /2) } } {3 a_G}$. For simplicity, we assume $\omega_0=\omega_1 = \omega$ and neglect the angular dependence in $H_D^\ell$ by letting $H_D^\ell$ be an unrotated Dirac Hamiltonian: $H_D^\ell (\vec{q}) = v_F \vec{\sigma} \cdot \vec{q}$, where $\vec{\sigma}=(\sigma_x,-\sigma_y)$ is the Pauli matrix.  
The zero-energy state of the Hamiltonian satisfies $\mathcal{H} \Psi =\sum_{j=1}^7 c_j \Psi_j = \vec{0}$, where $c_j$ is the column vectors of $\mathcal{H}$, and $\Psi_j$ is the $j$-th component of the spinor $\Psi$. Therefore, we obtain the following relation between components of $\Psi$
\begin{equation}
    \Psi_j = - (H_D^{\ell})^{-1} T^\dagger \psi_{20},
\end{equation}
where $j\neq 4$ ($\Psi_j$ is not a state on L2 or $\psi_{20}$). 
Using this, the effective Hamiltonian to the leading order in $\vec{q}$ is 
\begin{align}
    \bra{\Psi} \mathcal{H}^2_D (\vec{q}) \ket{\Psi} &=  \frac{v_F} {1+6(\alpha_{12}^2 + \alpha_{23}^2)} \psi_{20}^{\dagger} \Bigg\{ \vec{\sigma} \cdot \vec{q} + \sum_{n=1}^3 \Big[T_n^{12} (H_D^1 (\vec{q}+\vec{q}_n^{12}))^{-1} (\vec{\sigma} \cdot \vec{q}) (H_D^1 (\vec{q}+\vec{q}_n^{12}))^{-1} T^{12 \dagger}_n  \nonumber \\
    &\quad + T_n^{23} (H_D^3 (\vec{q}+\vec{q}_n^{23}))^{-1} (\vec{\sigma} \cdot \vec{q}) (H_D^3 (\vec{q}+\vec{q}^{23}_n))^{-1} T^{23 \dagger}_n\Big] \Bigg\} \psi_{20} \nonumber \\
    &= v_F^* \psi_{20}^{\dagger} \vec{\sigma} \cdot \vec{q} \psi_{20},
\end{align}
where the renormalized Fermi velocity $v^\star_F$ is 
\begin{equation}
    \frac{v_F^*}{v_F} = \frac{1-3(\alpha_{12}^2+\alpha_{23}^2)}{1+6(\alpha_{12}^2+\alpha_{23}^2)}.
\end{equation}
Figure~\ref{fig:vstar_vs_theta}(a) shows the $v_F^*$ to $v_F$ ratio as a function of $\theta_{12}$ at a few values of $\theta_{23}$. As $\theta_{23}$ increases, the $v_F^*/v_F$ ratio approaches the tBLG curve. Figure~\ref{fig:vstar_vs_theta}(b) shows $v_F^*/v_F$ for equal twist angles, which shows that perturbation theory predicts that $v_F^*$ can still go to zero at $1.72^\circ$. However, in our numerical calculation using the full Hamiltonian, we do not observe a complete flattening of bands at this twist angle.

\begin{figure}[ht!]
    \centering
    \includegraphics[width=0.5\linewidth]{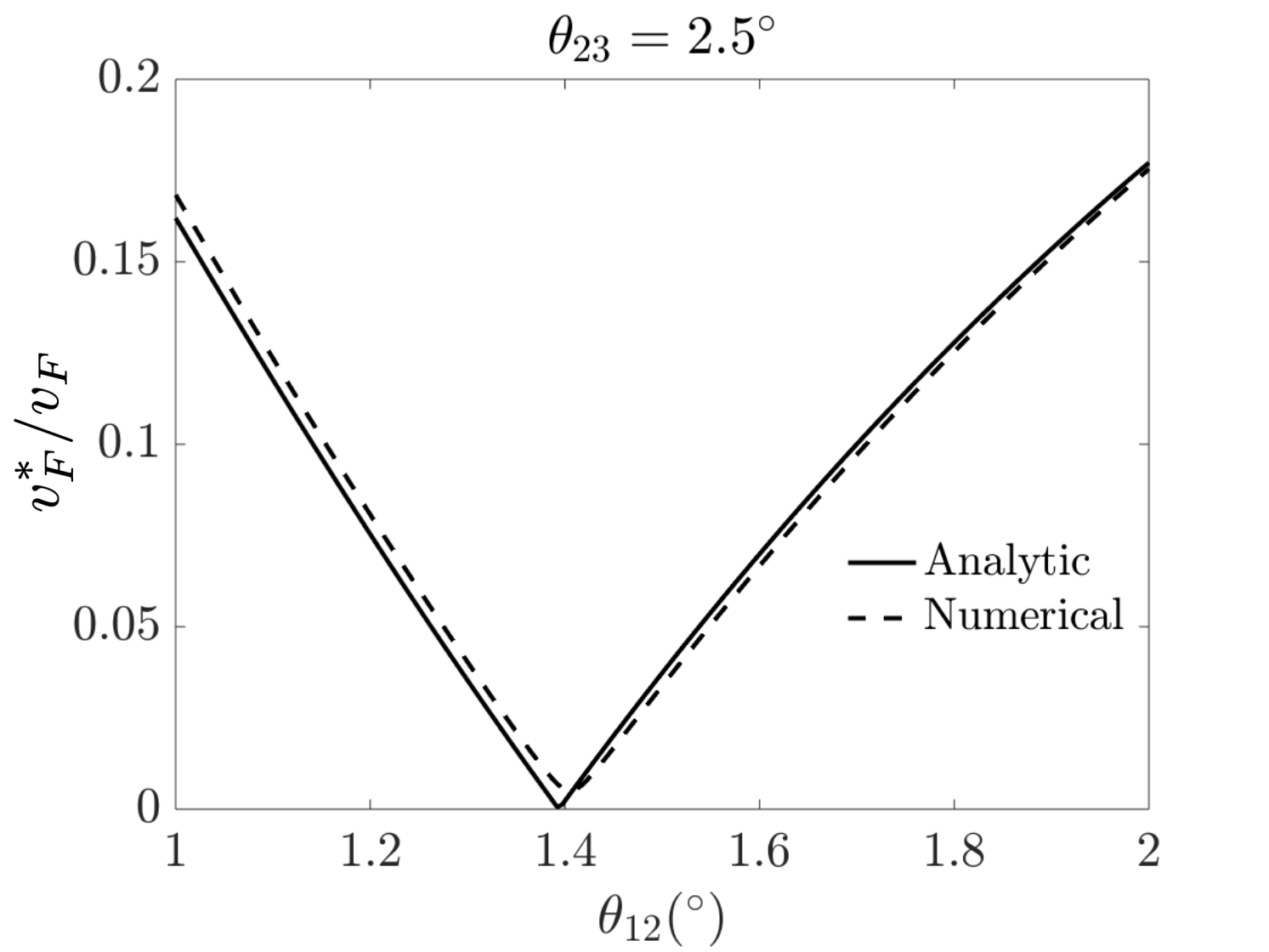}
    \caption{Comparison between the renormalized Fermi velocity $v_F^*$ of the Hamiltonian in Eq.~\eqref{eqn:heff} calculated analytically (solid line) and numerically (dashed line).}
    \label{fig:vstar_a_n}
\end{figure}

Finally, we show that our assumption in the analytic calculation of an unrotated Dirac Hamiltonian for the intralayer Hamiltonian and $\omega_0 = \omega_1$ does not significantly change the magic angle estimate. Figure~\ref{fig:vstar_a_n} compares the $v_F^*$ obtained analytically and numerically and show that the two curves and the magic angle do not differ significantly. In the numerical calculation, we diagonalize the $14\times14$ Hamiltonian with rotated Dirac equation for the intralayer terms and $\omega_0=0.07\,\mathrm{eV}, \omega_1=0.11\,\mathrm{eV}$ for the interlayer terms. At $\theta_{23} = 2.5^\circ$, the magic angle obtained analytically and numerically differ by 1.1\%.

\section{Comparison to other models}\label{sec:bernevig}

\begin{figure}[ht!]
\centering
\includegraphics[width=0.4\linewidth]{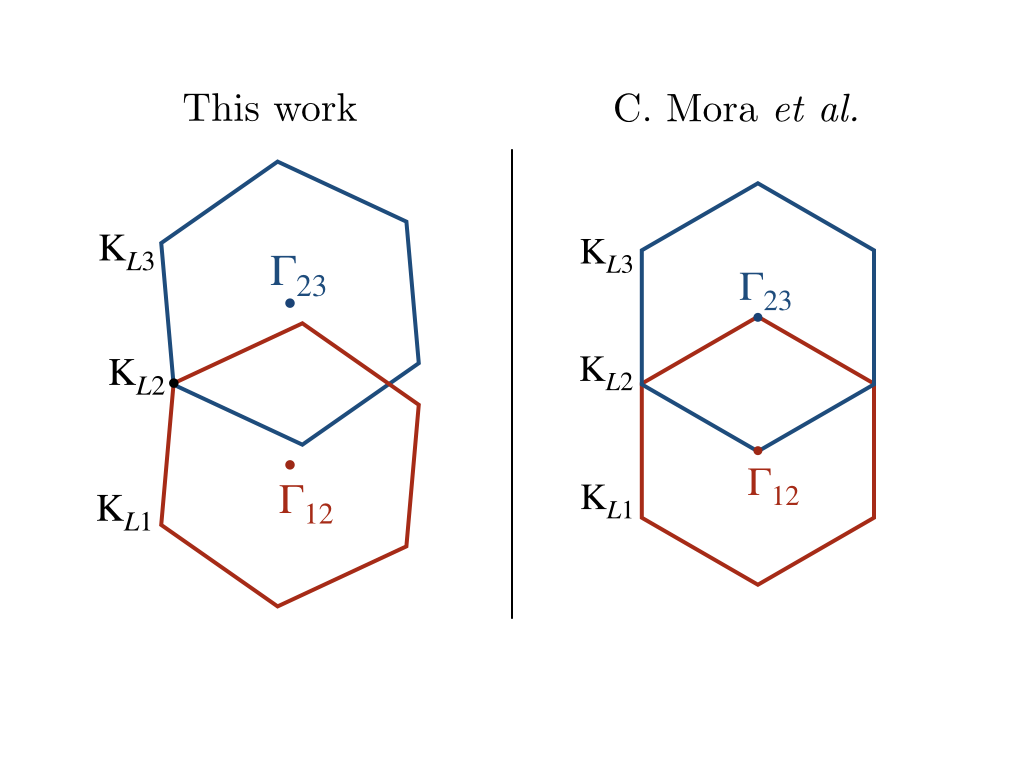}
\caption{Comparison of the bilayer moir\'e Brillouin zone geometry between our model and  \citet{mora2019flat} model with $\theta_{12}=\theta_{23}$. Left: two bilayer moir\'e Brillouin zones are misaligned by a small twist angle; right: the two bilayer moir\'e Brillouin zones are approximated to be aligned.}
\label{fig:geom_compare}
\end{figure}

\begin{figure}[ht!]
\centering
\includegraphics[width=0.7\linewidth]{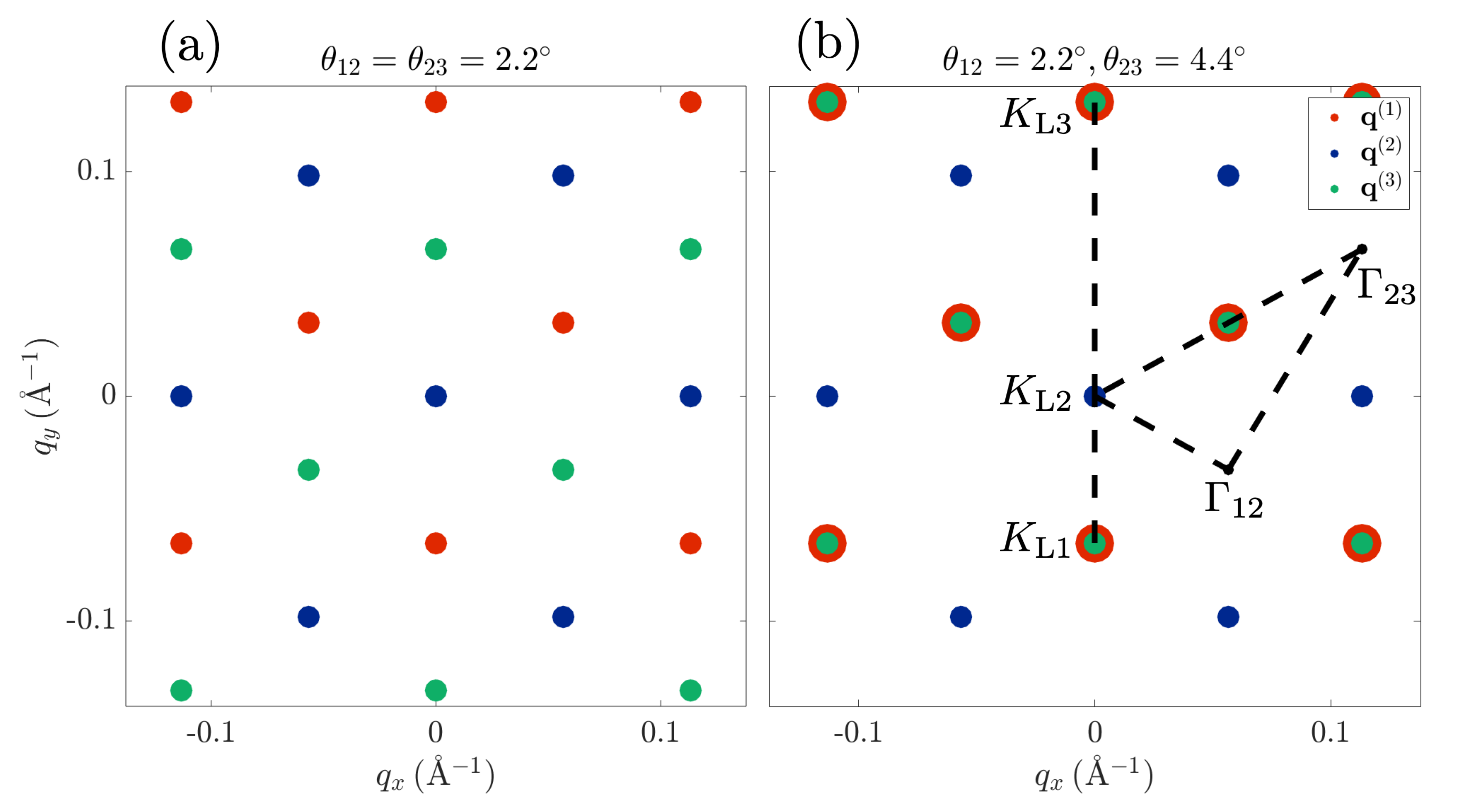}
\caption{The momentum degrees of freedom in the low-energy limit of the simplified model for (a) $\theta_{12}=\theta_{23}$ and (b) $2\theta_{12}=\theta_{23}$ with high symmetry points.}
\label{fig:kdof_ba}
\end{figure}

In this section, we compare our results to two other works~\cite{mora2019flat,amorim2018electronic}.
We first compare our results with the model proposed by Mora {\it et al}.~\cite{mora2019flat} and use it to gain further insights into our findings. In this alternate model, a different momentum-space basis is used by aligning the two bilayer moir\'e Brillouin zones [Fig.~\ref{fig:geom_compare}]. This approximation ignores the incommensurability of the system, making a two-dimensional momentum space crystal with the periodicity of the bilayer moir\'e Brillouin zone. As a result, the problem's complexity reduces to that of a bilayer. Formally, the Hamiltonian can still be written as the $3\times 3$ block as in Eq.~(1) in the main text, but the size of the basis is reduced to be on the same order as tBLG.
We implemented two cases: (1) $\theta_{12}=\theta_{23}$ and (2) $2 \theta_{12} = \theta_{23}$.
Figure~\ref{fig:kdof_ba} shows the momentum-space basis for these two cases. In case (2), the larger bilayer Brillouin zone (L1-L2) is folded onto the smaller Brillouin zone (L2-L3) in momentum space. This model essentially describes a system consisted of $2\times2$ L1-L2 moir\'e supercell and a L2-L3 moir\'e supercell. Figure~\ref{fig:ba_compare_dos_21} shows a comparison between the DOS obtained from the two models. 
We keep the values of $\omega_0$ and $\omega_1$ the same as our model and use the same approach to normalize the DOS for a direct comparison. We cut off the basis at the $4^{\mathrm{th}}$ shell and use a grid size $22\times 22$ for the density of states. The Gaussian FWHM we use is 5$\,$meV for $\theta < 2^\circ$ and 8$\,$meV for $\theta\geq2^\circ$, where $\theta$ is the twist angle that determines the size of the Brillouin zone. 

\begin{figure*}[ht!]
\centering
\includegraphics[width=0.8\textwidth]{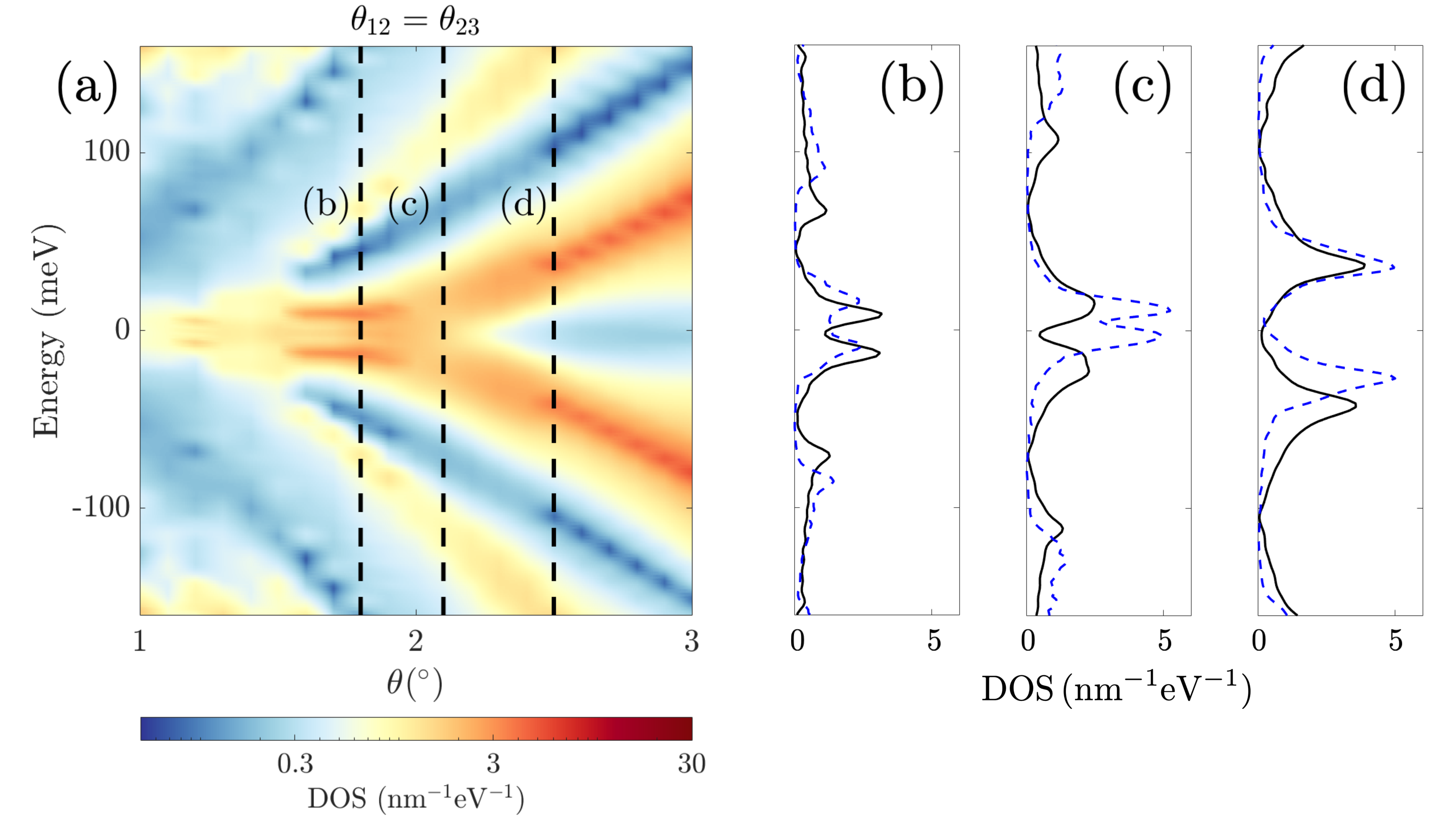}
\caption{DOS obtained with the \citet{mora2019flat} model DOS on a logarithmic color scale at equal twist angles (same color scale as Fig. 3(a) in the main text for a direct comparison) (b)-(d) DOS states along the black dashed line in (a) for (b) $\theta=1.8^\circ$, (c) $\theta=2.1^\circ$, (d) $\theta=2.5^\circ$, where the black solid lines are obtained using the \citet{mora2019flat} model, and the blue dashed lines are obtained using our full model.} 
\label{fig:bernevig}
\end{figure*}

\begin{figure*}[ht!]
\centering
\includegraphics[width=0.7\textwidth]{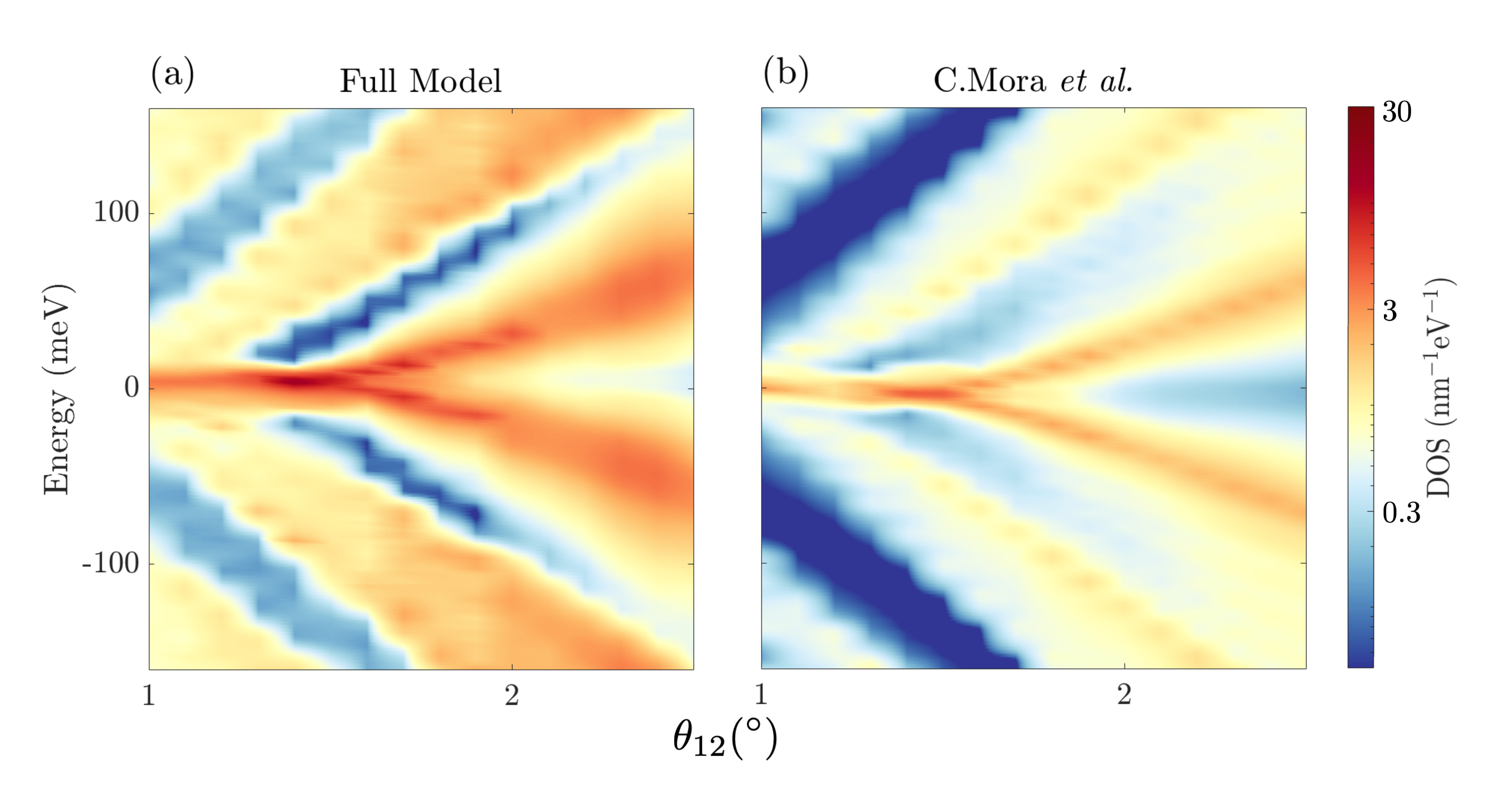}
\caption{DOS as a function of $\theta_{12}$ with $2\theta_{12} = \theta_{23}$ using (a) our full model and (b) the \citet{mora2019flat} model, both on a logarithmic color scale.}
\label{fig:ba_compare_dos_21}
\end{figure*}

For $\theta_{12} = \theta_{23}$, Fig.~\ref{fig:bernevig}(a) shows the DOS obtained with the simplified model, which agrees qualitatively with the DOS from our model [Fig.~3(a) of the main text].
However, here the DOS has the sharpest peak between $1.7^\circ$ and $1.8^\circ$, and at $2.1^\circ$ the VHS have a larger width compared to our model. Figure~\ref{fig:bernevig}(b)-(d) shows that the location of peaks away from the CNP are also very different from our model.

\begin{figure*}[ht!]
\centering
\includegraphics[width=0.8\textwidth]{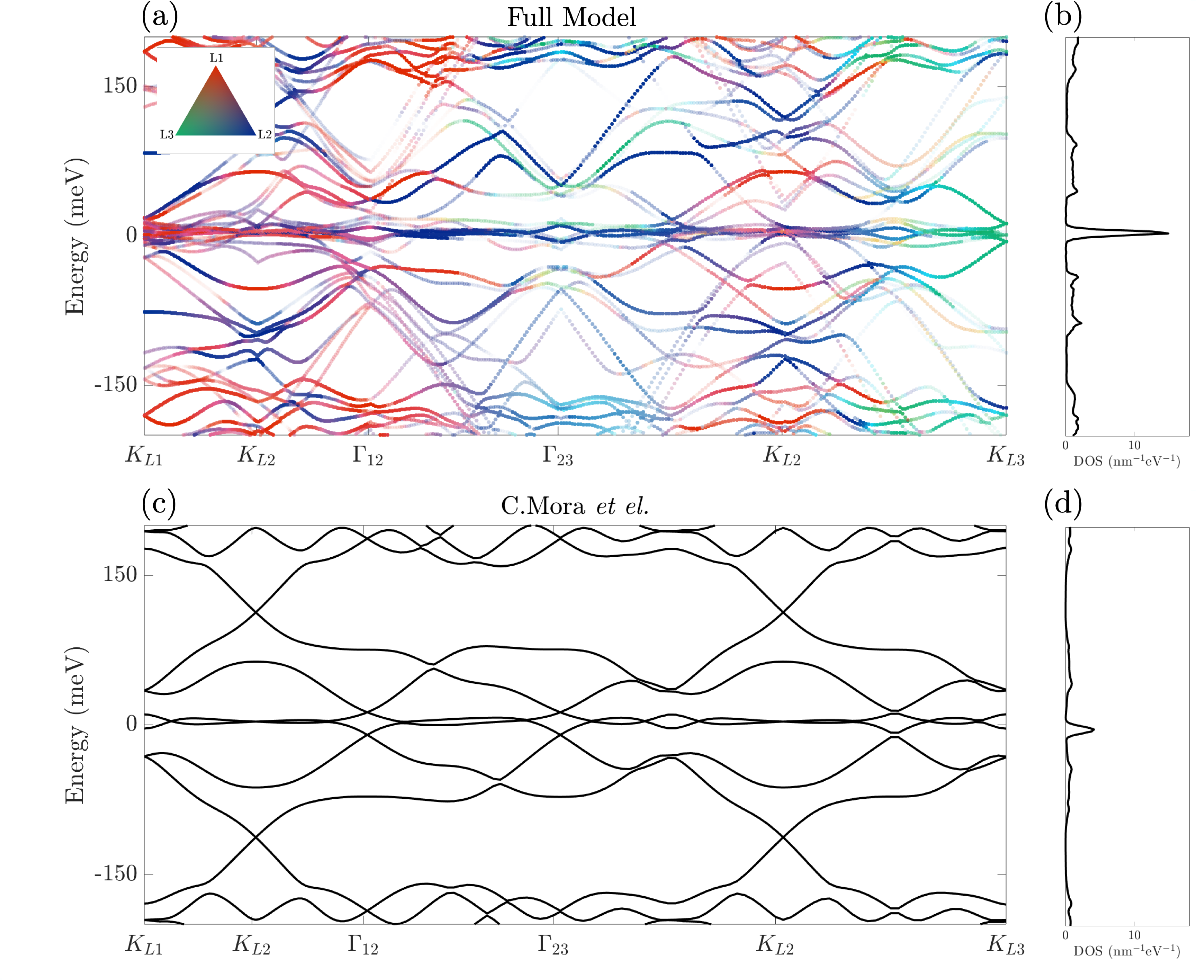}
\caption{Comparison of band structures and DOS at $\theta_{12}=1.4^\circ,\theta_{23}=2.8^\circ$ between our model (top) along the green dashed line in Fig.~\ref{fig:geom_demo}(b) and
the \citet{mora2019flat} model (bottom) along the black dashed line in Fig.~\ref{fig:kdof_ba}(b). In (a), colors represent the weight of the wavefunctions at the center site. Red, blue, and green represents weights purely from L1, L2, L3 respectively, and colors in between represent hybridization between different layers. A colormap is provided on the top left corner. The DOS from the two models are shown on the same scale. }
\label{fig:ba_compare_bands_21}
\end{figure*}

For $2\theta_{12}=\theta_{23}$, the two models predict similar trend for the VHS evolution, and the simplified model makes the right prediction for the magic angle. This is expected from perturbation theory, since the magic angle condition does not rely on the existence of a moir\'e of moir\'e cell [as was shown in Section~\ref{sec:mattlg}]. However, the magnitude of the DOS differs significantly between the two models. This is because there are two flat bands near the CNP in the simplified model, whereas in our full model, there is a large number of nearly overlapping flat bands due to incommensurability [Fig.~\ref{fig:ba_compare_bands_21}].
Figure~\ref{fig:ba_compare_bands_21} compares the band structure from our model and the simplified model. The two band structures are qualitatively similar but our model shows a large number of bands due to the lack of a periodic Brillouin zone. Furthermore, the aligned-bilayer approximation will exclude correlated phases that depend on band-hybridization or symmetries from the moir\'e of moir\'e length scale. Note that we do not plot the relative layer weights (color) of the band structure in the simplified model because of the way that the Brillouin zone is wrapped -- the L1 degrees of freedom are wrapped on top of the L3 degrees of freedom. Therefore, the wavefunction weights from the two models are not directly comparable for this particular high symmetry line cut.

We can use these results to further support our argument of bilayer moir\'e hybridization at equal twist angles. In this simplified model, sharpest VHS occur between $1.7^\circ$ and $1.8^\circ$, which is in better agreement with the magic angle prediction from perturbation theory. In our model, the sharpest peak and the narrowest width occurs at a larger angle ($2.1^\circ$). If this phenomenon is caused by moir\'e hybridization, the simplified model would not have it since it does not have the moir\'e of moir\'e scale. Indeed, the DOS from the two models differ most significantly at $2.1^\circ$ [see Fig.~\ref{fig:bernevig}(c)].

As we argue in the main text, adding electrons from the CNP at a low carrier concentration on the order of the tTLG moir\'e of moir\'e cells fills {\it one} flat band near the CNP in Fig.~\ref{fig:ba_compare_bands_21}(a) at a time. Injecting electrons at a carrier concentration comparable to the bilayer moir\'e cell density would fill {\it all} these flat bands near the CNP. The simplified model can again be used to understand this argument. The model also predicts some band flattening at certain twist angles, but there are only two flat bands near the CNP [Fig.~\ref{fig:ba_compare_bands_21}(c)]. Filling electrons to these two bands is equivalent to filling the bilayer moir\'e cell, since their momentum-space basis has the periodicity of bilayer moir\'e Brillouin zone and there is no moir\'e of moir\'e length in this model. These two flat bands near the CNP can be qualitatively considered as the limit where all the flat bands from our model overlap exactly on top of each other. Therefore, in terms of filling the supercell, filling the two flat bands from the simplified model is equivalent to filling all flat bands in the full model. 

In addition to its inability to make predictions about electronic behaviors at the moir\'e of moir\'e scale, another major limitation of the model is its difficulty to generalize to arbitrary twist angles. For each set of twist angles on a different $(m, n)$ harmonic, it requires the derivation of a new basis by folding the bilayer moir\'e Brillouin zone, while our model's basis is insensitive to the choice of angles and overcomes this limitation.
 
\begin{figure*}[ht!]
\centering
\includegraphics[width=0.7\textwidth]{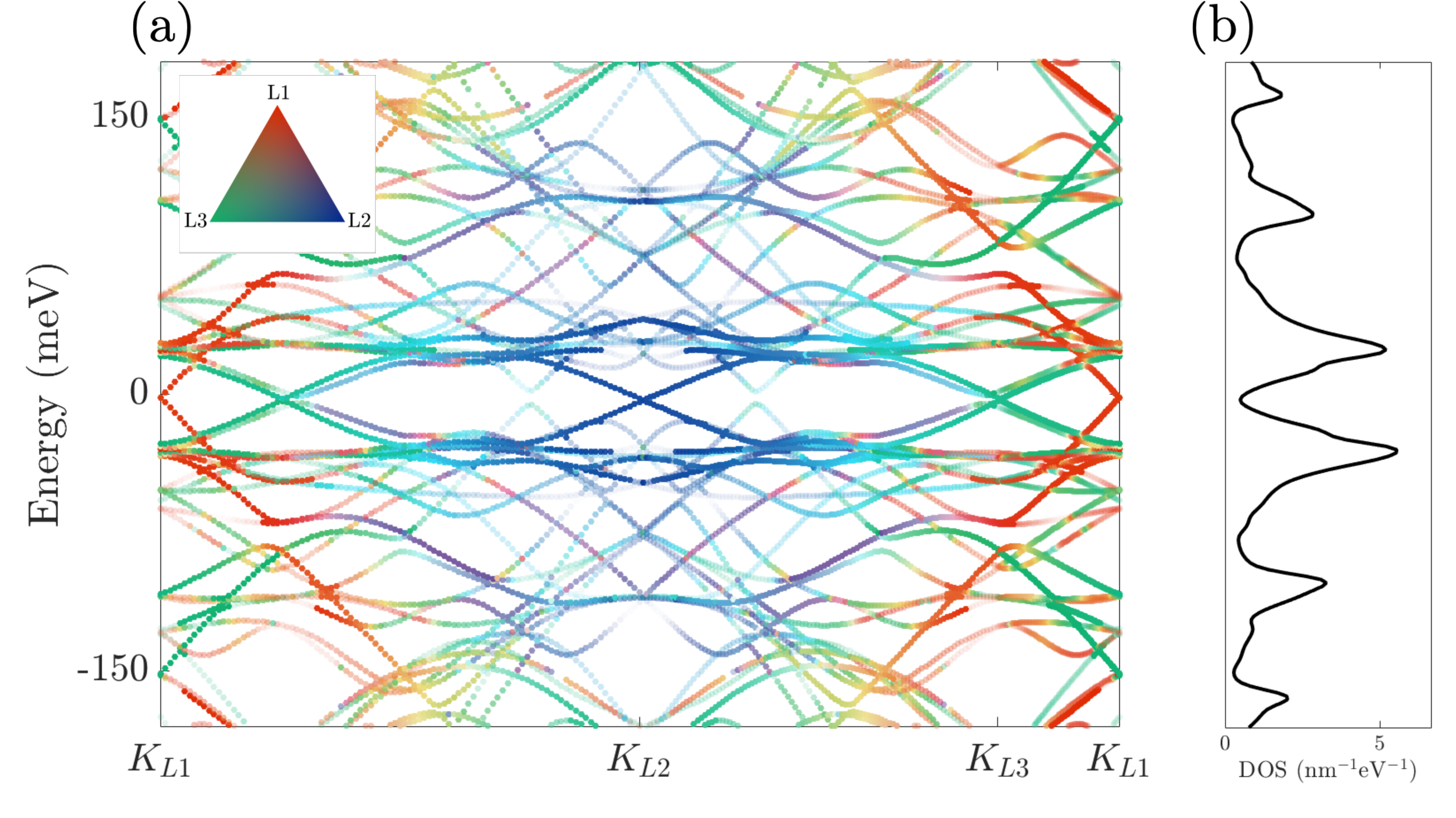}
\caption{(a) Band structure along a high symmetry line that connects the Dirac points of the three layers and (b) density of states at $\theta_{12}=-2.81^\circ, \theta_{23}=2.1^\circ$. The colormap in (a) is the same as in Fig.~\ref{fig:ba_compare_bands_21}(a).}
\label{fig:bands}
\end{figure*}

We can also use our model to study the case where L1 and L3 are twisted in the same direction (when $\theta_{12}$ and $\theta_{23}$ take opposite signs). This case has been studied theoretically by~\citet{amorim2018electronic} and its spectral properties have been investigated experimentally by~\citet{zuo2018scanning}. Unlike our model, \citet{amorim2018electronic} does not take the low-energy limit [see Section~\ref{sec:low_energy}]. Figure~\ref{fig:bands} shows the band structure and the corresponding DOS of $\theta_{12}=-2.81^\circ, \theta_{23} = 2.1^\circ$ obtained with our model, which is the same case as Figs. 1(a) and 2 presented in~\citet{amorim2018electronic}. The results from the two models show an agreement, with the same VHS positions. The difference in the band structure can be most likely attributed to the different ways of truncating the momentum-space bases between the two models. 

\bibliography{references}